\documentclass[
twocolumn,  
secnumarabic,
amssymb, 
nobibnotes,
aps, 
prd,
superscriptaddress,
nofootinbib,
]{revtex4-1}
\usepackage{algorithm}
\usepackage{placeins}

\newcommand{\diff}{{\rm d}}

\usepackage[ddmmyyyy]{datetime} 
\usepackage{amsmath} 
\usepackage[dvipsnames]{xcolor}
\usepackage{cancel} 
\usepackage{graphicx}
\usepackage{comment} 
\usepackage{soul,ulem} 
\usepackage{eqnarray}
\usepackage{multirow}
\soulregister\cite7
\soulregister\ref7 
\soulregister\eqref7 
\usepackage{bbding}
\usepackage{pifont}
\usepackage{wasysym}
\usepackage{amssymb}
\usepackage{textcomp} 
\usepackage{marginnote} 
\usepackage[colorinlistoftodos
, textsize=tiny]{todonotes} 
\usepackage{textcomp} 
\usepackage{lineno} 
\usepackage{hyperref}
\usepackage{cleveref}
\hypersetup{
    colorlinks=true,
    linkcolor= [rgb]{0.2,0.8,0.5},
    citecolor= [rgb]{0.2,0.5,0.8},
    filecolor=magenta,      
    urlcolor=[rgb]{0.2,0.5,0.8},
    urlbordercolor=red
}

\newcommand*\jcap{J. Cosmology Astropart. Phys.}

\newcommand*\pasp{PASP}

\newcommand{\ksM}{\text{km/s Mpc$^{-1} $}}
\newcommand{\Mpc}{{\rm Mpc}}
\newcommand{\Mp}{{ M_{\rm P}}}

\newcommand{\kb}{k_{\rm B}}

\newcommand{\Seff}{\mathcal{S}{\text{eff}}}
\newcommand{\LCDM}{{\Lambda \rm{CDM}}}

\newcommand{\Om}{\Omega_{\rm m}}
\newcommand{\Omzero}{\Omega_{\rm m0}}
\newcommand{\Or}{\Omega_{\rm \gamma}}
\newcommand{\Orzero}{\Omega_{\rm \gamma 0}}
\newcommand{\Ode}{\Omega_{\rm HDE}}
\newcommand{\Ogt}{\Omega_{\rm GT}}

\newcommand{\Hrec}{H_{\rm rec}}
\newcommand{\rd}{r_{\rm d}}

\newcommand{\Hzero}{H_{0}}
\newcommand{\wzero}{w_{0}}
\newcommand{\wa}{w_{\rm a}}
\newcommand{\AP}{A_{\rm P}}

\newcommand{\rhoDE}{\rho_{\rm HDE}}
\newcommand{\rhoGT}{\rho_{\rm GT}}
\newcommand{\rhorad}{\rho_{\rm \gamma}}
\newcommand{\rhom}{\rho_{\rm m}}
\newcommand{\Prad}{P_{\rm \gamma}}
\newcommand{\Pm}{P_{\rm m}}
\newcommand{\PDE}{P_{\rm DE}}
\newcommand{\PGT}{P_{\rm GT}}
\newcommand{\Rh}{R_{\rm h}}
\newcommand{\DL}{D_{\rm L}}
\newcommand{\Cp}{C}

\newcommand{\wDE}{\omega_{\rm HDE}}
\newcommand{\wGT}{\omega_{\rm GT}}

\newcommand{\DeltaBE}{\log{\mathcal{B}}}
\newcommand{\chibf}{\Delta \chi^{2}_{\rm b.f}}
\newcommand{\Panp}{Pan$^+$}

\newcommand{\ta}[1]{{\color{red} #1}}

\begin{document}

\title{ Holographic and Gravity-Thermodynamic Approaches in Entropic Cosmology: Bayesian Assessment using late-time Data}

\author{Udit K. Tyagi}
\email{udit19@iisertvm.ac.in}
\affiliation{School of Physics, Indian Institute of Science Education and Research Thiruvananthapuram, Maruthamala PO, Vithura, Thiruvananthapuram 695551, Kerala, India}

\author{Sandeep Haridasu}
 \email{sharidas@sissa.it}
 \affiliation{SISSA-International School for Advanced Studies, Via Bonomea 265, 34136 Trieste, Italy}
 \affiliation{IFPU, Institute for Fundamental Physics of the Universe, via Beirut 2, 34151 Trieste, Italy}
 \affiliation{INFN, Sezione di Trieste, Via Valerio 2, I-34127 Trieste, Italy}

\author{Soumen Basak} 
\email{sbasak@iisertvm.ac.in}
 \affiliation{School of Physics, Indian Institute of Science Education and Research Thiruvananthapuram, Maruthamala PO, Vithura, Thiruvananthapuram 695551, Kerala, India}
 
\begin{abstract}

{We investigate the cosmological implications of entropy-based approaches in the context of Holographic Dark Energy (HDE) and Gravity-Thermodynamics (GT) formalisms. We utilise the extended Barrow entropy form, with the index parameter $\Delta$, representing the fractal dimension of the horizon. We also test implementing different parameter ranges for $\Delta$, which can be extended to Tsallis' interpretation within the same formal cosmology. We perform a Bayesian analysis to constrain the cosmological parameters using the Pantheon+, more recent DESy5, DESI datasets. We find that the HDE model within almost all data combinations performs extremely well in comparison to the GT approach, which is usually strongly disfavored. Using the combination of DESy5+DESI alone, we find that the GT approaches are disfavored at $|\log \mathcal{B}| \sim 5.8$ and $|\log \mathcal{B}| \sim 6.2$ for the Barrow and Tsallis limits on $\Delta$, respectively,  wrt $\Lambda$CDM model. While the HDE approach is statistically equivalent to $\Lambda$CDM when comparing the Bayesian evidence. We also investigate the evolution of the dark energy equation of state and place limits on the same, consistent with quintessence-like behaviour in the HDE approaches.} 

\end{abstract}

\maketitle


\section{Introduction}\label{sec:Intro}
The discovery of the universe's accelerated expansion through supernova (SNe) observations \cite{Riess98} significantly transformed cosmological parameter estimation, necessitating the inclusion of the cosmological constant $\Lambda$. The $\Lambda$ Cold Dark Matter ($\LCDM$) model emerged as the concordance cosmology after surviving numerous subsequent scrutinies using updated SNe datasets \cite{Betoule:2014frx, Scolnic17, Scolnic:2021amr, Haridasu17a, DES:2024tys, Dam:2017xqs, Rubin:2016iqe} {and successfully explained} the Cosmic Microwave Background (CMB) \cite{Planck:2015lwi, Planck:2018vyg}. Despite its success, $\LCDM$ presents significant challenges in achieving a physical understanding, especially given unresolved tensions in parameter estimates. The most notable of these is the $H_0$ tension, which has reached a significance greater than $5\sigma$ \cite{Verde:2019ivm, Riess:2021jrx, Riess:2019qba} (see also, \cite{Riess:2019qba, Perivolaropoulos:2021jda, Efstathiou:2020wxn, Freedman:2021ahq, Verde:2023lmm, Abdalla:2022yfr}).  {Note however, that there are other measurements of $\Hzero$, either with improved methodology or different combinations of datasets finding tension of lesser significance \cite{Dominguez:2019jqc, Park:2019emi, Lin:2019zdn, Freedman:2020dne, Birrer:2020tax, Boruah:2020fhl, Wu:2021jyk, Cao:2022ugh, Chen:2024gnu, Efstathiou:2020wxn, Freedman:2021ahq}. }


This highlights the necessity to go beyond the standard model, with one promising avenue being the utilization of cosmological horizons. These are inspired by studies on black hole event horizons and their thermodynamic characteristics \cite{Wang:2016och, PhysRevD.83.123006, Saridakis_2018, Nojiri:2021iko, Nojiri:2005pu, Nojiri:2017opc}, a concept known as the holographic approach based on the Holographic Principle \cite{Wang:2016och, tHooft:1993dmi, Susskind:1994vu}. It is now widely recognized that black hole systems exhibit behaviors reminiscent of thermodynamic systems \cite{Bekenstein:1993dz, Nappi:1992as, Carlip:2014pma, Brown:1994sn, Oda:1994np, Ghosh:1994nz, Frolov:1994zi, Fursaev:1994pq, Louko:1994tv, Liberati:1996kt, Solodukhin:1996vx, Carlip:1996xa, Frolov:1996wd, Quevedo:2024fga}. This resemblance is encapsulated by the laws governing black hole thermodynamics, which include concepts such as the Bekenstein entropy and the Hawking temperature \cite{PhysRevD.7.2333, Bekenstein:1972tm, Hawking:1974rv, Hawking:1975vcx}.

The Bekenstein entropy, often termed ``area entropy'', is directly proportional to the area of a black hole's horizon and, thus, its radius squared \cite{PhysRevD.7.2333}. The holographic approach draws an analogy between a black hole horizon and the apparent/future cosmological horizon \cite{Hsu:2004ri, Oliveros:2022biu, Bhardwaj:2022uhf, Astashenok:2022pni, Astashenok:2023jfp, Tita:2024jzw} of the universe, suggesting that all the information contained within the universe is encoded on its horizon, akin to the event horizon of a black hole \cite{Saridakis_2018, Salzano:2017qac}.

Another intriguing approach to understanding cosmic acceleration and possibly resolving tensions draws from gravity-thermodynamics formalism \cite{Saridakis_2020}\cite{Leon_2021}, which is once again based on the foundational concepts in black hole physics developed by Hawking and Bekenstein \cite{Bekenstein:1972tm, Bekenstein:1993dz}. In this framework, we obtain the expansion history of the Universe implementing a thermodynamic perspective\cite{Cai:2005ra}, leveraging the relationship between entropy and horizon area, along with assumptions of local equilibrium conditions \cite{Asghari:2024sbu}. 

{In order to build cosmological models from the above-mentioned approaches, we can use different non-extensive entropy formulations. The most common and standard of these is the Bekenstein-Hawking entropy \cite{Bekenstein:1972tm, Bekenstein:1993dz, Hawking:1975vcx, PhysRevD.13.191, Hawking:1974rv}} given as,
\begin{equation}\label{eqn:S_BH}
S_{\rm BH} = \frac{A}{4 G \hbar}
\end{equation}
where A represents the area, $G,\,\hbar$ are the gravitational constant and reduced \textit{Planck} constant, respectively. 

{We use non-extensive entropies as in systems with long-range interactions such as Newtonian gravity, the standard Gibbs-Boltzmann entropy cannot be used \cite{Tsallis_2013} because, for such systems, the partition function diverges. Similarly, in the case of correlated systems such as strongly entangled particles and black holes (systems with entropy area law), the standard Gibbs extensive entropy does not apply. In these systems, the entropy is proportional to $L^{d-1}$ rather than $L^d$ (where $d$ represents the dimension of the system).} Inspired by the generalizations of Gibbs-Bekenstein entropy, generalizations of Bekenstein-Hawking entropy were put forward mainly to make the black-hole entropy additive while still keeping it non-extensive. 
Examples of such generalized non-extensive entropies include Tsallis entropy \cite{tsallis1988possible}, Rényi entropy \cite{Rényi}, Barrow entropy \cite{Barrow:2020tzx}, Sharma-Mittal entropy \cite{sharma1975new, sharma1977new}, and Kaniadakis entropy \cite{Kaniadakis:2005zk}. {See \cite{Nojiri:2022aof, Nojiri:2023bom} for a few generalized entropy forms and \cite{Ourabah:2023rsm}, where the entropy is derived starting with the gravitational potential.} Applying these entropy formulations gives rise to various so-called `Entropic Cosmology' models described and studied in \cite{Odintsov:2023vpj, Leon_2021, Naeem:2023tcu, Saridakis_2018, Nojiri:2022aof, Nakarachinda:2023jko, Fazlollahi:2022bgf, KordZangeneh:2023syq, Salehi:2023zqg, Nojiri:2021jxf, Nojiri:2019skr, Nojiri:2022dkr, Odintsov:2022qnn,Nojiri:2022nmu,Nojiri:2023nop, Nojiri:2023wzz,Manoharan:2024thb, Manoharan:2022qll}. However, our work primarily focuses on exploring Tsallis and Barrow entropies. Tsallis entropy was postulated to incorporate the concept of `multifractals' by scaling the probability of occurrence of microstates by a factor, while Barrow entropy was conceptualized to account for the effects of `quantum gravity spacetime foam' on the existing horizon structure \cite{Barrow_2020}. Although the physical principles underlying these two entropies are completely different, their formalisms are quite similar, with the only difference being the allowed ranges for a particular parameter (denoted as $\Delta$).

{Contrasting the formal implications of entropic gravity through both the holographic principle (HP) \cite{tHooft:1993dmi, Srednicki:1993im} implemented in \cite{Jacobson:2015hqa} and gravity thermodynamics in \cite{Jacobson:1995ab, Padmanabhan:2003gd}, \cite{Carroll:2016lku} have earlier concluded that the former is more appropriate than the latter, which fails to provide a self-consistent definition for entropy. In this work, we essentially contrast the application of HP to the dark energy problem using the formalism described in \cite{Li_2004, WANG20171}, and gravity-thermodynamics formalism described in \cite{Jacobson:1995ab, Cai:2005ra, Padmanabhan:2003gd, Padmanabhan:2009vy} and estimate the cosmological parameters by fitting the model against the cosmological observable that include recently published ``Pantheon+'', ``DESy5'' supernovae \cite{Scolnic:2021amr, DES:2024tys}, baryon acoustic oscillation ``DESI'' \cite{DESI:2024mwx} through Bayesian analysis. We constrain cosmological parameters and validate their consistency against the standard $\LCDM$ model. We further estimate the Bayesian evidence to determine which of the models is favored by the data as it is our primary objective to test which of the following approaches, namely Holographic or Gravity-Thermodynamics, is in better agreement with the data and how these two approaches fare against the $\Lambda$CDM. As a supplementary analysis, we also assess the dark energy equation of the state and its dynamic behavior within these extended models. }

The paper is structured as follows: \Cref{sec:modelling} delves into the cosmological modeling, \Cref{sec:HDE} and \Cref{sec:GT} address the Barrow Holographic Dark Energy and Barrow gravity-thermodynamics approaches, respectively. The observational dataset is presented in \Cref{sec:Data}, followed by our methodology and results in \Cref{sec:Results}, and conclusions in \Cref{sec:Conclusions}. Unless otherwise mentioned, we write all expressions in natural units $\kb = c = \hbar = 1$. 

\section{Modelling }\label{sec:modelling}
{In this section}, we briefly describe the two entropy-based approaches, namely the Holographic and gravity-thermodynamics approaches. For this, we utilize the Barrow entropy formalism \cite{Barrow:2020tzx}, which is mathematically equivalent to the Tsallis entropy formalism\cite{Tsallis_2013} as both formalisms share the same parametric form, however, they allow for varied parameter spaces and have different physical interpretations.

The Barrow entropy proposal for the black hole surface area is based on introducing a fractal structure for the horizon geometry. This model consists of a three-dimensional spherical analog of a 'Koch Snowflake', employing an infinite diminishing hierarchy of touching spheres around the Schwarzschild event horizon. Through this method, a fractal structure for the horizon is created with finite volume and infinite area \cite{BARROW2020135643,Denkiewicz:2023hyj,Nojiri:2022aof}. Consequently, this model predicts an entropy different from Bekenstein-Hawking, where the entropy is formalized as

\begin{equation}
\label{eqn:SB}
S_{\rm B} = \left(\frac{A}{\AP}\right)^{1 + \frac{\Delta}{2}}
\end{equation}

Here, $A$ is the horizon area, $\AP$ is the Planck area, and $\Delta$ is the free parameter representing the fractal dimension, bounded by $0 \le \Delta \le 1$. We keep the introduction to the modeling brief as it has been studied and discussed extensively \cite{Odintsov:2023vpj,Leon_2021,Naeem:2023tcu, Saridakis_2018,Nojiri:2022aof,Nakarachinda:2023jko,Fazlollahi:2022bgf,KordZangeneh:2023syq,Salehi:2023zqg}. 

\subsection{Holographic dark energy}
\label{sec:HDE}

{According to the Holographic Principle, the dark energy density of the Universe is given by $\rhoDE = \mathcal{S}_{\text{eff}} L^{-4}$ \cite{WANG20171, Li_2004}, where $\mathcal{S}_{\text{eff}}$ is the effective entropy and $L$ is the cosmological length scale, which can be interpreted as either the Hubble horizon, particle horizon, or future horizon. The dark energy density for the Barrow HDE model (using \cref{eqn:SB}), can be expressed as
\begin{equation}
\label{rho_DE}
\rhoDE = B L^{\Delta - 2},
\end{equation}
where $B$ is a free parameter with dimensions $[L]^{-(\Delta + 2)}$. In the limiting case, where $\Delta = 0$, this formalism reduces to the standard Holographic Dark Energy (HDE) \cite{WANG20171, Li_2004},
\begin{equation}\label{B}
\rhoDE = 3 C^{2} \Mp^{2} L^{-2}, \qquad B = 3 C^{2} \Mp^{2}.
\end{equation}
}

{While the choice of the cosmological length scale is still an open issue, it has been argued in\citep{Hsu_2004} that Hubble horizon ($H^{-1}$) cannot be used in the Holographic Dark energy case as it gives rise to inconsistencies, see also \cite{Davis:2003ad,Dabrowski:2020atl}. Following the approaches used in\cite{Li_2004, Denkiewicz:2023hyj, Salzano:2017qac} we utilize the future horizon as the cosmological length scale, which is given as,}
\begin{equation}
\label{eqn:future_horizon}
\Rh = a \int_{t}^{\infty} \frac{\diff t}{a} \equiv a \int_{a}^{\infty} \frac{\diff a}{H a^2}
\end{equation}
Consequently, the dark energy density $\rhoDE$ can now be written as 
\begin{equation}
\label{eqn:rhoDE}
\rhoDE = B \Rh^{\Delta-2}.
\end{equation}

The Friedmann and Raychaudhuri equations in a flat universe with dark energy($\rhoDE$), matter($\rhom$), and radiation($\rhorad$) are, 
\begin{eqnarray}
\label{eqn:Friedmann}
H^2 &=& \frac{1}{3 \Mp^2}\left(\rhom + \rhoDE + \rhorad\right) \\
-\dot H &=& \frac{1}{ 2 \Mp^2}\left( \rhom + \rhoDE + \PDE + \rhorad + \Prad \right)
\end{eqnarray}
where $P_{\rm I =\{HDE, \gamma\}}$ are the pressure terms for dark energy and radiation, respectively. We have already assumed pressureless matter ($\Pm = 0$). The radiation pressure and energy density are related as $P_{r} = \frac{\rho_{r}}{3} $. One can now introduce the convenient fractional energy density parameters as $\Omega_{\rm i} = \frac{\rho_{\rm i}}{3 \Mp^2 H^2} $,  where $\rm{i \in \{m, HDE, \gamma\}}$ can be substituted for matter, dark energy, and radiation components, respectively. Substituting $\rhoDE$ from \cref{eqn:rhoDE} for the fractional energy density and using the future horizon definition in \cref{eqn:future_horizon} yields,
\begin{equation}
\label{eqn:integral}
a \int_{a}^{\infty} \frac{\diff a}{H a^{2}} = \left( \frac{3 \Mp^2 H^2 \Ode}{B} \right)^\frac{1}{\Delta-2}
\end{equation}
Finally, $\rhom$ = ${\rho_{\rm m 0}}/{a^3}$ and $\rhorad$ = ${\rho_{\rm{ \gamma 0}}}/{a^4}$, with $\rho_{\rm m 0}$ and $\rho_{\rm \gamma 0}$  as the value of the matter and radiation-energy density at the present epoch $(a = 1)$. Therefore, $\Om = \Omzero \Hzero^2 /{a^3 H^2} $ and $\Or = \Orzero \Hzero^2 /{a^4 H^2} $, where the closure condition ensures $\Omzero + \Ode(a=1) + \Orzero= 1$. Using the definitions of $\Om$, $\Or$ as mentioned above and the closure condition, we can write the expansion rate as,

\begin{equation}
\label{eqn:Hubble}
H(a) = \Hzero \sqrt{\frac{\Omzero a^{-3} + \Orzero a^{-4} }{1-\Ode(a)}}
\end{equation}
Utilizing \cref{eqn:Hubble} in \cref{eqn:integral} and differentiating both sides w.r.t $a$ gives the evolution of $\Ode$ as,

\begin{align}
  \label{eqn:dOmegadeda}
  \begin{split}
  \frac{\Ode'(a)}{[1 - \Ode(a)]} &= \frac{\left(2 + {\Delta}\right)\Orzero a^{-4} + (1 + \Delta)\Omzero a^{-3}}{\Omzero a^{-3} + \Orzero a^{-4}}  +\\ & \left[1 - \Ode(a)\right]^{\frac{\Delta/2}{\Delta - 2}}\Ode(a)^{\frac{1}{2- \Delta}} Q(a)
  \end{split}
\end{align}
where,
\begin{align*}
\label{eqn:F_r}
Q(a) &= 2 \left(1 - \frac{\Delta}{2}\right)\left(\Hzero \sqrt{\frac{\Omzero}{ a^{3}} + \frac{\Orzero}{a^{4}}}\right)^{\frac{\Delta}{2 - \Delta}} \left(\frac{B}{3 \Mp^2}\right)^\frac{2}{{\Delta} -2},
\end{align*}
and $\Ode'(a) = \diff \ln(\Ode)/\diff \ln a$. Solving the above \cref{eqn:dOmegadeda} provides us with the expansion rate in \cref{eqn:Hubble}. Establishing the Hubble rate, one can now formulate the {dark energy} Equation of State (EoS), $\wDE = {\PDE}/{\rhoDE} $ for the Holographic Dark energy using the conservation law, 
\begin{equation}
\dot \rho_{\rm HDE} + 3 H \rhoDE(1 + \wDE) =  0, 
\end{equation}
where,
\begin{equation}
\label{eqn:rho_dot}
\dot \rho_{\rm HDE} = B(\Delta-2)\left(\frac{\rhoDE}{B}\right)^{\frac{\Delta-3}{\Delta-2}}\left[H\left(\frac{\rhoDE}{B}\right)^{\frac{1}{\Delta-2}} - 1\right]
\end{equation}
we obtain the EoS parameter as 
\begin{equation}{\label{EoS}}
1+\wDE = \left[\frac{B(2-\Delta)}{3 H \rhoDE}\left(\frac{\rhoDE}{B}\right)^{\frac{\Delta-3}{\Delta-2}}\left[H\left(\frac{\rhoDE}{B}\right)^{\frac{1}{\Delta-2}} - 1\right]\right]
\end{equation}
and given the values of $B$ (\cref{B}), $\rhoDE$ (\cref{eqn:dOmegadeda}) and $H$ (\cref{eqn:Hubble}), one can easily evaluate the DE EoS within the current Holographic dark energy approach. 

Note that the current formalism is a generalization of the standard Holographic Dark Energy model, assuming the future horizon as the cosmological length scale. {As aforementioned}, a similar approach can be taken for the Hubble or particle horizon as well, whose implications have been discussed in \cite{Li_2004, Hsu:2004ri,Pavon:2005yx}. In our current implementation, we remain with the standard assumption. 

\subsection{Gravity-Thermodynamics }
\label{sec:GT}

In the gravity-thermodynamic approach, one can derive the Friedmann equations from entropy formulations by using the first law of thermodynamics \cite{Jacobson:1995ab}. It is interesting to note that by using the standard Bekenstein-Hawking entropy (\cref{eqn:S_BH}) we obtain the standard $\LCDM$ model, where $\Lambda$ emerges as an integration constant \cite{Cai:2005ra, Padmanabhan:2003gd, Padmanabhan:2009vy}. {As shown in \cite{Nojiri:2021iko, Nojiri:2022aof, Leon_2021}, this formalism can be easily extended to generalized entropy functions to obtain modified Friedmann equations. In this context, aside from naturally having the cosmological constant at late times, one can obtain additional non-zero corrections to the early-time radiation epoch \cite{Denkiewicz:2023hyj}.} Contrary to the Holographic approach, within this formalism, we choose Hubble horizon as the cosmological length scale to maintain consistency with the previous works done in the gravity-thermodynamics approach \cite{Nojiri:2022aof,Leon_2021}, 
\begin{equation}
\Rh = \frac{1}{H}
\end{equation}
In order to derive the Friedmann equation for this approach, we first calculate the change in energy ($\diff E/\diff Q$) contained in the region within the cosmological horizon, 
\begin{equation}{\label{dQ}}
\diff Q = -\diff E = -\frac{4 \pi}{3} \Rh^{3} \dot \rho \diff t 
\end{equation}
where $\rho$ represents the energy density. Further dQ can also be obtained from the first law of thermodynamics as,
\begin{equation}
  {\label{eqn:first_law_ther}}
  \diff Q = T \diff \Seff
\end{equation}
where $T$ is the Gibbons-Hawking temperature \ta{\cite{Hawking:1974rv}} and is defined as $T = \frac{\displaystyle{1}}{\displaystyle{2 \pi \Rh}}$.
$\Seff$ in this case corresponds to \cref{eqn:SB}. 
Substituting the above expressions in \cref{eqn:first_law_ther} and further simplifying it gives,
\begin{equation}\label{i1}
-\frac{4 \pi}{3}\dot \rho  = \frac{1}{\pi \Rh^{3}}\left(1+\frac{\Delta}{2}\right)\left(\frac{4 \pi}{\AP}\right)^{1+\frac{\displaystyle{\Delta}}{\displaystyle{{2}}}} \Rh^{\Delta}\dot \Rh
\end{equation}
Integrating both sides of \cref{i1} and rearranging the terms gives us a modified Friedmann equation.
\begin{equation}
  {\label{eqn:gt-friedmann}}
\begin{split}
H^{2} &= \frac{1}{3 \Mp^2}\left(\rhom + \rhorad + \rhoGT \right) , \\ 
\rhoGT &= \frac{3}{\Mp^{-2}}\left[\frac{\Lambda}{3} + H^{2}\left(1 - \left(\frac{\pi}{G}\right)^{\frac{\Delta}{2}} \left(\frac{2+\Delta}{2-\Delta}\right)H^{-\Delta}\right)
 \right]
\end{split}
\end{equation}

Following the work done in \cite{Nojiri:2022aof}, we have set $\AP = 4 G$ to derive the above equation. However, in our analysis, we leave it as a free parameter \cite{Leon_2021}. 

Inserting the expression of $\rhoGT$ into the Friedmann equation yields the following relation between the {density parameters implying the closure equation as},
\begin{equation}
\Om + \Or + \Ogt = 1
\end{equation}
where dark energy parameter \(\Ogt\) is given by, 
\begin{equation}
\Ogt = 1 + \Omega_{\Lambda} -\left(\frac{\pi}{G} \right)^{\frac{\Delta}{2}}\left(\frac{2+\Delta}{2 - \Delta}\right) H^{-\Delta}.
\end{equation}
In the limit $\Delta$ tends to $0$, we recover the equations in the standard \(\LCDM\) model. Similarly, the expression of the expansion rate can be written in terms of the density parameters at the present epoch as, 
\begin{widetext}
\begin{equation}
  {\label{eqn:Hubble_barrow_gravity}}
E(z) = \frac{H}{H_{0}} = \left[\beta\left(\frac{2-\Delta}{2+\Delta}\right)\left[ \Omzero (1+z)^{3} + \Orzero (1+z)^{4} + \Omega_{\Lambda}\right]\right]^{\frac{\displaystyle{1}}{\displaystyle{2-\Delta}}}
\end{equation}
\end{widetext}

where the parameter $\beta$ is defined as, 
\begin{equation}
\beta = H_0^{\Delta} {\left(\frac{G}{\pi}\right)}^{\frac{\displaystyle{\Delta}}{\displaystyle{2}}}
\end{equation}

Note that, from \cref{eqn:Hubble_barrow_gravity} the standard $\Lambda$CDM model is recovered for parameter values $\Delta = 0$ and $\beta = 1$. 
Using the Raychaudhuri equation, similar to the approach in \Cref{sec:HDE}, and substituting for energy densities and pressure terms (calculated using the conservation law  $\dot \rho + 3 H(\rho + P) = 0$) \cite{Leon_2021} one can determine the Dark energy EoS ($\wGT = {\PGT}/{\rhoGT}$) as,
\begin{equation}
\label{eqn:EoSGT}
1+\wGT =  \frac{2(1+z)E(z)E'(z)[1-(1+\frac{\Delta}{2})(E(z) \beta)^{-\Delta}}{3 \Omega_\Lambda + 3 E^{2}(z) \left[ 1 - \left(\frac{2+\Delta}{2-\Delta}\right) (E(z) \beta)^{-\Delta}\right]}
\end{equation}
where E'(z) is defined as the derivative of E wrt z. 

\subsection{Model interpretation}
\label{sec:model_interpretation}
We now discuss the difference in the interpretation between the two approaches and the corresponding limits of the parameter $\Delta$ for the two formalisms. Within the Barrow entropy formalism, the allowed range for the values of the index ($\Delta$) is strictly limited to $\Delta \in \{0, 1\}$, based on the physical reasoning of the fractal nature of the corrections on the surface of the black hole horizon. On the contrary, in the Tsallis interpretation, there exist no immediate limits on the values of $\Delta$ and can extend beyond the above-mentioned range \cite{Denkiewicz:2023hyj}. Therefore, for the sake of analysis, we assume a larger yet consistent range of parameter space $\Delta \in \{-3,3\}$, which encompasses the range allowed for the Barrow interpretation of the entropy. Note that these limits become extremely important from a model selection point of view, allowing one to assess the viability of the models in a Bayesian comparison. Needless to say, while the larger allowed parameter space in Tsallis' case can clearly yield a better fitting of the data in terms of the $\chi^2$ it cannot be taken for granted in terms of the Bayesian evidence, which takes into account the allowed prior volume that the data allows within the posteriors. Note also that arbitrarily increasing the prior volume in the case of Tsallis will penalize the model even more. Therefore, we remain with a conservative limit mentioned earlier, which scales the value of the index $S\sim A^{\delta}$\footnote{Note that $\delta$ is the usual notation utilized in writing the Tsallis entropy \cite{Tsallis_2013}. The values of $\delta\leq 0$ are already beyond physicality of the model, with $\delta\to 1$, retrieving the standard BH entropy in \cref{eqn:S_BH}.} as $\delta \in \{-0.5, 2.5\}$ (see \cref{tab:priors}).

Alongside the physical limits of the index, the current formalisms heavily rely on the assumption of the Horizon length scale. For instance, while the HDE approach usually assumes the future horizon following \cite{Li_2004}, it is also possible to obtain a consistent cosmology using the Hubble horizon \cite[see also \cite{Davis:2003ad}]{Pavon:2005yx, Dabrowski:2020atl}. Similarly, the use of Hubble horizon in the GT approach eases the formalism, providing simpler analytical formalism. In the current line of investigation, i.e., utilizing data to assess preference for different entropic approaches, it could be of utmost importance to assess the assumptions of the horizon as well, which we intend for dedicated future work. 

\section{Methodology and Data sets}\label{sec:Data}
{Our work employs Supernovae (SNe) and Baryon Acoustic Oscillations (BAO) data sets. The SNe dataset is currently the best representative of standard candles, while the BAO is the best representative of the standard rulers. Being low-redshift datasets, they are particularly effective at constraining dark energy, as the observations indicate that dark energy has only recently ($z\lesssim 0.6$) \cite{Haridasu18, Gomez-Valent18, Farooq:2013hq, Farooq:2016zwm, Yu:2017iju} begun to dominate the evolution of the Universe. }

{\bf\textit{SNe}}: { The luminosity distance of a distant source such as Type Ia supernovae is estimated for a given cosmological model assuming the FLRW metric for a flat Universe as,
\begin{equation}
\DL(z) = (1 + z) \frac{c}{H_0}\int_0^z \frac{\diff \bar{z}}{E(\bar{z})},
\end{equation}
where \(H\) is the expansion history of the universe, see \cref{eqn:Hubble,eqn:Hubble_barrow_gravity}. Given the luminosity distance \(\DL\), the corresponding distance modulus is given by $\mu(z) = m_{\rm B}(z) - M_{\rm B}$, where \(M_{\rm B}\) is the absolute magnitude of Type Ia supernovae and \(m_{\rm B}(z)\) is the apparent magnitude at the redshift $z$,

\begin{equation}
m_{\rm B}(z) = 5\log_{10}\left(\frac{\DL(z)}{1\, \Mpc}\right) + 25 + M_{\rm B},
\end{equation}

{The cosmological parameters are constrained by minimizing a \(\chi^2\) likelihood,
\[
-2 \ln \mathcal{L_{SN}} = \chi^2_{\mathcal{SN}} = \Delta \mathbf{\mu}^T (C_{\text{SN}})^{-1} \Delta \mathbf{\mu}
\]
where \(\Delta \mu\) is the difference between the observed distance modulus and theoretical distance modulus, and the corresponding covariance matrix \(C_{\text{SN}}\) of the measurements comprises both statistical and systematic errors.}

We utilize the well-established Pantheon+(\Panp) dataset \cite{Scolnic:2021amr} alongside the more recent DESy5\cite{DES:2024tys} dataset.} {\Panp dataset is a compilation of spectroscopically confirmed Type-${I \rm{a}}$ SNe, which has been widely utilized and well studied,} { and includes $1550$ SNe observed over the redshift range from z = 0.001 to $z = 2.26$. We adopt the publicly available likelihood presented in \cite{Brout:2022vxf} with a lower redshift bound of $z>0.01$\footnote{This limits the dataset to 1590 correlated measurements taking into account both the systematic and statistical uncertainties.}, to reduce the effect of peculiar velocities on the more local SNe (see \cite{Peterson:2021hel})}. {When utilising the \Panp data alone (corresponding to the first block of \cref{tab:table1}), we also impose a prior on the absolute magnitude $M_{\rm b} = -19.253 \pm 0.029$\footnote{We do not perform a complete SH0ES+\Panp analysis as it is equivalent to the simple $M_{\rm b}$ prior imposed here and is anyhow neither are suitable for the joint analysis with BAO data. }, which is equivalent to the Cepheid calibration presented in \cite{Scolnic:2021amr}.} On the other hand, the more recent DESy5 dataset presents the collection of supernovae from the five-year DES Supernova program. To classify these supernovae, the DES Supernova program uses a machine learning algorithm applied to their light curves in four photometric bands \cite{DES:2024haq}. In this survey, 1830 Type Ia supernovae were identified within a redshift range of $0.02\lesssim z \lesssim 1.13$. While the DESy5 SNe has not been scrutinized to the extent of the earlier \Panp dataset, we intend to update the constraints and draw parallels to the recent dynamical dark energy claims in \cite{DESI:2024mwx}. {It should be noted that in the joint analysis, both \Panp and DESy5\footnote{DESy5 dataset is provided by analytically marginalising upon the absolute magnitude calibration and cannot be utilised to infer $H_0$ \cite{DES:2024haq}. } are uncalibrated.}

{\bf\textit{BAO}:} { We utilize a compilation of the most recent BAO \cite{DESI:2024mwx} dataset for our analysis. 
{Dark Energy Spectroscopic Instrument (DESI) survey provides us with observables $D_H/r_d$ and $D_M/r_d$.
The observations are available across seven uncorrelated redshift bins utilizing over 6 million extra-galactic objects within the redshift range $0.1 < z < 4.2$, see Table 1. of \cite{DESI:2024mwx}. The likelihood for the BAO data, $\mathcal{L_{BAO}}$, is written taking into account both 
observables and their corresponding correlation at each redshift (please see Section 2.1 in \cite{DESI:2024mwx} for more details). }}{In summary, we use the uncalibrated BAO dataset, wherein the sound horizon $\rd$ remains a free parameter and is set/calibrated using the CMB priors, as elaborated in the following paragraph.} 

The inclusion of the Cosmic Microwave Background (CMB) data/priors, as was earlier done in \cite{Denkiewicz:2023hyj}, can immensely aid in parameter estimation. However, we refrain from using strong CMB priors 
such as the reduced CMB likelihood \cite{Wang06, Wang07}, which is heavily reliant on the assumed cosmology in obtaining the so-called shift parameters in the $\LCDM$ model. In contrast, we utilize less stringent priors on $\{\rd, \, \Hrec\}$\footnote{The value of $\Hrec$ is evaluated at recombination redshift of $z_{\rm rec} = 1089$, which is independent of most common late-time modifications \cite{Verde17}.} which are known to be independent of late-time cosmology \cite{Verde17}. We adopt the values and their corresponding covariance from \cite{Haridasu:2020pms}, which were obtained utilizing the \textit{Planck} 2018 likelihoods \cite{Planck:2018vyg}. We mainly intend this addition as the inverse distance ladder analysis of the BAO data, only mildly aiding the constraints on the late-time parameters {and providing a value of $\Hzero$.} Finally, we also assume a prior on the radiation density $\Orzero h^2 = 4.18343 \times 10^{-5}$, following present-day CMB temperature of $T_{\rm CMB} = 2.7255$ \cite{Fixsen09, Planck:2018vyg}.

A full Bayesian joint analysis is performed utilizing the publicly available \texttt{emcee}\footnote{\href{http://dfm.io/emcee/current/}{http://dfm.io/emcee/current/}} package \citep{Foreman-Mackey13} which implements an affine-invariant ensemble sampler. We analyse the generated MCMC samples using \texttt{corner}\footnote{\href{https://corner.readthedocs.io/en/latest/}{https://corner.readthedocs.io/en/latest/}} and/or \texttt{GetDist} \footnote{\href{https://getdist.readthedocs.io/}{https://getdist.readthedocs.io/}} \cite{Lewis:2019xzd} packages. Finally, we compute the Bayesian Evidence $\Delta \mathcal{B}$ \cite{Trotta:2017wnx, Trotta:2008qt, Haridasu17_bao}. For this purpose, we utilise \texttt{MCEvidence}\footnote{\href{https://github.com/yabebalFantaye/MCEvidence}{https://github.com/yabebalFantaye/MCEvidence}} \cite{Heavens:2017afc}. In our comparison, we utilise the convention that a positive value of $\Delta \log\mathcal{B} = \log\mathcal{B}_{\rm Ref=\LCDM} -\log\mathcal{B}_{\rm I}$  would imply that reference model ($\LCDM$) is preferred over model-I. As is the usual practice we refer to Jefferys' scale \cite{jeffreys1998theory, Kass95} (see also \cite{Trotta:2008qt, Nesseris:2012cq}) for assessing the strength at which a given model is preferred. In \cref{tab:priors} we summarize the priors utilised on the parameters within the Bayesian analysis. It is indeed the priors on the parameter $\Delta$ that distinguish the Barrow and Tsallis models, which we denote hereafter as Barrow Holographic dark energy (BHDE) or Barrow Gravity-Thermodynamics (BGT) and similarly THDE/TGT, respectively.

{\renewcommand{\arraystretch}{1.4}
\setlength{\tabcolsep}{6pt}
\begin{table}[h!]
\centering
\caption{Priors used in the Bayesian analysis.}
\label{tab:priors}
\begin{tabular}{ccc}
\hline
\textbf{Parameter} & Model & \textbf{Prior} \\
\hline
\hline
$\Omzero$ & & $[0.1, 0.5]$ \\
$\Hzero$ & & $[40.0, 100.0]$ \\
\multirow{2}{*}{$\Delta$} & Tsallis &$[-3.0, 3.0]$ \\
 & Barrow & $[0.0, 1.0]$ \\

$\Cp$\footnote{HDE normalization parameter, see \cref{sec:Prior_effects} for a discussion on the effects of the priors on $\Cp$.} & & $[0.0, 10.0]$ \\
$\beta$ & & $[0.5, 1.5]$.\\
$\rd$ & & $[125, 160]$.\\
$M_{\rm b}$ & & $[-20.5, -18.5]$.\\
\hline
\end{tabular}
\end{table}
}

\begin{figure*}
\label{fig:HDE_GT}
    \centering
    \includegraphics[scale=0.42]{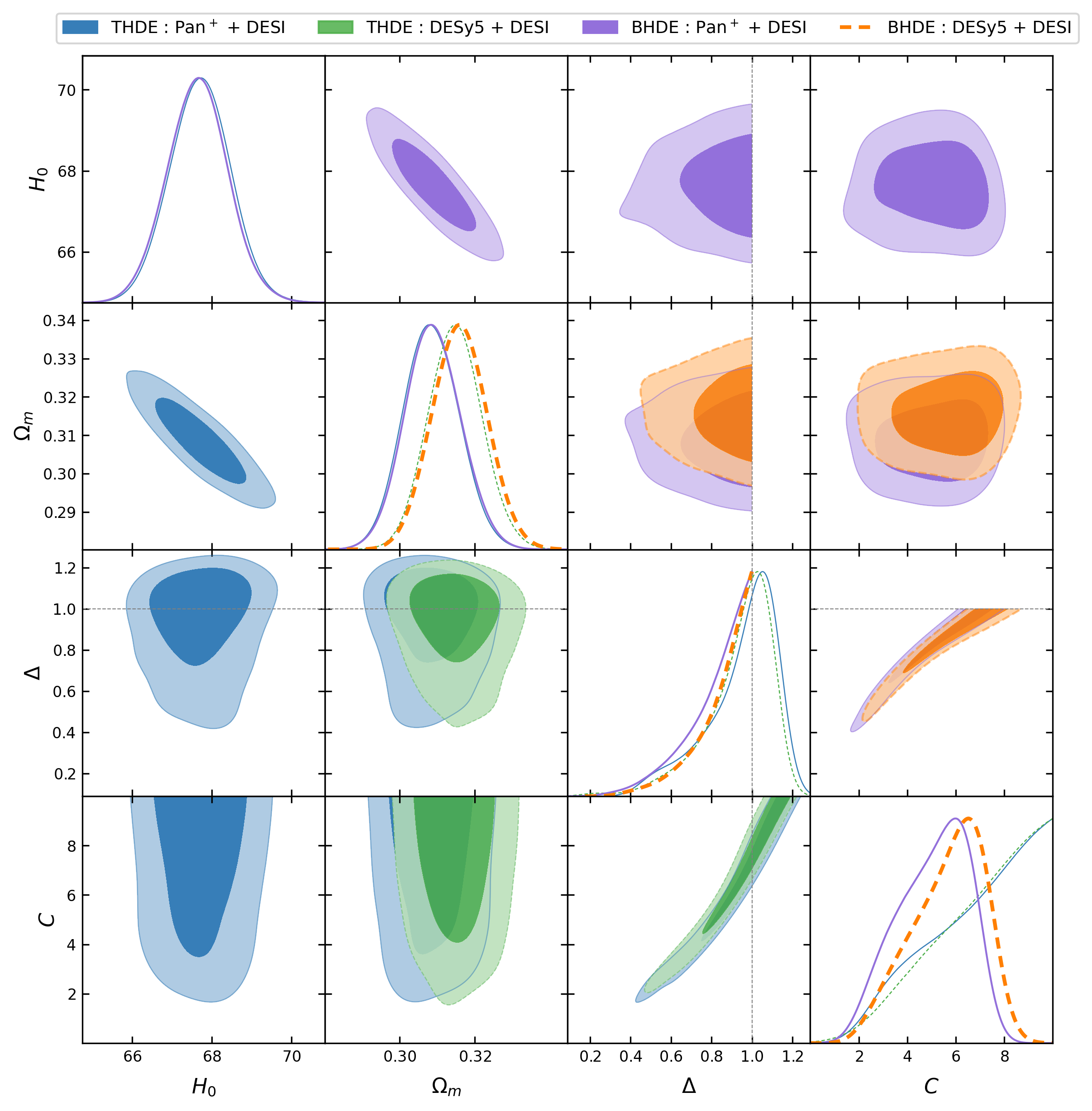}
    \includegraphics[scale=0.42]{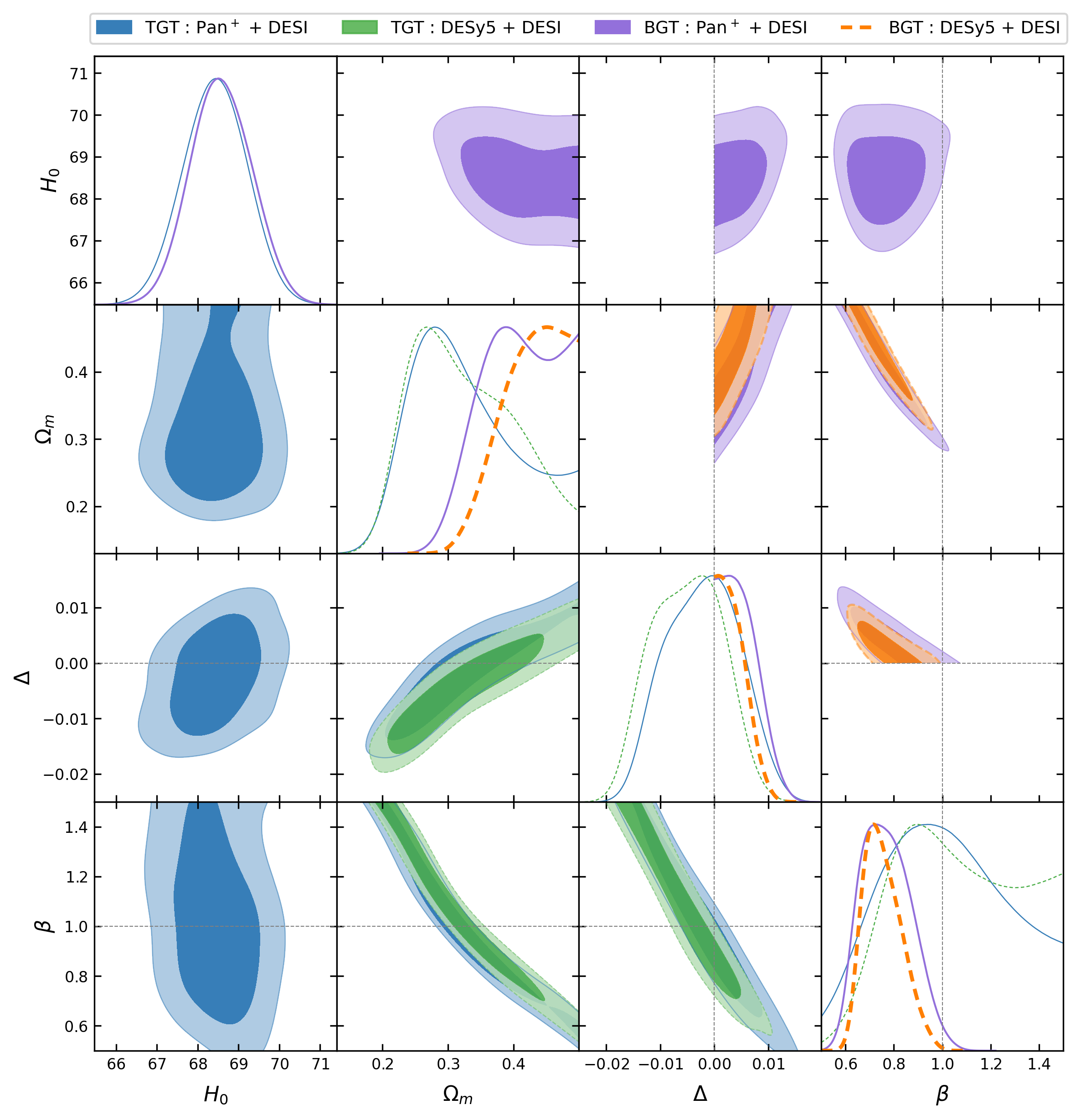}
    \caption{Contours for the \(68\%\) and \(95\%\) confidence limits on the Holographic Dark Energy approach (\textit{left}) and the Gravity Thermodynamics approach (\textit{right}), for all data combinations under considerations. The upper and lower triangle plots show the constraints when imposing the Barrow and Tsallis limits on the parameter \(\Delta\), respectively.
}
    \label{fig:HDE_GT}
\end{figure*}

\section{Results and discussion}\label{sec:Results}
We begin by presenting our results for the constraints obtained for the extended models and then proceed to discuss the model selection based on Bayesian evidence. As we have described in the \Cref{sec:modelling}, the essential difference between the Barrow or the Tsallis interpretation of the functional model is solely the prior ranges which define the physical viability of the model. 

\subsection{Constraints}\label{sec:cons}
All the parameter constraints are summarized in \Cref{tab:table1}, which are obtained utilising combinations of Pan$^+$ and DESy5 datasets, respectively. Although we present the results using both \Panp and DESy5 the supernovae compilations, we mainly refer to the former for our final results, while utilising the latter for a comparative discussion as it is the most recent and latest dataset available.  We first discuss the results obtained using the Holographic principle, followed by a discussion of the same using gravity-thermodynamics formalism, before presenting a comparative analysis.

{\bf\textit{Holographic principle:}} We find that our constraints are consistent with the limits quoted in earlier works for the BHDE \cite{Saridakis_2020, Denkiewicz:2023hyj, Dabrowski:2020atl}. Our results also align with other works in the context of THDE \cite{Saridakis_2018, Mendoza-Martinez:2024gbx, DAgostino:2019wko}, where the HDE normalization parameter \(\Cp = 1\) \footnote{Note that there is no immediate physical argument to fix \(\Cp = 1\).} is sometimes assumed. As seen in \cref{fig:HDE_GT}, when \(\Cp = 1\) is imposed, our results converge to the constraints presented in these works, providing extremely tight constraints on the index parameter \(\Delta\). Note that our constraints are slightly less stringent, being \(\Delta > 0.48\) at \(2\sigma\) C.L., compared to \(\Delta > 0.63\) \cite{Denkiewicz:2023hyj}, as we only use SNe and BAO datasets (even with the more recent DESI) and do not include very strong CMB constraints. However, it is interesting and validating that the inclusion of additional datasets {in \cite{Denkiewicz:2023hyj}} such as Cosmic Chronometers \cite{Moresco16a, Moresco16} and Strong lensing datasets \cite{Birrer:2018vtm} only mildly aid the joint constraints, while the latter prefers a slightly larger value of \(\Omzero\).

However, the above interpretation of consistency comes with the caveat that the posteriors are subject to the assumed prior volume. This is evident as the sampling effects\footnote{Such sampling effects are common in cases where the posterior tends to converge to the limits of the prior volume, where the MCMC-based sampling tends to provide incorrect confidence limits. Please see \cref{sec:Prior_effects} for a detailed discussion.} on the parameter space when the prior volume is increased to that of the Tsallis bounds. On the other hand, we find an upper limit on \(\Delta\) completely driven by the upper limit of \(\Cp < 10\). In turn, we find that \(\Delta = 0\) is a disfavored scenario within both the Barrow and Tsallis priors, as presented in \cite{Denkiewicz:2023hyj}, thus deviating from the standard Holographic dark energy \cite{Li_2004}. It should be noted that only the BAO datasets are able to provide strong limits on \(\Delta\), being constrained by the combination of both distance and expansion rate.

The constraints obtained on the parameters \(\{\Hzero, \Omzero\}\) are completely consistent with those obtained in the \(\Lambda\)CDM model. While the SNe datasets alone are unable to constrain the matter density parameter \(\Omzero\), the BAO dataset along with the inverse distance ladder priors from CMB is able to provide a tight constraint on the parameter. Additionally, the inclusion of SH0ES calibration (\(M_{\rm b}\)-prior) to the SNe dataset or the \(\rd\) prior to the BAO dataset does not influence the constraints on the holographic entropy parameters. Interestingly, the HDE approach is able to solve the \(\Hzero\)-tension \cite{Riess:2021jrx} while satisfying the inverse distance ladder prior, aided by mildly lower values of \(\Omzero \sim 0.28\). However, this is disfavored when the SNe is included in the analysis, showing no advantage in solving the \(\Hzero\)-tension. Furthermore, the most recent dataset combinations of DESI+DESy5 and DESI+Pan\(^+\) show no difference in the constraints on the HDE parameters. However, the inclusion of the most recent DESy5 SNe dataset provides strong implications for model selection, which is indeed the primary focus of our investigation.

{In this context, it is interesting to note that earlier \cite{Colgain:2022nlb}, have shown that there are discrepancies within the constraints of the $\Delta$ parameter when estimated using SNe and BAO datasets, essentially when $\Cp =1$ is fixed. }

{\bf\textit{Gravity-Thermodynamics}:} 
The joint analysis indicates that the cosmological parameters \(\Hzero\) and \(\Omzero\) are consistent with the concordance cosmology. However, there is a strong degeneracy between the matter density (\(\Omzero\)) and the normalization parameter \(\beta\), as seen in the right panel of \cref{fig:HDE_GT}. The range of \(\beta > 1.0\) prefers low values of \(\Omzero\), and the opposite is observed for \(\beta < 1\). In comparison to the results presented in \cite{Leon_2021}, where strong CMB priors are imposed on \(\Omzero\) to obtain strong constraints on \(\beta\), we do not find such constraints in our analysis. As anticipated in \cref{sec:GT}, \(\Delta \rightarrow 0\) and \(\beta \to 1\) indicate that the model converges to \(\Lambda\)CDM cosmology. In our analysis, we find that \(\beta \leq 1\) at least at a \(\sim 2\sigma\) C.L. when \(\Delta \to 0\), indicating a deviation from the $\LCDM$ model, which is also reflected in the Bayesian evidence (see \cref{sec:BE}). This deviation is evident in the constraints with \(\beta \sim 0.75\), as seen in \cref{tab:table1} and the upper triangle of the left panel in \cref{fig:HDE_GT}, when the Barrow limits are adopted.  When the limits on \(\Delta\) are extended to the Tsallis model, \(\beta \sim 1\) does obtain a posterior peak, although it remains completely unconstrained while retaining the deviation from the $\LCDM$ case at \(\sim 2\sigma\). We find the significance of this deviation in BHDE limits to be at \(1.23\sigma, 2.05\sigma, 3.45\sigma\) using the DESI, \Panp+DESI, DESy5+DESI datasets, respectively. The latter of the three data combinations is a significant deviation, being equivalent to the deviation from \(\Lambda\)CDM quoted in \cite{DESI:2024mwx} using phenomenological modeling of dynamical dark energy, namely the CPL \cite{Chevallier:2000qy, Linder:2002et} model considerations. Further discussion on this topic is provided in \cref{sec:DE}.

In \cite{Barrow:2020kug, Ghoshal:2021ief, Jizba:2023fkp}, limits on the exponent ($\Delta$) were placed considering the Big Bang Nucleosynthesis at early times \cite[see also][for limits based on inflation]{Luciano:2023roh, Luciano:2022hhy}. In the context of the equivalent GT approach as explored here, extremely tight bounds of $\Delta \leq 1.4\times10^{-4}$ were presented over the allowed values  \cite{Barrow:2020kug}. More recently, in \cite{Luciano:2023roh}, a less stringent limit of $\Delta \sim 0.008$ is reported, considering a similar analysis. {These limits are subject to the cosmological approach followed using the $\Seff$ and are valid only within the GT approach}. Also, the normalization parameter $\beta$ is usually set to unity in most of the aforementioned analyses. When imposing the Barrow limits of $\Delta > 0$, we find the $95\%$ C.L. upper limit to be $\Delta< 0.011$ and $\Delta < 0.008$ using the \Panp+DESI and DESy5+DESI, respectively. We find that our constraints here are in very good agreement with the limits set in the aforementioned analyses, which consider completely different epochs. When imposing the Tsallis limits, we find that the median of the posteriors tends to $\Delta<0$ while being consistent with $\Delta\to 0$ within $1-2\sigma$ C.L. for all dataset combinations.

{Comparing the HDE and GT approaches, we find $\Omzero$ is better constrained in the former than in the latter, where the upper limit on the same is weakened in almost all the data combinations. Interestingly, the BGT with tighter bounds of $\Delta$ tends to push the $\Omzero$ values to be larger than in the TGT case with wider $\Delta$ priors, which in turn provides no upper bound. While the HDE approach usually constrains $\Omzero$, to be larger than the $\LCDM$ value.}

\begin{figure}
    \centering
    \includegraphics[scale = 0.25]{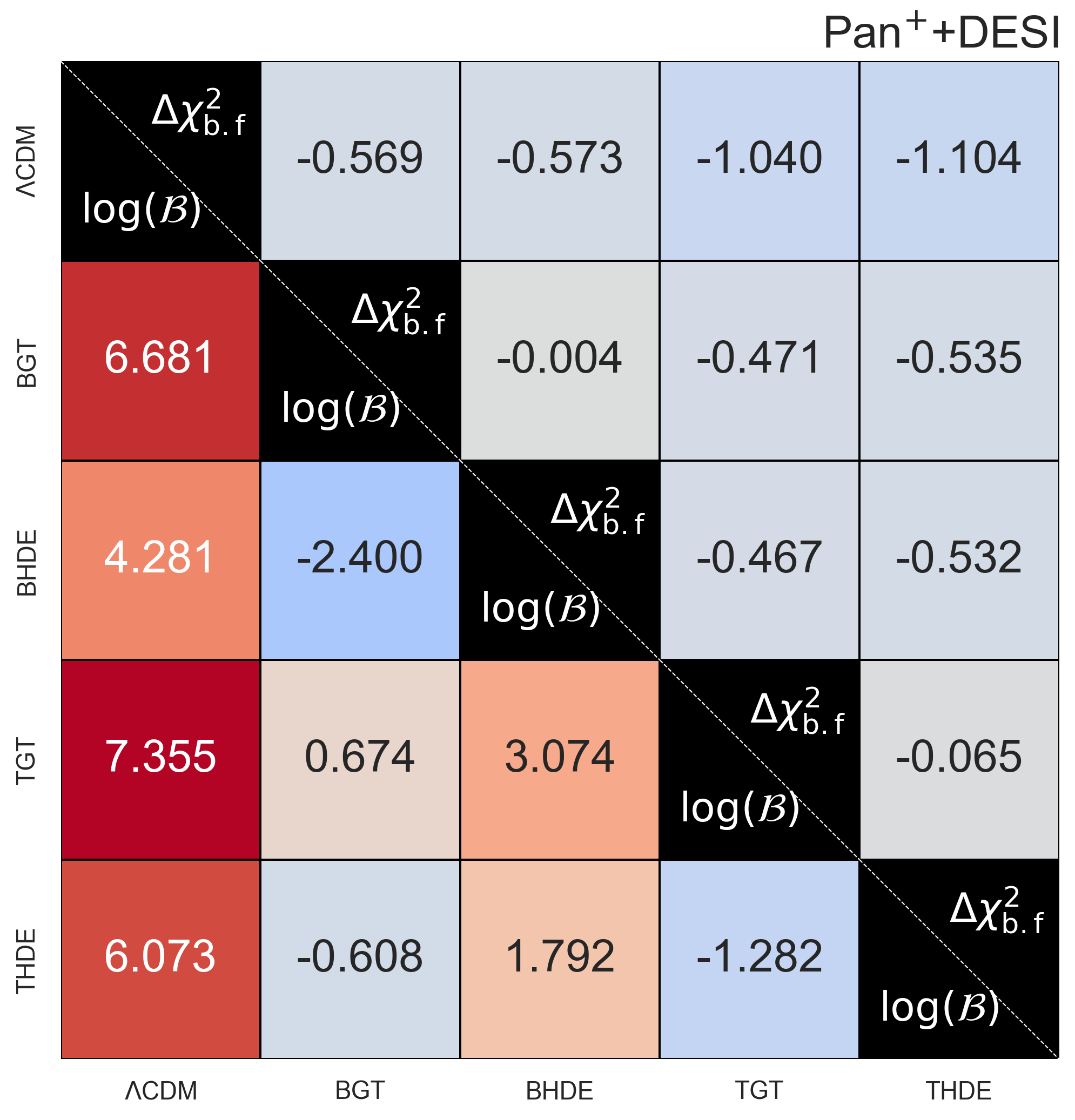}
    \includegraphics[scale = 0.25]{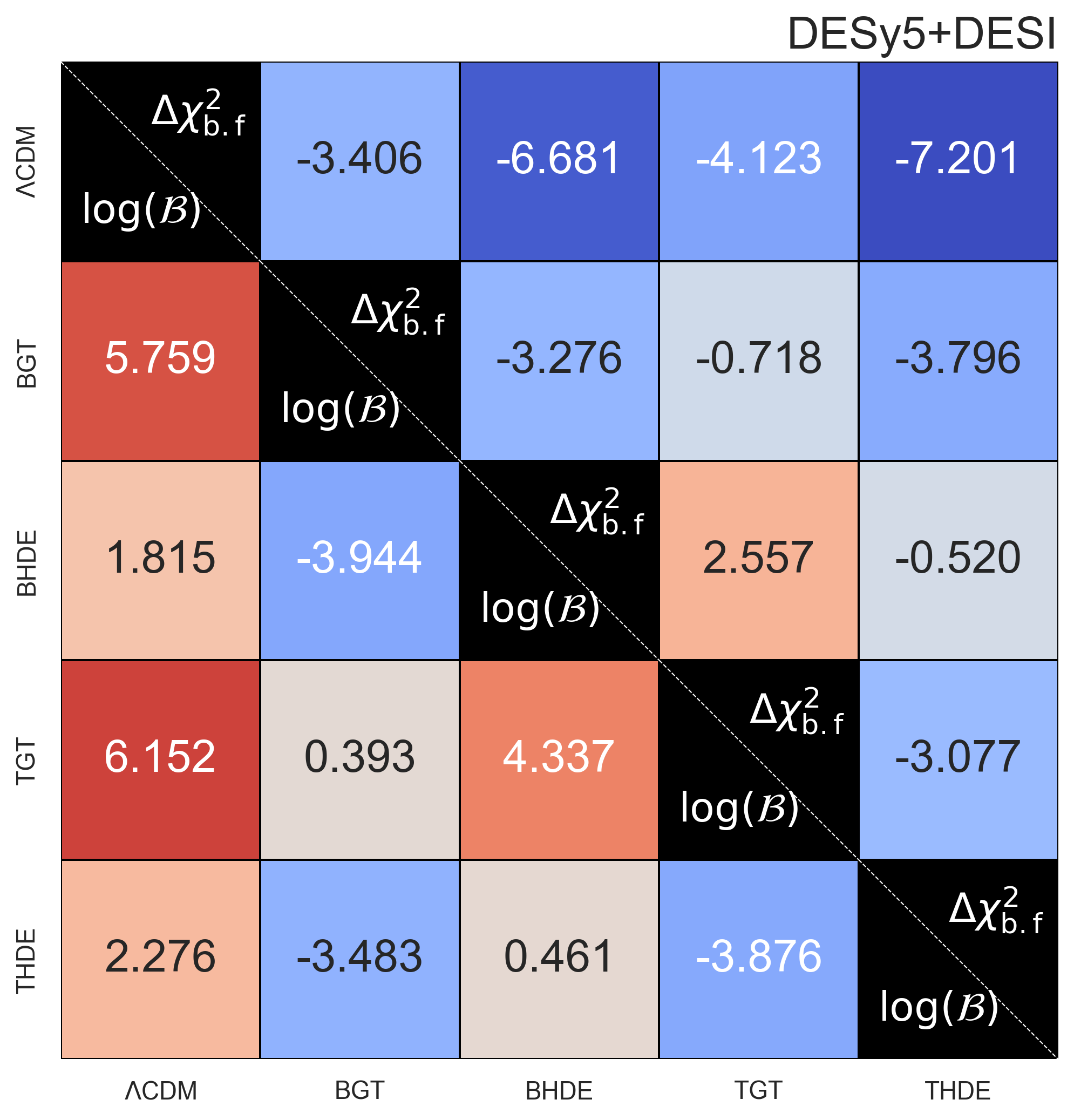}
    
    \caption{Comparison of the Bayesian Evidence and the best-fit $\Delta \chi^2$ for all the combinations of models under consideration. The {left} panel shows the data combination \Panp + DESI and the right panel is shown for the more recent DESy5+DESI. }
    \label{fig:BE}
\end{figure}

\subsection{Bayesian Evidence}\label{sec:BE}
We estimate the Bayesian evidence comparing against the $\LCDM$ model as presented in the last column of \cref{tab:table1} and \cref{fig:BE}. As anticipated, we find that the $\LCDM$ model is almost always preferred over the extended HDE and GT approaches both within the Barrow and Tsallis limits. However, we mainly focus on our primary motive of assessing the performance of HDE and GT Entropic approaches against each other. As can be seen from the SNe-only analysis, neither the \Panp nor the DESy5 is able to strongly constrain the parameters nor provide a strong preference for the $\LCDM$ model. SNe datasets provide maximum Bayesian evidence against the THDE model of the order $\DeltaBE \sim 3.3$ using the \Panp compilation, which is moderate evidence disfavoring the model. On the other hand, the DESI dataset is consistently able to disfavor all the extended models with almost a strong significance of $\DeltaBE \gtrsim 5$. However, in this case, the GT approaches are more strongly disfavored than the HDE approaches. 

{Having established the preferences of individual datasets, we now turn to the joint analysis of the SNe and BAO datasets. In \cref{fig:BE}, within each heatmap, we show the comparison of the $\DeltaBE$ and $\chibf$, in the lower and upper triangles, respectively. The values in each cell are computed as the difference between the value corresponding to the model on the x-axis wrt the respective model on the y-axis. For instance, the first column shows the difference between the Evidence for the models on the y-axis wrt $\LCDM$ on the x-axis, where we see that the $\LCDM$ is always preferred over the extension. Similarly, the top row shows that all the models on the x-axis have better $\chi^2_{\rm b.f}$ than that in the $\LCDM$ model.} { This is indeed what one would expect, with increasing number of parameters in the extended models they will fit the data better in terms of $\chi^2$. However, since the number of parameters in GT and HDE models is the same, even $\chi^2$ can be used to compare these two approaches. We find that these results are in line with the Bayesian evidence results i.e. $\chi^2_{\rm b.f}$ of the HDE model is lower than the $\chibf$ in the GT approach. }

We find that the \Panp+DESI dataset once again disfavors all the extended models strongly, i.e., $\DeltaBE\gtrsim 5$. The BHDE is disfavored wrt the $\LCDM$ model having $\DeltaBE\sim 4.3$, which is equivalent significance reported in \cite{Denkiewicz:2023hyj}, while using a different set of data\footnote{{As mentioned in \cref{sec:Data}, we have compared our results with the earlier completed SDSS + \Panp dataset combination finding that the BHDE model is disfavored wrt $\LCDM$ at $\DeltaBE \sim 4.8$, which is a similar inference made in \cite{Denkiewicz:2023hyj}(see Table 1. therein).}}. However, very interestingly now replacing the \Panp with the more recent DESy5 SNe we find a clear preference for the HDE models over the GT approaches. The aforementioned significance of $\DeltaBE\sim 4.3$ is now reduced to $\DeltaBE\sim 1.8$ using the more recent DESI+ DESy5 dataset, implying BHDE performs very well. We find that the newer DESy5+DESI data compilations fit the HDE models with better $\chibf \sim -7$ for both the Barrow and Tsallis bounds. This, in turn, reflects positively in terms of Evidence as the HDE models can be interpreted to fit almost equivalently to the $\LCDM$ model.  On the other hand, the GT approaches, while having slightly better $\chibf \sim -3.4$ (Barrow) and $\chibf \sim -4.1$ (Tsallis), fare extremely poorly being disfavored with $\DeltaBE \sim 5.8$ and $\DeltaBE \sim 6.2$ in terms of Bayesian evidence, respectively. The THDE and BHDE models, on the other hand, perform equally well as the $\LCDM$ model. In this context, we almost most definitely find that the HDE approach performs better than the GT approach. 

{In a similar way, comparing the GT approaches with the BHDE model, we find that the TGT and BGT approaches are disfavored at $\DeltaBE \sim 4.0$ and $\DeltaBE \sim 4.3$, respectively, when using DESy5+DESI data. With the \Panp+DESI we find a moderate preference of $\DeltaBE\sim -3.1$ for the BHDE over the TGT model.}
{From the evidence analysis, we can also note that generally, the Barrow entropic models are mildly favored over their corresponding Tsallis counterparts for all the combinations of datasets.}

\begin{figure}
    
    \includegraphics[scale=0.85]{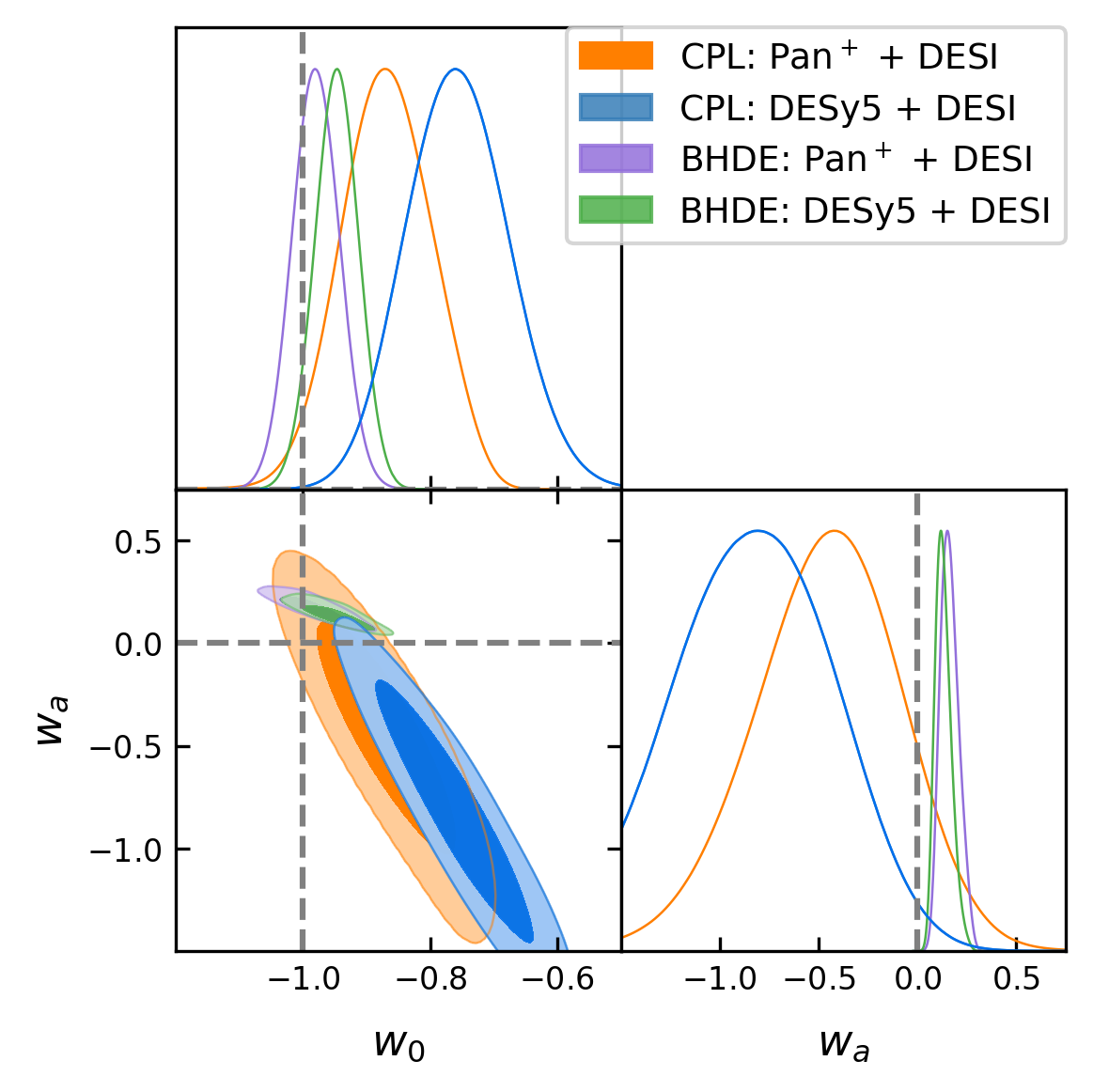}
    \caption{Contours for Barrow Holographic Dark Energy (BHDE) cosmology obtain the dataset combinations of SNe + DESI. The phenomenological CPL model is also shown for comparison.}
    \label{fig:EoS:w0wa}
\end{figure}

\begin{figure*}
    
    \includegraphics[scale=0.57]{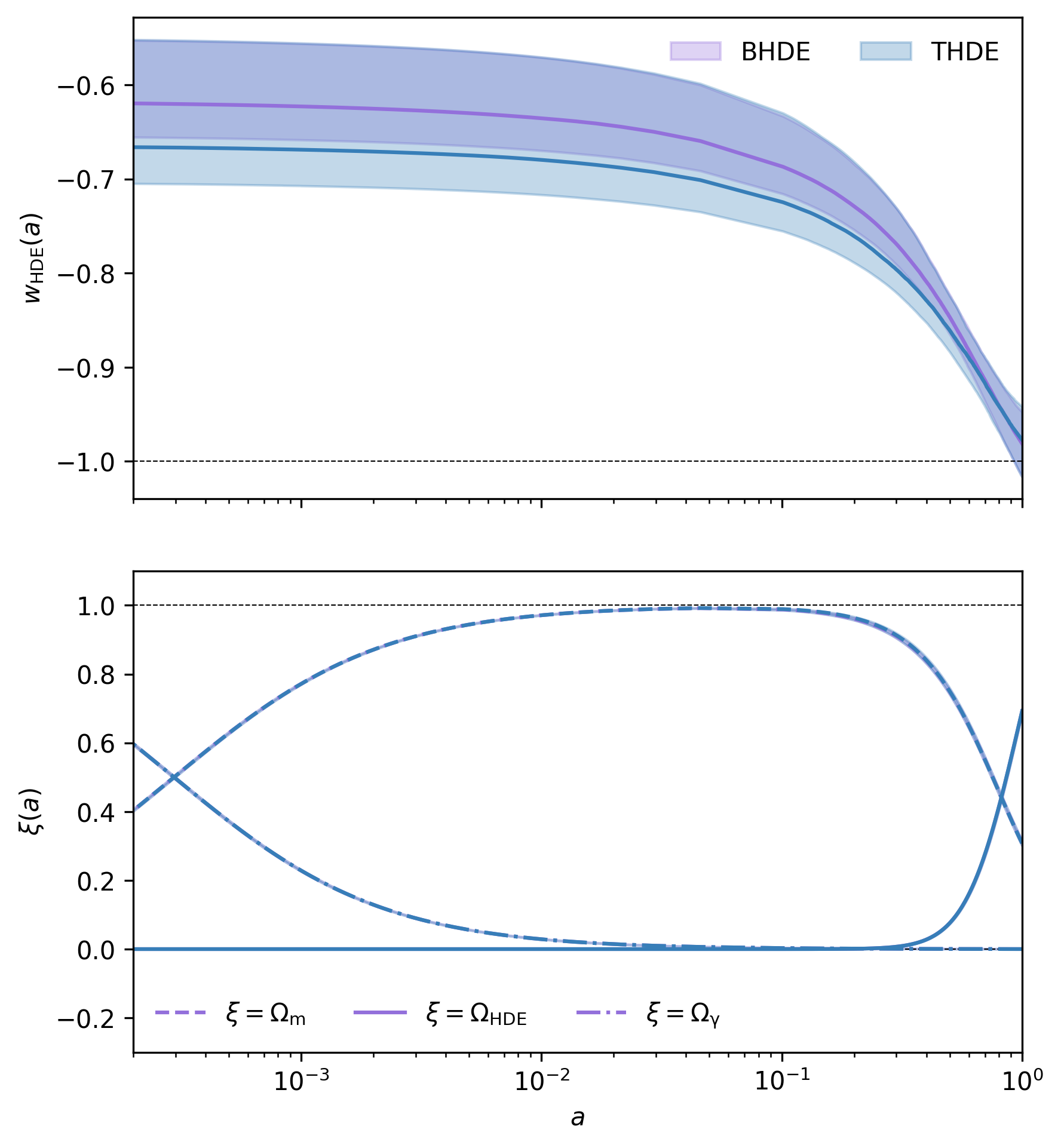}
    \includegraphics[scale=0.57]{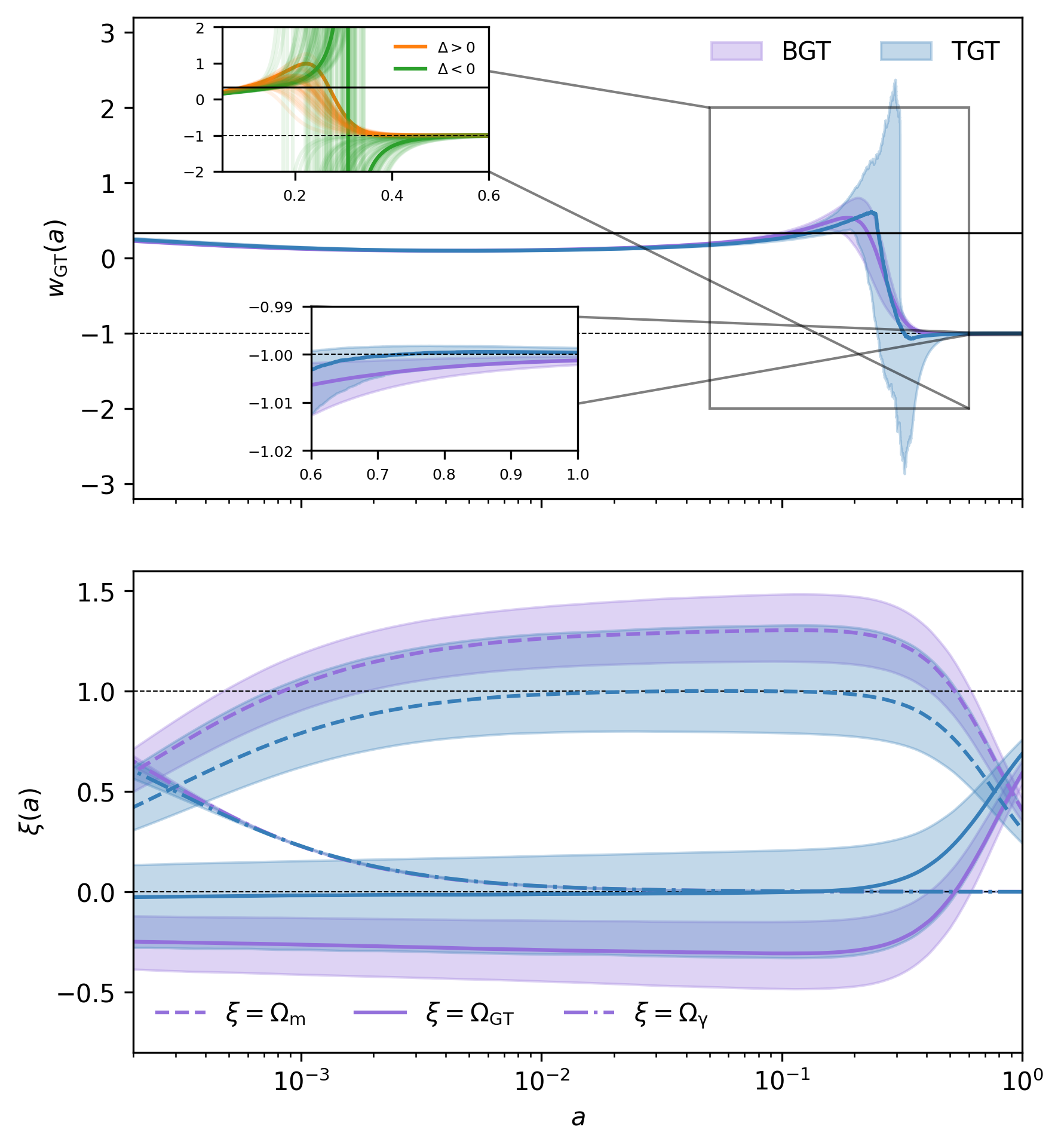}
    \caption{We show the evolution of different density contributions (\textit{bottom}) and corresponding dark energy EoS (\textit{top}) in both the formalisms, Gravity-Thermodynamics (\textit{right}) and Holographic approach (\textit{left}). The shaded region indicates the $68\%$ C.L. regions obtained utilizing the BAO (DESI) + SNe (\Panp) datasets. In the top-right plot showing the DE EoS for the GT approach, we highlight the constraints for $\wGT$ in the range $a\in \{0.6, 1\}$ and the singularity around $a\sim \{0.05, 0.6\}$. }
    \label{fig:EoS:wofz}
\end{figure*}

\subsection{Dark Energy implications}\label{sec:DE}
{We now turn to estimate the dark energy equation of state, $w(z)$ for the two entropic formalisms. As is very well known, the HDE and GT approaches provide dynamical behavior to the dark energy density ($\Ode, \Ogt$). We construct the $w(z)$ and also compute the derivative of the same to construct the $\{\wzero, \wa\}$\footnote{Here $\wa$ is the derivative of the $w(z)$ wrt scale factor $a$ today ($a\to1$) and $\wzero$ $ = w(z = 0)$.}\footnote{\label{labelCPL} In this formalism, the dynamical dark energy is modeled through the EoS $w(z) = w_0 + w_{\rm a} (1-a)$, and is usually depicted as \cref{fig:EoS:w0wa}.} parameter space.} This assessment is also necessary to better understand the constraints obtained in the parameter space formally utilized when assessing deviations from the standard \(\LCDM\) scenario. For ease of comparison, we display both the functional evolution of \(w(z)\) in \cref{fig:EoS:wofz} and the \(\wzero\) vs \(\wa\) parameter space (see \cref{fig:EoS:w0wa}). {As we show in \cref{fig:EoS:w0wa}, the BHDE models are constrained in similar parameter ranges as the phenomenological CPL parameters. The larger posteriors of CPL parameter space are completely in agreement with the much tighter posteriors within the HDE model. }

As shown in the top panels of \cref{fig:EoS:wofz}, we find that the dark energy equation of state ($\wGT$) in the TGT approach shows a singularity at $z \sim 1.8$ transitioning from the $\wGT(z \lesssim 1.8) \sim -1$ to $\wGT(z \gtrsim 1.8) \sim 0$, consistent with $\LCDM$ at late-times and transitioning to a radiation-like behavior at higher redshifts. This is a formal assessment of the argument presented in \cite{Denkiewicz:2023hyj} that the GT approach can mimic the $\LCDM$ model at late times while only being a radiation correction to the standard $\LCDM$ model at higher redshifts. Note that the singularity present in the TGT model is not present in the BGT model, which shows a smooth transition to $\wGT\to 0$. {This difference is driven by the fact that posterior for $\Delta <0 $ is allowed in the TGT models, while it is forced to be positive in the BGT limits. This changes the nature of the denominator in \cref{eqn:EoSGT}, which allows for some of the parameter space to provide the singular behavior. In top-right panel of \cref{fig:EoS:wofz}, we show in the inset highlighting the range $a\in \{0.05, 0.6\}$, the curves obtained for $\wGT$ for $\Delta> 0$ (orange) and $\Delta< 0$ (green).}

Our results for the BGT models here are in good agreement with those presented in \cite{Leon_2021}, where a simple quintessence-like EoS was obtained, which is also complemented by their inclusion of strong $\LCDM$ based CMB priors on $\Omzero$ assumed therein. As found in \cite{Leon_2021}, and shown in the lower-right panel of \cref{fig:EoS:wofz} for the BGT model, we find possible negative values of $\Ogt$, which is not observed in the TGT approach. {This is driven by the larger values of $\Om$ obtained within the BGT model (see \cref{tab:table1}), in contrast to lower values of $\Om$ and corresponding negative values of $\Delta$. Several recent works \cite{Sen:2021wld, Akarsu:2023mfb}, have projected a possibility of a transition of dark energy density from a negative to a positive value. We recover a similar behavior in the BGT model, however, requiring the closure equation to be satisfied, which yields no additional advantage in addressing cosmological ($\Hzero$) tensions \cite{Verde:2023lmm}. Also, note that we find the $\Ogt$ to be consistent with zero within $\sim 2\sigma$ C.L. in contrast to the $\sim 1\sigma$ level reported in \cite{Leon_2021}. }

On the other hand, the HDE shows a more quintessence-like behavior mimicking a freezing field \cite{Linder06} with $\wDE \gtrsim -1$ at all redshifts. Note that this is in contrast to the recent claims of dynamical dark energy at higher redshifts\cite{DESI:2024mwx} {(see also \cite{Park:2024jns})}, where a late-time quintessence with phantom-crossing was observed using the standard Chavelier-Polarski-Linder (CPL)\footref{labelCPL} parameterization \cite{Chevallier:2000qy, Linder:2002et}.  While we do not formally compare the CPL model and our HDE model, it is evident that the HDE model is consistent with the dynamical dark energy claims within $2\sigma$ limits. 
We find that the $\wzero$ estimates in all the entropic scenarios are consistent within $1\sigma$ with cosmological constant.

\section{Conclusions}\label{sec:Conclusions}
{We have implemented and assessed the viability of widely used cosmological models based on Entropic approaches implementing the "Holographic Principle" (HDE)\cite{Li_2004} and "Gravity - Thermodynamics" (GT) \cite{Padmanabhan:2003gd, Jacobson:1995ab} using the most recent late-time cosmological observables, namely SNe from {\Panp\cite{Scolnic:2021amr}} and DESy5 \cite{DES:2024tys} and BAO from DESI \cite{DESI:2024mwx}. This is then followed by our primary objective, which is to perform model selection utilizing the Bayesian evidence to asses which of the two approaches is preferred by the data. We summarize our final results as follows:

\begin{itemize}
    \item We find the SNe datasets, both \Panp and DESy5, are unable to strongly disfavor the entropic approaches, performing equivalently as the $\LCDM$ model. 
    \item The more recent BAO (DESI) dataset is able to provide strong evidence against the GT models while moderately disfavouring the HDE approach. 
    \item We find that when using the combination of most recent datasets, DESI(BAO)+Pan$^+$(SNe), we find that all the models are strongly disfavored, except the BHDE model, which is moderately disfavored.
    \item However, interestingly, using the more recent DESy5 dataset in combination with the recent DESI data, we find that the GT approaches are strongly disfavored $\DeltaBE \gtrsim 6$. On the other hand,  the HDE approach performs equivalently to the  $\LCDM$ model. 
    \item This establishes a clear result that amongst the entropic-based approaches to late-time cosmology, the HDE and hence the holographic principle is the preferred direction to explore given the most recent data. 
    \item Alongside assessing the viability of the models, we also present the dark energy constraints for the HDE and GT approaches. 
    \item We find the HDE approaches to be consistent with simple quintessence-like behavior without phantom crossing. The GT approach, on the other hand, does present mild phantom crossing with radiation-like behavior at high redshift. 

\end{itemize}

Entropic approaches provide a simple yet natural extension to standard cosmology based on cosmological constant. As indicated earlier in \cite{Carroll:2016lku}, the implementation through the Holographic principle and gravity thermodynamics conjecture differ formally. However, given the diversity in the existing possibilities, it is important to assess the possible directions that need to be explored in the future. In this context, we have attempted to contrast the HP-based applications to dark energy and modifications to cosmological constant through GT conjecture, finding a clear preference for the former. We intend to extend our current analysis to the application of more general entropy forms \cite{Nojiri:2022aof} to cosmology. Also, in light of the more precise late-time data from EUCLID \cite{Amendola:2016saw, Euclid:2024yrr}, LSST \cite{Ivezic:2008fe} to arrive, it is both necessary and imminent to establish viable directions to explore within the vast landscape of available entropic approaches. } 

\section*{Acknowledgments}
We thank the anonymous referee for useful suggestions helping to improve the presentation of the work. We are grateful to Eoin Colgain and Manosh Manoharan for useful comments. BSH is supported by the INFN INDARK grant and acknowledges support from the COSMOS project of the Italian Space Agency (cosmosnet.it).

\appendix
\section{Effect of Priors on the Constraints}\label{sec:Prior_effects}

\begin{figure}
    \centering
    \includegraphics[scale = 0.85]{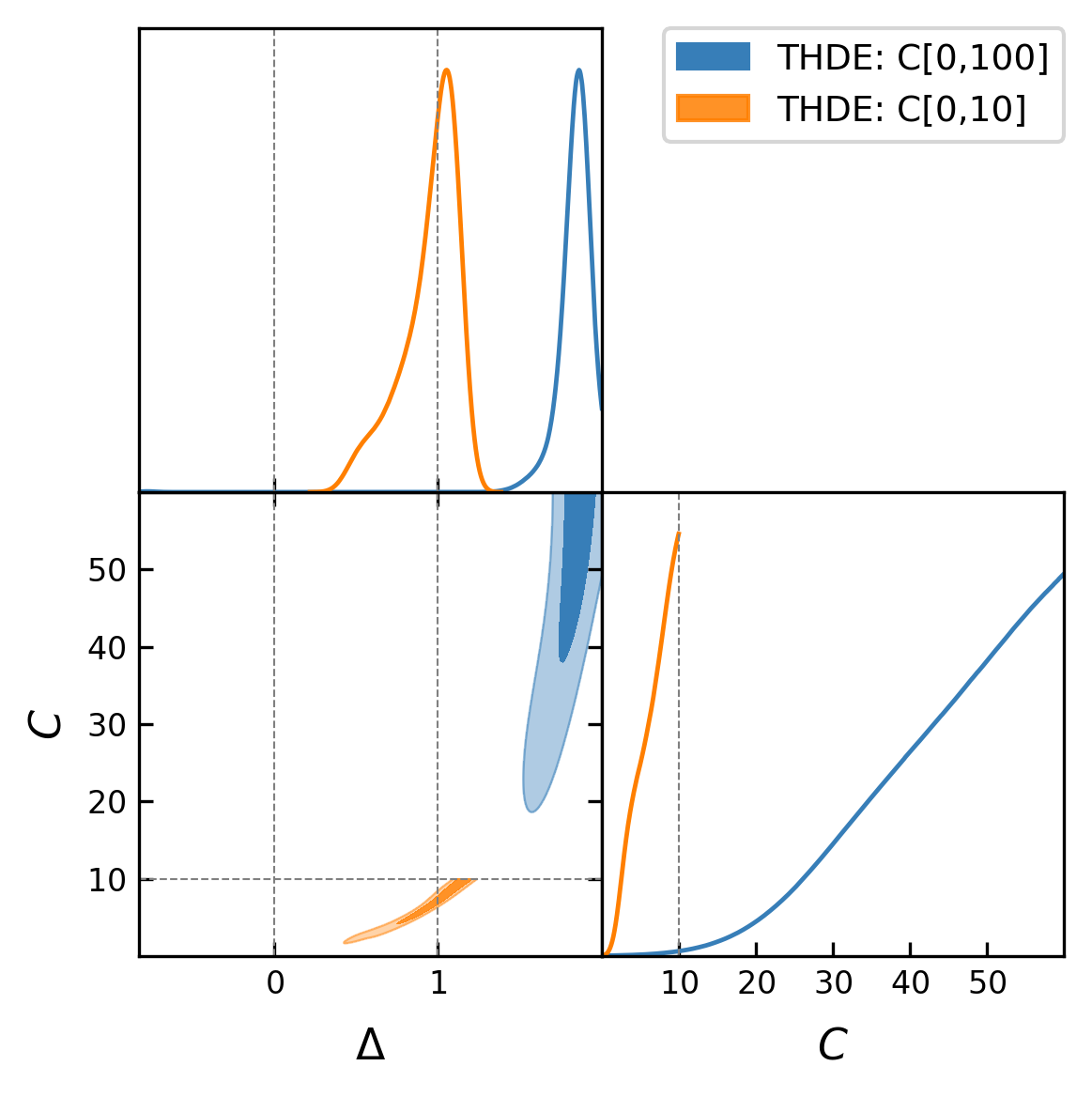}
    \caption{Contours for $68\%, 95\%$ C.L., in the THDE model with varied limits on the normalization parameter (\(\Cp\)). The posteriors are obtained utilizing the datasets BAO (DESI) + SNe (\Panp). }
    
    \label{fig:samp_effect}
\end{figure}

{As we have earlier mentioned in \cref{tab:priors}, we have utilized $\Cp \in \{0, 10\}$, in our main analysis for both THDE and BHDE models. However, as can be seen in \cref{fig:samp_effect} when the limits are increased to $\Cp\in \{0, 100\}$, the posterior in $\Cp \, \text{vs.} \, \Delta$ parameter space is completely beyond the limits presented in the main analysis. The case of BHDE allows for the possibility of constraining $\Cp$ as well, guided by the limit of $\Delta<1.0$. This, however, is not true for the THDE model. Conservatively, we maintain the same limits as those used for the BHDE, which also maintains consistency in the context of comparing the Bayesian Evidence. Increasing the prior volume of THDE can additionally penalize the model while providing better $\chibf$. However, this choice of the prior does not influence our final inference that the HDE approach performs better than the gravity-thermodynamics. For instance, when using the \Panp + DESI dataset, we find THDE with extended priors on $\Cp$ to be disfavor moderately at $\DeltaBE \sim 3.48$, in contrast to the strong evidence of $\DeltaBE \sim 6.07$ (see \cref{tab:table1}) obtained against the model in the main analysis. }

\section{Tables of constraints and contour plots}\label{sec:Tables}
For brevity in the main text, we show our tables of constraints, here in the appendix. 

{\renewcommand{\arraystretch}{1.8}
\setlength{\tabcolsep}{4pt}
\begin{table*}[h!]

    \caption{Constraints (68\% C.L.) on the free model parameters of the entropic models and the $\LCDM$ model.  The parameter $H_0$ is measured in units of $\ksM$. In the last column, we show the Bayesian evidence for each model computed against the $\LCDM$ model. We remind that in our comparison a positive value of $\DeltaBE$ implies that the model in comparison is disfavored wrt $\LCDM$ model.}
    \label{tab:table1}    
    \begin{tabular}{ccccccccc} 
      \hline
      \textbf{Dataset} & Model &\textbf{$\Omega_{\rm m}$} & \textbf{$H_{0}$} & \textbf{$\Delta$} & \textbf{$C/\beta$} & \textbf{$M_{b}$} & \textbf{$r_{d}$} & $\DeltaBE$ \\
      
      \hline
      \hline
      
      \multirow{3}{*}{Pan$^{+}$} & $\LCDM$ & $0.332_{-0.018}^{+0.018}$ & $73.250_{-0.995}^{+1.021}$ & $-$ & $-$  & $-19.252_{-0.029}^{+0.029}$ & $-$ & $0$  \\
                            & BGT & $>0.243$ & $73.309_{-0.941}^{+1.011}$ & $<0.627$ & $>0.725$ & $-19.253_{-0.027}^{+0.029}$ & $-$ & $2.154$  \\
                            & BHDE & $0.249_{-0.062}^{+0.057}$ & $73.154_{-1.041}^{+1.040}$ & $>0.282$ & $>2.09$ & $-19.250_{-0.029}^{+0.030}$ & $-$ & $1.936$  \\
                            & TGT & $--$ & $73.136_{-0.980}^{+0.986}$ & $-0.411_{-0.441}^{+0.473}$ & $--$ & $-19.254_{-0.028}^{+0.029}$ & $-$ & $2.354$  \\
                            & THDE & $0.253_{-0.070}^{+0.062}$ & $73.020_{-0.955}^{+1.039}$ & $0.674_{-0.459}^{+0.281}$ & $--$ &$-19.255_{-0.027}^{+0.029}$ & $-$ &  $3.329$ \\
                            
        \cline{1-2} 
                                
          \multirow{3}{*}{DESy5} & $\LCDM$ & $0.353_{-0.017}^{+0.017}$ & $69.481_{-19.853}^{+20.840}$ & $-$ & $-$  & $-$ & $-$ & $0$  \\
                            & BGT & $>0.256$ & $69.693_{-20.261}^{+20.854}$ & $<0.776$ & $>0.775$ & $-$ & $-$ & $1.842$  \\
                            & BHDE & $<0.324$ & $69.128_{-19.716}^{+20.347}$ & $--$ & $>1.9$ & $-$ & $-$ & $0.932$  \\

                            & TGT & $--$ & $70.558_{-20.380}^{+20.001}$ & $-0.502_{-0.519}^{+0.570}$ & $--$ & $-$ & $-$ & $2.018$   \\
                            & THDE & $<0.332$ & $66.345_{-18.780}^{+22.919}$ & $0.573_{-0.546}^{+0.386}$ & $--$ &$-$ & $-$ & $2.568$ \\

      \cline{1-2}
      
      \multirow{2}{*}{DESI} & $\LCDM$ & $0.296_{-0.007}^{+0.007}$ & $69.187_{-0.615}^{+0.649}$ & $-$ & $-$ &  $-$ & $147.263_{-0.323}^{+0.334}$ & $0$  \\
      & BGT & $>0.263$ & $69.743_{-0.948}^{+0.941}$ & $<0.016$ & $0.762_{-0.128}^{+0.194}$ & $-$& $147.254_{-0.336}^{+0.329}$ & $6.453$  \\
                                & BHDE & $0.280_{-0.024}^{+0.022}$ & $71.065_{-2.724}^{+3.355}$ & $>0.241$ & $3.777_{-1.961}^{+1.655}$ & $-$ & $147.284_{-0.312}^{+0.348}$ & $4.129$    \\ 
                                & TGT & $0.320_{-0.085}^{+0.101}$ & $69.525_{-0.973}^{+0.983}$ & $0.003_{-0.009}^{+0.008}$ & $--$ & $-$ &  $147.236_{-0.339}^{+0.340}$ &   $7.471$ \\
                                & THDE & $0.278_{-0.025}^{+0.021}$ & $71.365_{-2.640}^{+3.489}$ & $0.960_{-0.352}^{+0.208}$ & $--$ & $-$ & $147.280_{-0.364}^{+0.344}$ &  $5.022$ \\

    \cline{1-2}
      \multirow{2}{*}{Pan$^{+}$+DESI} & $\LCDM$ & $0.301_{-0.006}^{+0.006}$ & $68.825_{-0.610}^{+0.583}$ & $-$ & $-$ & $-19.398_{-0.018}^{+0.017}$ & $147.149_{-0.317}^{+0.323}$ & $0$  \\

      & BGT & $>0.314$ & $68.566_{-0.740}^{+0.781}$ & $<0.011$ & $0.764_{-0.106}^{+0.115}$ & $-19.405_{-0.020}^{+0.021}$ & $147.228_{-0.332}^{+0.331}$ & $6.681$  \\
                                & BHDE & $0.309_{-0.007}^{+0.007}$ & $67.673_{-0.778}^{+0.694}$ & $>0.502$ & $5.253_{-1.984}^{+1.386}$ & $-19.427_{-0.020}^{+0.018}$ & $147.330_{-0.339}^{+0.357}$ & $4.281$   \\
                                & TGT & $>0.218$ & $68.446_{-0.794}^{+0.777}$ & $-0.002_{-0.008}^{+0.007}$ & $--$ & $-19.407_{-0.022}^{+0.021}$ & $147.258_{-0.340}^{+0.341}$ &   $7.355$\\

                                & THDE & $0.308_{-0.007}^{+0.008}$ & $67.698_{-0.794}^{+0.715}$ & $0.988_{-0.356}^{+0.124}$ & $>0.279$ &$-19.426_{-0.020}^{+0.019}$ & $147.328_{-0.343}^{+0.319}$ & $6.073$   \\ 

    \cline{1-2}
      \multirow{2}{*}{DESy5+DESI} & $\LCDM$ & $0.305_{-0.006}^{+0.006}$ & $69.488_{-0.568}^{+0.589}$ & $-$ & $-$ & $-$ & $147.067_{-0.329}^{+0.326}$ & $0$  \\
      & BGT & $>0.346$ & $67.893_{-0.692}^{+0.720}$ & $<0.008$ & $0.743_{-0.068}^{+0.094}$ & $-$ & $147.251_{-0.334}^{+0.343}$ & $5.759$  \\
                                & BHDE & $0.316_{-0.007}^{+0.007}$ & $66.891_{-0.729}^{+0.700}$ & $>0.565$ & $5.901_{-2.055}^{+1.290}$ & $-$ & $147.352_{-0.337}^{+0.324}$ & $1.815$   \\ 

                                & TGT & $0.311_{-0.073}^{+0.098}$ & $67.722_{-0.716}^{+0.726}$ & $-0.005_{-0.007}^{+0.007}$ & $>0.709$ & $-$ & $147.225_{-0.338}^{+0.339}$ & $6.152$   \\
                                & THDE & $0.315_{-0.007}^{+0.007}$ & $67.028_{-0.726}^{+0.715}$ & $0.987_{-0.212}^{+0.104}$ & $>3.12$ &$-$ & $147.319_{-0.311}^{+0.339}$ &   $2.276$ \\

      \hline
    \end{tabular}
\end{table*}
}

\bibliography{references}

\begin{thebibliography}{148}%
\makeatletter
\providecommand \@ifxundefined [1]{%
 \@ifx{#1\undefined}
}%
\providecommand \@ifnum [1]{%
 \ifnum #1\expandafter \@firstoftwo
 \else \expandafter \@secondoftwo
 \fi
}%
\providecommand \@ifx [1]{%
 \ifx #1\expandafter \@firstoftwo
 \else \expandafter \@secondoftwo
 \fi
}%
\providecommand \natexlab [1]{#1}%
\providecommand \enquote  [1]{``#1''}%
\providecommand \bibnamefont  [1]{#1}%
\providecommand \bibfnamefont [1]{#1}%
\providecommand \citenamefont [1]{#1}%
\providecommand \href@noop [0]{\@secondoftwo}%
\providecommand \href [0]{\begingroup \@sanitize@url \@href}%
\providecommand \@href[1]{\@@startlink{#1}\@@href}%
\providecommand \@@href[1]{\endgroup#1\@@endlink}%
\providecommand \@sanitize@url [0]{\catcode `\\12\catcode `\$12\catcode `\&12\catcode `\#12\catcode `\^12\catcode `\_12\catcode `\%12\relax}%
\providecommand \@@startlink[1]{}%
\providecommand \@@endlink[0]{}%
\providecommand \url  [0]{\begingroup\@sanitize@url \@url }%
\providecommand \@url [1]{\endgroup\@href {#1}{\urlprefix }}%
\providecommand \urlprefix  [0]{URL }%
\providecommand \Eprint [0]{\href }%
\providecommand \doibase [0]{http://dx.doi.org/}%
\providecommand \selectlanguage [0]{\@gobble}%
\providecommand \bibinfo  [0]{\@secondoftwo}%
\providecommand \bibfield  [0]{\@secondoftwo}%
\providecommand \translation [1]{[#1]}%
\providecommand \BibitemOpen [0]{}%
\providecommand \bibitemStop [0]{}%
\providecommand \bibitemNoStop [0]{.\EOS\space}%
\providecommand \EOS [0]{\spacefactor3000\relax}%
\providecommand \BibitemShut  [1]{\csname bibitem#1\endcsname}%
\let\auto@bib@innerbib\@empty
\bibitem [{\citenamefont {Riess}\ \emph {et~al.}(1998)\citenamefont {Riess} \emph {et~al.}}]{Riess98}%
  \BibitemOpen
  \bibfield  {author} {\bibinfo {author} {\bibfnamefont {A.~G.}\ \bibnamefont {Riess}} \emph {et~al.} (\bibinfo {collaboration} {Supernova Search Team}),\ }\href {\doibase 10.1086/300499} {\bibfield  {journal} {\bibinfo  {journal} {Astron. J.}\ }\textbf {\bibinfo {volume} {116}},\ \bibinfo {pages} {1009} (\bibinfo {year} {1998})},\ \Eprint {http://arxiv.org/abs/astro-ph/9805201} {arXiv:astro-ph/9805201 [astro-ph]} \BibitemShut {NoStop}%
\bibitem [{\citenamefont {Betoule}\ \emph {et~al.}(2014)\citenamefont {Betoule} \emph {et~al.}}]{Betoule:2014frx}%
  \BibitemOpen
  \bibfield  {author} {\bibinfo {author} {\bibfnamefont {M.}~\bibnamefont {Betoule}} \emph {et~al.} (\bibinfo {collaboration} {SDSS}),\ }\href {\doibase 10.1051/0004-6361/201423413} {\bibfield  {journal} {\bibinfo  {journal} {Astron. Astrophys.}\ }\textbf {\bibinfo {volume} {568}},\ \bibinfo {pages} {A22} (\bibinfo {year} {2014})},\ \Eprint {http://arxiv.org/abs/1401.4064} {arXiv:1401.4064 [astro-ph.CO]} \BibitemShut {NoStop}%
\bibitem [{\citenamefont {Scolnic}\ \emph {et~al.}(2018)\citenamefont {Scolnic} \emph {et~al.}}]{Scolnic17}%
  \BibitemOpen
  \bibfield  {author} {\bibinfo {author} {\bibfnamefont {D.~M.}\ \bibnamefont {Scolnic}} \emph {et~al.},\ }\href {\doibase 10.3847/1538-4357/aab9bb} {\bibfield  {journal} {\bibinfo  {journal} {Astrophys. J.}\ }\textbf {\bibinfo {volume} {859}},\ \bibinfo {pages} {101} (\bibinfo {year} {2018})},\ \Eprint {http://arxiv.org/abs/1710.00845} {arXiv:1710.00845 [astro-ph.CO]} \BibitemShut {NoStop}%
\bibitem [{\citenamefont {Scolnic}\ \emph {et~al.}(2022)\citenamefont {Scolnic} \emph {et~al.}}]{Scolnic:2021amr}%
  \BibitemOpen
  \bibfield  {author} {\bibinfo {author} {\bibfnamefont {D.}~\bibnamefont {Scolnic}} \emph {et~al.},\ }\href {\doibase 10.3847/1538-4357/ac8b7a} {\bibfield  {journal} {\bibinfo  {journal} {Astrophys. J.}\ }\textbf {\bibinfo {volume} {938}},\ \bibinfo {pages} {113} (\bibinfo {year} {2022})},\ \Eprint {http://arxiv.org/abs/2112.03863} {arXiv:2112.03863 [astro-ph.CO]} \BibitemShut {NoStop}%
\bibitem [{\citenamefont {Haridasu}\ \emph {et~al.}(2018{\natexlab{a}})\citenamefont {Haridasu}, \citenamefont {Lukovi{\'c}},\ and\ \citenamefont {Vittorio}}]{Haridasu17a}%
  \BibitemOpen
  \bibfield  {author} {\bibinfo {author} {\bibfnamefont {B.~S.}\ \bibnamefont {Haridasu}}, \bibinfo {author} {\bibfnamefont {V.~V.}\ \bibnamefont {Lukovi{\'c}}}, \ and\ \bibinfo {author} {\bibfnamefont {N.}~\bibnamefont {Vittorio}},\ }\href {\doibase 10.1088/1475-7516/2018/05/033} {\bibfield  {journal} {\bibinfo  {journal} {\jcap}\ }\textbf {\bibinfo {volume} {5}},\ \bibinfo {eid} {033} (\bibinfo {year} {2018}{\natexlab{a}})},\ \Eprint {http://arxiv.org/abs/1711.03929} {arXiv:1711.03929} \BibitemShut {NoStop}%
\bibitem [{\citenamefont {Abbott}\ \emph {et~al.}(2024)\citenamefont {Abbott} \emph {et~al.}}]{DES:2024tys}%
  \BibitemOpen
  \bibfield  {author} {\bibinfo {author} {\bibfnamefont {T.~M.~C.}\ \bibnamefont {Abbott}} \emph {et~al.} (\bibinfo {collaboration} {DES}),\ }\href@noop {} {\  (\bibinfo {year} {2024})},\ \Eprint {http://arxiv.org/abs/2401.02929} {arXiv:2401.02929 [astro-ph.CO]} \BibitemShut {NoStop}%
\bibitem [{\citenamefont {Dam}\ \emph {et~al.}(2017)\citenamefont {Dam}, \citenamefont {Heinesen},\ and\ \citenamefont {Wiltshire}}]{Dam:2017xqs}%
  \BibitemOpen
  \bibfield  {author} {\bibinfo {author} {\bibfnamefont {L.~H.}\ \bibnamefont {Dam}}, \bibinfo {author} {\bibfnamefont {A.}~\bibnamefont {Heinesen}}, \ and\ \bibinfo {author} {\bibfnamefont {D.~L.}\ \bibnamefont {Wiltshire}},\ }\href {\doibase 10.1093/mnras/stx1858} {\bibfield  {journal} {\bibinfo  {journal} {Mon. Not. Roy. Astron. Soc.}\ }\textbf {\bibinfo {volume} {472}},\ \bibinfo {pages} {835} (\bibinfo {year} {2017})},\ \Eprint {http://arxiv.org/abs/1706.07236} {arXiv:1706.07236 [astro-ph.CO]} \BibitemShut {NoStop}%
\bibitem [{\citenamefont {Rubin}\ and\ \citenamefont {Hayden}(2016)}]{Rubin:2016iqe}%
  \BibitemOpen
  \bibfield  {author} {\bibinfo {author} {\bibfnamefont {D.}~\bibnamefont {Rubin}}\ and\ \bibinfo {author} {\bibfnamefont {B.}~\bibnamefont {Hayden}},\ }\href {\doibase 10.3847/2041-8213/833/2/L30} {\bibfield  {journal} {\bibinfo  {journal} {Astrophys. J. Lett.}\ }\textbf {\bibinfo {volume} {833}},\ \bibinfo {pages} {L30} (\bibinfo {year} {2016})},\ \Eprint {http://arxiv.org/abs/1610.08972} {arXiv:1610.08972 [astro-ph.CO]} \BibitemShut {NoStop}%
\bibitem [{\citenamefont {Ade}\ \emph {et~al.}(2016)\citenamefont {Ade} \emph {et~al.}}]{Planck:2015lwi}%
  \BibitemOpen
  \bibfield  {author} {\bibinfo {author} {\bibfnamefont {P.~A.~R.}\ \bibnamefont {Ade}} \emph {et~al.} (\bibinfo {collaboration} {Planck}),\ }\href {\doibase 10.1051/0004-6361/201525833} {\bibfield  {journal} {\bibinfo  {journal} {Astron. Astrophys.}\ }\textbf {\bibinfo {volume} {594}},\ \bibinfo {pages} {A24} (\bibinfo {year} {2016})},\ \Eprint {http://arxiv.org/abs/1502.01597} {arXiv:1502.01597 [astro-ph.CO]} \BibitemShut {NoStop}%
\bibitem [{\citenamefont {Aghanim}\ \emph {et~al.}(2020)\citenamefont {Aghanim} \emph {et~al.}}]{Planck:2018vyg}%
  \BibitemOpen
  \bibfield  {author} {\bibinfo {author} {\bibfnamefont {N.}~\bibnamefont {Aghanim}} \emph {et~al.} (\bibinfo {collaboration} {Planck}),\ }\href {\doibase 10.1051/0004-6361/201833910} {\bibfield  {journal} {\bibinfo  {journal} {Astron. Astrophys.}\ }\textbf {\bibinfo {volume} {641}},\ \bibinfo {pages} {A6} (\bibinfo {year} {2020})},\ \bibinfo {note} {[Erratum: Astron.Astrophys. 652, C4 (2021)]},\ \Eprint {http://arxiv.org/abs/1807.06209} {arXiv:1807.06209 [astro-ph.CO]} \BibitemShut {NoStop}%
\bibitem [{\citenamefont {Verde}\ \emph {et~al.}(2019)\citenamefont {Verde}, \citenamefont {Treu},\ and\ \citenamefont {Riess}}]{Verde:2019ivm}%
  \BibitemOpen
  \bibfield  {author} {\bibinfo {author} {\bibfnamefont {L.}~\bibnamefont {Verde}}, \bibinfo {author} {\bibfnamefont {T.}~\bibnamefont {Treu}}, \ and\ \bibinfo {author} {\bibfnamefont {A.~G.}\ \bibnamefont {Riess}},\ }in\ \href {\doibase 10.1038/s41550-019-0902-0} {\emph {\bibinfo {booktitle} {{Nature Astronomy 2019}}}}\ (\bibinfo {year} {2019})\ \Eprint {http://arxiv.org/abs/1907.10625} {arXiv:1907.10625 [astro-ph.CO]} \BibitemShut {NoStop}%
\bibitem [{\citenamefont {Riess}\ \emph {et~al.}(2022)\citenamefont {Riess} \emph {et~al.}}]{Riess:2021jrx}%
  \BibitemOpen
  \bibfield  {author} {\bibinfo {author} {\bibfnamefont {A.~G.}\ \bibnamefont {Riess}} \emph {et~al.},\ }\href {\doibase 10.3847/2041-8213/ac5c5b} {\bibfield  {journal} {\bibinfo  {journal} {Astrophys. J. Lett.}\ }\textbf {\bibinfo {volume} {934}},\ \bibinfo {pages} {L7} (\bibinfo {year} {2022})},\ \Eprint {http://arxiv.org/abs/2112.04510} {arXiv:2112.04510 [astro-ph.CO]} \BibitemShut {NoStop}%
\bibitem [{\citenamefont {Riess}(2019)}]{Riess:2019qba}%
  \BibitemOpen
  \bibfield  {author} {\bibinfo {author} {\bibfnamefont {A.~G.}\ \bibnamefont {Riess}},\ }\href {\doibase 10.1038/s42254-019-0137-0} {\bibfield  {journal} {\bibinfo  {journal} {Nature Rev. Phys.}\ }\textbf {\bibinfo {volume} {2}},\ \bibinfo {pages} {10} (\bibinfo {year} {2019})},\ \Eprint {http://arxiv.org/abs/2001.03624} {arXiv:2001.03624 [astro-ph.CO]} \BibitemShut {NoStop}%
\bibitem [{\citenamefont {Perivolaropoulos}\ and\ \citenamefont {Skara}(2022)}]{Perivolaropoulos:2021jda}%
  \BibitemOpen
  \bibfield  {author} {\bibinfo {author} {\bibfnamefont {L.}~\bibnamefont {Perivolaropoulos}}\ and\ \bibinfo {author} {\bibfnamefont {F.}~\bibnamefont {Skara}},\ }\href {\doibase 10.1016/j.newar.2022.101659} {\bibfield  {journal} {\bibinfo  {journal} {New Astron. Rev.}\ }\textbf {\bibinfo {volume} {95}},\ \bibinfo {pages} {101659} (\bibinfo {year} {2022})},\ \Eprint {http://arxiv.org/abs/2105.05208} {arXiv:2105.05208 [astro-ph.CO]} \BibitemShut {NoStop}%
\bibitem [{\citenamefont {Efstathiou}(2020)}]{Efstathiou:2020wxn}%
  \BibitemOpen
  \bibfield  {author} {\bibinfo {author} {\bibfnamefont {G.}~\bibnamefont {Efstathiou}},\ }\href@noop {} {\  (\bibinfo {year} {2020})},\ \Eprint {http://arxiv.org/abs/2007.10716} {arXiv:2007.10716 [astro-ph.CO]} \BibitemShut {NoStop}%
\bibitem [{\citenamefont {Freedman}(2021)}]{Freedman:2021ahq}%
  \BibitemOpen
  \bibfield  {author} {\bibinfo {author} {\bibfnamefont {W.~L.}\ \bibnamefont {Freedman}},\ }\href {\doibase 10.3847/1538-4357/ac0e95} {\bibfield  {journal} {\bibinfo  {journal} {Astrophys. J.}\ }\textbf {\bibinfo {volume} {919}},\ \bibinfo {pages} {16} (\bibinfo {year} {2021})},\ \Eprint {http://arxiv.org/abs/2106.15656} {arXiv:2106.15656 [astro-ph.CO]} \BibitemShut {NoStop}%
\bibitem [{\citenamefont {Verde}\ \emph {et~al.}(2023)\citenamefont {Verde}, \citenamefont {Sch\"oneberg},\ and\ \citenamefont {Gil-Mar\'\i{}n}}]{Verde:2023lmm}%
  \BibitemOpen
  \bibfield  {author} {\bibinfo {author} {\bibfnamefont {L.}~\bibnamefont {Verde}}, \bibinfo {author} {\bibfnamefont {N.}~\bibnamefont {Sch\"oneberg}}, \ and\ \bibinfo {author} {\bibfnamefont {H.}~\bibnamefont {Gil-Mar\'\i{}n}},\ }\href@noop {} {\  (\bibinfo {year} {2023})},\ \Eprint {http://arxiv.org/abs/2311.13305} {arXiv:2311.13305 [astro-ph.CO]} \BibitemShut {NoStop}%
\bibitem [{\citenamefont {Abdalla}\ \emph {et~al.}(2022)\citenamefont {Abdalla} \emph {et~al.}}]{Abdalla:2022yfr}%
  \BibitemOpen
  \bibfield  {author} {\bibinfo {author} {\bibfnamefont {E.}~\bibnamefont {Abdalla}} \emph {et~al.},\ }\href {\doibase 10.1016/j.jheap.2022.04.002} {\bibfield  {journal} {\bibinfo  {journal} {JHEAp}\ }\textbf {\bibinfo {volume} {34}},\ \bibinfo {pages} {49} (\bibinfo {year} {2022})},\ \Eprint {http://arxiv.org/abs/2203.06142} {arXiv:2203.06142 [astro-ph.CO]} \BibitemShut {NoStop}%
\bibitem [{\citenamefont {Dom\'\i{}nguez}\ \emph {et~al.}(2019)\citenamefont {Dom\'\i{}nguez}, \citenamefont {Wojtak}, \citenamefont {Finke}, \citenamefont {Ajello}, \citenamefont {Helgason}, \citenamefont {Prada}, \citenamefont {Desai}, \citenamefont {Paliya}, \citenamefont {Marcotulli},\ and\ \citenamefont {Hartmann}}]{Dominguez:2019jqc}%
  \BibitemOpen
  \bibfield  {author} {\bibinfo {author} {\bibfnamefont {A.}~\bibnamefont {Dom\'\i{}nguez}}, \bibinfo {author} {\bibfnamefont {R.}~\bibnamefont {Wojtak}}, \bibinfo {author} {\bibfnamefont {J.}~\bibnamefont {Finke}}, \bibinfo {author} {\bibfnamefont {M.}~\bibnamefont {Ajello}}, \bibinfo {author} {\bibfnamefont {K.}~\bibnamefont {Helgason}}, \bibinfo {author} {\bibfnamefont {F.}~\bibnamefont {Prada}}, \bibinfo {author} {\bibfnamefont {A.}~\bibnamefont {Desai}}, \bibinfo {author} {\bibfnamefont {V.}~\bibnamefont {Paliya}}, \bibinfo {author} {\bibfnamefont {L.}~\bibnamefont {Marcotulli}}, \ and\ \bibinfo {author} {\bibfnamefont {D.}~\bibnamefont {Hartmann}},\ }\href {\doibase 10.3847/1538-4357/ab4a0e} {\  (\bibinfo {year} {2019}),\ 10.3847/1538-4357/ab4a0e},\ \Eprint {http://arxiv.org/abs/1903.12097} {arXiv:1903.12097 [astro-ph.CO]} \BibitemShut {NoStop}%
\bibitem [{\citenamefont {Park}\ and\ \citenamefont {Ratra}(2020)}]{Park:2019emi}%
  \BibitemOpen
  \bibfield  {author} {\bibinfo {author} {\bibfnamefont {C.-G.}\ \bibnamefont {Park}}\ and\ \bibinfo {author} {\bibfnamefont {B.}~\bibnamefont {Ratra}},\ }\href {\doibase 10.1103/PhysRevD.101.083508} {\bibfield  {journal} {\bibinfo  {journal} {Phys. Rev. D}\ }\textbf {\bibinfo {volume} {101}},\ \bibinfo {pages} {083508} (\bibinfo {year} {2020})},\ \Eprint {http://arxiv.org/abs/1908.08477} {arXiv:1908.08477 [astro-ph.CO]} \BibitemShut {NoStop}%
\bibitem [{\citenamefont {Lin}\ and\ \citenamefont {Ishak}(2021)}]{Lin:2019zdn}%
  \BibitemOpen
  \bibfield  {author} {\bibinfo {author} {\bibfnamefont {W.}~\bibnamefont {Lin}}\ and\ \bibinfo {author} {\bibfnamefont {M.}~\bibnamefont {Ishak}},\ }\href {\doibase 10.1088/1475-7516/2021/05/009} {\bibfield  {journal} {\bibinfo  {journal} {JCAP}\ }\textbf {\bibinfo {volume} {05}},\ \bibinfo {pages} {009} (\bibinfo {year} {2021})},\ \Eprint {http://arxiv.org/abs/1909.10991} {arXiv:1909.10991 [astro-ph.CO]} \BibitemShut {NoStop}%
\bibitem [{\citenamefont {Freedman}\ \emph {et~al.}(2020)\citenamefont {Freedman}, \citenamefont {Madore}, \citenamefont {Hoyt}, \citenamefont {Jang}, \citenamefont {Beaton}, \citenamefont {Lee}, \citenamefont {Monson}, \citenamefont {Neeley},\ and\ \citenamefont {Rich}}]{Freedman:2020dne}%
  \BibitemOpen
  \bibfield  {author} {\bibinfo {author} {\bibfnamefont {W.~L.}\ \bibnamefont {Freedman}}, \bibinfo {author} {\bibfnamefont {B.~F.}\ \bibnamefont {Madore}}, \bibinfo {author} {\bibfnamefont {T.}~\bibnamefont {Hoyt}}, \bibinfo {author} {\bibfnamefont {I.~S.}\ \bibnamefont {Jang}}, \bibinfo {author} {\bibfnamefont {R.}~\bibnamefont {Beaton}}, \bibinfo {author} {\bibfnamefont {M.~G.}\ \bibnamefont {Lee}}, \bibinfo {author} {\bibfnamefont {A.}~\bibnamefont {Monson}}, \bibinfo {author} {\bibfnamefont {J.}~\bibnamefont {Neeley}}, \ and\ \bibinfo {author} {\bibfnamefont {J.}~\bibnamefont {Rich}},\ }\href {\doibase 10.3847/1538-4357/ab7339} {\  (\bibinfo {year} {2020}),\ 10.3847/1538-4357/ab7339},\ \Eprint {http://arxiv.org/abs/2002.01550} {arXiv:2002.01550 [astro-ph.GA]} \BibitemShut {NoStop}%
\bibitem [{\citenamefont {Birrer}\ \emph {et~al.}(2020)\citenamefont {Birrer} \emph {et~al.}}]{Birrer:2020tax}%
  \BibitemOpen
  \bibfield  {author} {\bibinfo {author} {\bibfnamefont {S.}~\bibnamefont {Birrer}} \emph {et~al.},\ }\href {\doibase 10.1051/0004-6361/202038861} {\bibfield  {journal} {\bibinfo  {journal} {Astron. Astrophys.}\ }\textbf {\bibinfo {volume} {643}},\ \bibinfo {pages} {A165} (\bibinfo {year} {2020})},\ \Eprint {http://arxiv.org/abs/2007.02941} {arXiv:2007.02941 [astro-ph.CO]} \BibitemShut {NoStop}%
\bibitem [{\citenamefont {Boruah}\ \emph {et~al.}(2021)\citenamefont {Boruah}, \citenamefont {Hudson},\ and\ \citenamefont {Lavaux}}]{Boruah:2020fhl}%
  \BibitemOpen
  \bibfield  {author} {\bibinfo {author} {\bibfnamefont {S.~S.}\ \bibnamefont {Boruah}}, \bibinfo {author} {\bibfnamefont {M.~J.}\ \bibnamefont {Hudson}}, \ and\ \bibinfo {author} {\bibfnamefont {G.}~\bibnamefont {Lavaux}},\ }\href {\doibase 10.1093/mnras/stab2320} {\bibfield  {journal} {\bibinfo  {journal} {Mon. Not. Roy. Astron. Soc.}\ }\textbf {\bibinfo {volume} {507}},\ \bibinfo {pages} {2697} (\bibinfo {year} {2021})},\ \Eprint {http://arxiv.org/abs/2010.01119} {arXiv:2010.01119 [astro-ph.CO]} \BibitemShut {NoStop}%
\bibitem [{\citenamefont {Wu}\ \emph {et~al.}(2022)\citenamefont {Wu}, \citenamefont {Zhang},\ and\ \citenamefont {Wang}}]{Wu:2021jyk}%
  \BibitemOpen
  \bibfield  {author} {\bibinfo {author} {\bibfnamefont {Q.}~\bibnamefont {Wu}}, \bibinfo {author} {\bibfnamefont {G.-Q.}\ \bibnamefont {Zhang}}, \ and\ \bibinfo {author} {\bibfnamefont {F.-Y.}\ \bibnamefont {Wang}},\ }\href {\doibase 10.1093/mnrasl/slac022} {\bibfield  {journal} {\bibinfo  {journal} {Mon. Not. Roy. Astron. Soc.}\ }\textbf {\bibinfo {volume} {515}},\ \bibinfo {pages} {L1} (\bibinfo {year} {2022})},\ \bibinfo {note} {[Erratum: Mon.Not.Roy.Astron.Soc. 531, L8 (2024)]},\ \Eprint {http://arxiv.org/abs/2108.00581} {arXiv:2108.00581 [astro-ph.CO]} \BibitemShut {NoStop}%
\bibitem [{\citenamefont {Cao}\ and\ \citenamefont {Ratra}(2022)}]{Cao:2022ugh}%
  \BibitemOpen
  \bibfield  {author} {\bibinfo {author} {\bibfnamefont {S.}~\bibnamefont {Cao}}\ and\ \bibinfo {author} {\bibfnamefont {B.}~\bibnamefont {Ratra}},\ }\href {\doibase 10.1093/mnras/stac1184} {\bibfield  {journal} {\bibinfo  {journal} {Mon. Not. Roy. Astron. Soc.}\ }\textbf {\bibinfo {volume} {513}},\ \bibinfo {pages} {5686} (\bibinfo {year} {2022})},\ \Eprint {http://arxiv.org/abs/2203.10825} {arXiv:2203.10825 [astro-ph.CO]} \BibitemShut {NoStop}%
\bibitem [{\citenamefont {Chen}\ \emph {et~al.}(2024)\citenamefont {Chen}, \citenamefont {Kumar}, \citenamefont {Ratra},\ and\ \citenamefont {Xu}}]{Chen:2024gnu}%
  \BibitemOpen
  \bibfield  {author} {\bibinfo {author} {\bibfnamefont {Y.}~\bibnamefont {Chen}}, \bibinfo {author} {\bibfnamefont {S.}~\bibnamefont {Kumar}}, \bibinfo {author} {\bibfnamefont {B.}~\bibnamefont {Ratra}}, \ and\ \bibinfo {author} {\bibfnamefont {T.}~\bibnamefont {Xu}},\ }\href {\doibase 10.3847/2041-8213/ad2e97} {\bibfield  {journal} {\bibinfo  {journal} {Astrophys. J. Lett.}\ }\textbf {\bibinfo {volume} {964}},\ \bibinfo {pages} {L4} (\bibinfo {year} {2024})},\ \Eprint {http://arxiv.org/abs/2401.13187} {arXiv:2401.13187 [astro-ph.CO]} \BibitemShut {NoStop}%
\bibitem [{\citenamefont {Wang}\ \emph {et~al.}(2017{\natexlab{a}})\citenamefont {Wang}, \citenamefont {Wang},\ and\ \citenamefont {Li}}]{Wang:2016och}%
  \BibitemOpen
  \bibfield  {author} {\bibinfo {author} {\bibfnamefont {S.}~\bibnamefont {Wang}}, \bibinfo {author} {\bibfnamefont {Y.}~\bibnamefont {Wang}}, \ and\ \bibinfo {author} {\bibfnamefont {M.}~\bibnamefont {Li}},\ }\href {\doibase 10.1016/j.physrep.2017.06.003} {\bibfield  {journal} {\bibinfo  {journal} {Phys. Rept.}\ }\textbf {\bibinfo {volume} {696}},\ \bibinfo {pages} {1} (\bibinfo {year} {2017}{\natexlab{a}})},\ \Eprint {http://arxiv.org/abs/1612.00345} {arXiv:1612.00345 [astro-ph.CO]} \BibitemShut {NoStop}%
\bibitem [{\citenamefont {del Campo}\ \emph {et~al.}(2011)\citenamefont {del Campo}, \citenamefont {Fabris}, \citenamefont {Herrera},\ and\ \citenamefont {Zimdahl}}]{PhysRevD.83.123006}%
  \BibitemOpen
  \bibfield  {author} {\bibinfo {author} {\bibfnamefont {S.}~\bibnamefont {del Campo}}, \bibinfo {author} {\bibfnamefont {J.~C.}\ \bibnamefont {Fabris}}, \bibinfo {author} {\bibfnamefont {R.}~\bibnamefont {Herrera}}, \ and\ \bibinfo {author} {\bibfnamefont {W.}~\bibnamefont {Zimdahl}},\ }\href {\doibase 10.1103/PhysRevD.83.123006} {\bibfield  {journal} {\bibinfo  {journal} {Phys. Rev. D}\ }\textbf {\bibinfo {volume} {83}},\ \bibinfo {pages} {123006} (\bibinfo {year} {2011})}\BibitemShut {NoStop}%
\bibitem [{\citenamefont {Saridakis}\ \emph {et~al.}(2018)\citenamefont {Saridakis}, \citenamefont {Bamba}, \citenamefont {Myrzakulov},\ and\ \citenamefont {Anagnostopoulos}}]{Saridakis_2018}%
  \BibitemOpen
  \bibfield  {author} {\bibinfo {author} {\bibfnamefont {E.~N.}\ \bibnamefont {Saridakis}}, \bibinfo {author} {\bibfnamefont {K.}~\bibnamefont {Bamba}}, \bibinfo {author} {\bibfnamefont {R.}~\bibnamefont {Myrzakulov}}, \ and\ \bibinfo {author} {\bibfnamefont {F.~K.}\ \bibnamefont {Anagnostopoulos}},\ }\href {\doibase 10.1088/1475-7516/2018/12/012} {\bibfield  {journal} {\bibinfo  {journal} {Journal of Cosmology and Astroparticle Physics}\ }\textbf {\bibinfo {volume} {2018}},\ \bibinfo {pages} {012} (\bibinfo {year} {2018})}\BibitemShut {NoStop}%
\bibitem [{\citenamefont {Nojiri}\ \emph {et~al.}(2021)\citenamefont {Nojiri}, \citenamefont {Odintsov},\ and\ \citenamefont {Paul}}]{Nojiri:2021iko}%
  \BibitemOpen
  \bibfield  {author} {\bibinfo {author} {\bibfnamefont {S.}~\bibnamefont {Nojiri}}, \bibinfo {author} {\bibfnamefont {S.~D.}\ \bibnamefont {Odintsov}}, \ and\ \bibinfo {author} {\bibfnamefont {T.}~\bibnamefont {Paul}},\ }\href {\doibase 10.3390/sym13060928} {\bibfield  {journal} {\bibinfo  {journal} {Symmetry}\ }\textbf {\bibinfo {volume} {13}},\ \bibinfo {pages} {928} (\bibinfo {year} {2021})},\ \Eprint {http://arxiv.org/abs/2105.08438} {arXiv:2105.08438 [gr-qc]} \BibitemShut {NoStop}%
\bibitem [{\citenamefont {Nojiri}\ and\ \citenamefont {Odintsov}(2006)}]{Nojiri:2005pu}%
  \BibitemOpen
  \bibfield  {author} {\bibinfo {author} {\bibfnamefont {S.}~\bibnamefont {Nojiri}}\ and\ \bibinfo {author} {\bibfnamefont {S.~D.}\ \bibnamefont {Odintsov}},\ }\href {\doibase 10.1007/s10714-006-0301-6} {\bibfield  {journal} {\bibinfo  {journal} {Gen. Rel. Grav.}\ }\textbf {\bibinfo {volume} {38}},\ \bibinfo {pages} {1285} (\bibinfo {year} {2006})},\ \Eprint {http://arxiv.org/abs/hep-th/0506212} {arXiv:hep-th/0506212} \BibitemShut {NoStop}%
\bibitem [{\citenamefont {Nojiri}\ and\ \citenamefont {Odintsov}(2017)}]{Nojiri:2017opc}%
  \BibitemOpen
  \bibfield  {author} {\bibinfo {author} {\bibfnamefont {S.}~\bibnamefont {Nojiri}}\ and\ \bibinfo {author} {\bibfnamefont {S.~D.}\ \bibnamefont {Odintsov}},\ }\href {\doibase 10.1140/epjc/s10052-017-5097-x} {\bibfield  {journal} {\bibinfo  {journal} {Eur. Phys. J. C}\ }\textbf {\bibinfo {volume} {77}},\ \bibinfo {pages} {528} (\bibinfo {year} {2017})},\ \Eprint {http://arxiv.org/abs/1703.06372} {arXiv:1703.06372 [hep-th]} \BibitemShut {NoStop}%
\bibitem [{\citenamefont {'t~Hooft}(1993)}]{tHooft:1993dmi}%
  \BibitemOpen
  \bibfield  {author} {\bibinfo {author} {\bibfnamefont {G.}~\bibnamefont {'t~Hooft}},\ }\href@noop {} {\bibfield  {journal} {\bibinfo  {journal} {Conf. Proc. C}\ }\textbf {\bibinfo {volume} {930308}},\ \bibinfo {pages} {284} (\bibinfo {year} {1993})},\ \Eprint {http://arxiv.org/abs/gr-qc/9310026} {arXiv:gr-qc/9310026} \BibitemShut {NoStop}%
\bibitem [{\citenamefont {Susskind}(1995)}]{Susskind:1994vu}%
  \BibitemOpen
  \bibfield  {author} {\bibinfo {author} {\bibfnamefont {L.}~\bibnamefont {Susskind}},\ }\href {\doibase 10.1063/1.531249} {\bibfield  {journal} {\bibinfo  {journal} {J. Math. Phys.}\ }\textbf {\bibinfo {volume} {36}},\ \bibinfo {pages} {6377} (\bibinfo {year} {1995})},\ \Eprint {http://arxiv.org/abs/hep-th/9409089} {arXiv:hep-th/9409089} \BibitemShut {NoStop}%
\bibitem [{\citenamefont {Bekenstein}(1994)}]{Bekenstein:1993dz}%
  \BibitemOpen
  \bibfield  {author} {\bibinfo {author} {\bibfnamefont {J.~D.}\ \bibnamefont {Bekenstein}},\ }\href {\doibase 10.1103/PhysRevD.49.1912} {\bibfield  {journal} {\bibinfo  {journal} {Phys. Rev. D}\ }\textbf {\bibinfo {volume} {49}},\ \bibinfo {pages} {1912} (\bibinfo {year} {1994})},\ \Eprint {http://arxiv.org/abs/gr-qc/9307035} {arXiv:gr-qc/9307035} \BibitemShut {NoStop}%
\bibitem [{\citenamefont {Nappi}\ and\ \citenamefont {Pasquinucci}(1992)}]{Nappi:1992as}%
  \BibitemOpen
  \bibfield  {author} {\bibinfo {author} {\bibfnamefont {C.~R.}\ \bibnamefont {Nappi}}\ and\ \bibinfo {author} {\bibfnamefont {A.}~\bibnamefont {Pasquinucci}},\ }\href {\doibase 10.1142/S021773239200272X} {\bibfield  {journal} {\bibinfo  {journal} {Mod. Phys. Lett. A}\ }\textbf {\bibinfo {volume} {7}},\ \bibinfo {pages} {3337} (\bibinfo {year} {1992})},\ \Eprint {http://arxiv.org/abs/gr-qc/9208002} {arXiv:gr-qc/9208002} \BibitemShut {NoStop}%
\bibitem [{\citenamefont {Carlip}(2014)}]{Carlip:2014pma}%
  \BibitemOpen
  \bibfield  {author} {\bibinfo {author} {\bibfnamefont {S.}~\bibnamefont {Carlip}},\ }\href {\doibase 10.1142/S0218271814300237} {\bibfield  {journal} {\bibinfo  {journal} {Int. J. Mod. Phys. D}\ }\textbf {\bibinfo {volume} {23}},\ \bibinfo {pages} {1430023} (\bibinfo {year} {2014})},\ \Eprint {http://arxiv.org/abs/1410.1486} {arXiv:1410.1486 [gr-qc]} \BibitemShut {NoStop}%
\bibitem [{\citenamefont {Brown}(1994)}]{Brown:1994sn}%
  \BibitemOpen
  \bibfield  {author} {\bibinfo {author} {\bibfnamefont {J.~D.}\ \bibnamefont {Brown}},\ }\href@noop {} {\  (\bibinfo {year} {1994})},\ \Eprint {http://arxiv.org/abs/gr-qc/9404006} {arXiv:gr-qc/9404006} \BibitemShut {NoStop}%
\bibitem [{\citenamefont {Oda}(1994)}]{Oda:1994np}%
  \BibitemOpen
  \bibfield  {author} {\bibinfo {author} {\bibfnamefont {I.}~\bibnamefont {Oda}},\ }\href {\doibase 10.1016/0370-2693(94)91361-7} {\bibfield  {journal} {\bibinfo  {journal} {Phys. Lett. B}\ }\textbf {\bibinfo {volume} {338}},\ \bibinfo {pages} {165} (\bibinfo {year} {1994})},\ \Eprint {http://arxiv.org/abs/hep-th/9405185} {arXiv:hep-th/9405185} \BibitemShut {NoStop}%
\bibitem [{\citenamefont {Ghosh}\ and\ \citenamefont {Mitra}(1996)}]{Ghosh:1994nz}%
  \BibitemOpen
  \bibfield  {author} {\bibinfo {author} {\bibfnamefont {A.}~\bibnamefont {Ghosh}}\ and\ \bibinfo {author} {\bibfnamefont {P.}~\bibnamefont {Mitra}},\ }\href {\doibase 10.1142/S0217732396001247} {\bibfield  {journal} {\bibinfo  {journal} {Mod. Phys. Lett. A}\ }\textbf {\bibinfo {volume} {11}},\ \bibinfo {pages} {1231} (\bibinfo {year} {1996})},\ \Eprint {http://arxiv.org/abs/gr-qc/9408040} {arXiv:gr-qc/9408040} \BibitemShut {NoStop}%
\bibitem [{\citenamefont {Frolov}(1995)}]{Frolov:1994zi}%
  \BibitemOpen
  \bibfield  {author} {\bibinfo {author} {\bibfnamefont {V.~P.}\ \bibnamefont {Frolov}},\ }\href {\doibase 10.1103/PhysRevLett.74.3319} {\bibfield  {journal} {\bibinfo  {journal} {Phys. Rev. Lett.}\ }\textbf {\bibinfo {volume} {74}},\ \bibinfo {pages} {3319} (\bibinfo {year} {1995})},\ \Eprint {http://arxiv.org/abs/gr-qc/9406037} {arXiv:gr-qc/9406037} \BibitemShut {NoStop}%
\bibitem [{\citenamefont {Fursaev}(1995)}]{Fursaev:1994pq}%
  \BibitemOpen
  \bibfield  {author} {\bibinfo {author} {\bibfnamefont {D.~V.}\ \bibnamefont {Fursaev}},\ }\href {\doibase 10.1142/S0217732395000697} {\bibfield  {journal} {\bibinfo  {journal} {Mod. Phys. Lett. A}\ }\textbf {\bibinfo {volume} {10}},\ \bibinfo {pages} {649} (\bibinfo {year} {1995})},\ \Eprint {http://arxiv.org/abs/hep-th/9408066} {arXiv:hep-th/9408066} \BibitemShut {NoStop}%
\bibitem [{\citenamefont {Louko}\ and\ \citenamefont {Whiting}(1995)}]{Louko:1994tv}%
  \BibitemOpen
  \bibfield  {author} {\bibinfo {author} {\bibfnamefont {J.}~\bibnamefont {Louko}}\ and\ \bibinfo {author} {\bibfnamefont {B.~F.}\ \bibnamefont {Whiting}},\ }\href {\doibase 10.1103/PhysRevD.51.5583} {\bibfield  {journal} {\bibinfo  {journal} {Phys. Rev. D}\ }\textbf {\bibinfo {volume} {51}},\ \bibinfo {pages} {5583} (\bibinfo {year} {1995})},\ \Eprint {http://arxiv.org/abs/gr-qc/9411017} {arXiv:gr-qc/9411017} \BibitemShut {NoStop}%
\bibitem [{\citenamefont {Liberati}(1997)}]{Liberati:1996kt}%
  \BibitemOpen
  \bibfield  {author} {\bibinfo {author} {\bibfnamefont {S.}~\bibnamefont {Liberati}},\ }\href@noop {} {\bibfield  {journal} {\bibinfo  {journal} {Nuovo Cim. B}\ }\textbf {\bibinfo {volume} {112}},\ \bibinfo {pages} {405} (\bibinfo {year} {1997})},\ \Eprint {http://arxiv.org/abs/gr-qc/9601032} {arXiv:gr-qc/9601032} \BibitemShut {NoStop}%
\bibitem [{\citenamefont {Solodukhin}(1996)}]{Solodukhin:1996vx}%
  \BibitemOpen
  \bibfield  {author} {\bibinfo {author} {\bibfnamefont {S.~N.}\ \bibnamefont {Solodukhin}},\ }\href {\doibase 10.1103/PhysRevD.54.3900} {\bibfield  {journal} {\bibinfo  {journal} {Phys. Rev. D}\ }\textbf {\bibinfo {volume} {54}},\ \bibinfo {pages} {3900} (\bibinfo {year} {1996})},\ \Eprint {http://arxiv.org/abs/hep-th/9601154} {arXiv:hep-th/9601154} \BibitemShut {NoStop}%
\bibitem [{\citenamefont {Carlip}(1997)}]{Carlip:1996xa}%
  \BibitemOpen
  \bibfield  {author} {\bibinfo {author} {\bibfnamefont {S.}~\bibnamefont {Carlip}},\ }\href {\doibase 10.1007/978-94-011-5812-1_5} {\bibfield  {journal} {\bibinfo  {journal} {Astrophys. Space Sci. Libr.}\ }\textbf {\bibinfo {volume} {211}},\ \bibinfo {pages} {35} (\bibinfo {year} {1997})},\ \Eprint {http://arxiv.org/abs/gr-qc/9603049} {arXiv:gr-qc/9603049} \BibitemShut {NoStop}%
\bibitem [{\citenamefont {Frolov}\ \emph {et~al.}(1996)\citenamefont {Frolov}, \citenamefont {Fursaev},\ and\ \citenamefont {Zelnikov}}]{Frolov:1996wd}%
  \BibitemOpen
  \bibfield  {author} {\bibinfo {author} {\bibfnamefont {V.~P.}\ \bibnamefont {Frolov}}, \bibinfo {author} {\bibfnamefont {D.~V.}\ \bibnamefont {Fursaev}}, \ and\ \bibinfo {author} {\bibfnamefont {A.~I.}\ \bibnamefont {Zelnikov}},\ }\href {\doibase 10.1016/0370-2693(96)00661-2} {\bibfield  {journal} {\bibinfo  {journal} {Phys. Lett. B}\ }\textbf {\bibinfo {volume} {382}},\ \bibinfo {pages} {220} (\bibinfo {year} {1996})},\ \Eprint {http://arxiv.org/abs/hep-th/9603175} {arXiv:hep-th/9603175} \BibitemShut {NoStop}%
\bibitem [{\citenamefont {Quevedo}\ \emph {et~al.}(2024)\citenamefont {Quevedo}, \citenamefont {Quevedo},\ and\ \citenamefont {Sanchez}}]{Quevedo:2024fga}%
  \BibitemOpen
  \bibfield  {author} {\bibinfo {author} {\bibfnamefont {H.}~\bibnamefont {Quevedo}}, \bibinfo {author} {\bibfnamefont {M.~N.}\ \bibnamefont {Quevedo}}, \ and\ \bibinfo {author} {\bibfnamefont {A.}~\bibnamefont {Sanchez}},\ }\href@noop {} {\  (\bibinfo {year} {2024})},\ \Eprint {http://arxiv.org/abs/2405.04474} {arXiv:2405.04474 [gr-qc]} \BibitemShut {NoStop}%
\bibitem [{\citenamefont {Bekenstein}(1973)}]{PhysRevD.7.2333}%
  \BibitemOpen
  \bibfield  {author} {\bibinfo {author} {\bibfnamefont {J.~D.}\ \bibnamefont {Bekenstein}},\ }\href {\doibase 10.1103/PhysRevD.7.2333} {\bibfield  {journal} {\bibinfo  {journal} {Phys. Rev. D}\ }\textbf {\bibinfo {volume} {7}},\ \bibinfo {pages} {2333} (\bibinfo {year} {1973})}\BibitemShut {NoStop}%
\bibitem [{\citenamefont {Bekenstein}(1972)}]{Bekenstein:1972tm}%
  \BibitemOpen
  \bibfield  {author} {\bibinfo {author} {\bibfnamefont {J.~D.}\ \bibnamefont {Bekenstein}},\ }\href {\doibase 10.1007/BF02757029} {\bibfield  {journal} {\bibinfo  {journal} {Lett. Nuovo Cim.}\ }\textbf {\bibinfo {volume} {4}},\ \bibinfo {pages} {737} (\bibinfo {year} {1972})}\BibitemShut {NoStop}%
\bibitem [{\citenamefont {Hawking}(1974)}]{Hawking:1974rv}%
  \BibitemOpen
  \bibfield  {author} {\bibinfo {author} {\bibfnamefont {S.~W.}\ \bibnamefont {Hawking}},\ }\href {\doibase 10.1038/248030a0} {\bibfield  {journal} {\bibinfo  {journal} {Nature}\ }\textbf {\bibinfo {volume} {248}},\ \bibinfo {pages} {30} (\bibinfo {year} {1974})}\BibitemShut {NoStop}%
\bibitem [{\citenamefont {Hawking}(1975)}]{Hawking:1975vcx}%
  \BibitemOpen
  \bibfield  {author} {\bibinfo {author} {\bibfnamefont {S.~W.}\ \bibnamefont {Hawking}},\ }\href {\doibase 10.1007/BF02345020} {\bibfield  {journal} {\bibinfo  {journal} {Commun. Math. Phys.}\ }\textbf {\bibinfo {volume} {43}},\ \bibinfo {pages} {199} (\bibinfo {year} {1975})},\ \bibinfo {note} {[Erratum: Commun.Math.Phys. 46, 206 (1976)]}\BibitemShut {NoStop}%
\bibitem [{\citenamefont {Hsu}(2004{\natexlab{a}})}]{Hsu:2004ri}%
  \BibitemOpen
  \bibfield  {author} {\bibinfo {author} {\bibfnamefont {S.~D.~H.}\ \bibnamefont {Hsu}},\ }\href {\doibase 10.1016/j.physletb.2004.05.020} {\bibfield  {journal} {\bibinfo  {journal} {Phys. Lett. B}\ }\textbf {\bibinfo {volume} {594}},\ \bibinfo {pages} {13} (\bibinfo {year} {2004}{\natexlab{a}})},\ \Eprint {http://arxiv.org/abs/hep-th/0403052} {arXiv:hep-th/0403052} \BibitemShut {NoStop}%
\bibitem [{\citenamefont {Oliveros}\ \emph {et~al.}(2022)\citenamefont {Oliveros}, \citenamefont {Sabogal},\ and\ \citenamefont {Acero}}]{Oliveros:2022biu}%
  \BibitemOpen
  \bibfield  {author} {\bibinfo {author} {\bibfnamefont {A.}~\bibnamefont {Oliveros}}, \bibinfo {author} {\bibfnamefont {M.~A.}\ \bibnamefont {Sabogal}}, \ and\ \bibinfo {author} {\bibfnamefont {M.~A.}\ \bibnamefont {Acero}},\ }\href {\doibase 10.1140/epjp/s13360-022-02994-z} {\bibfield  {journal} {\bibinfo  {journal} {Eur. Phys. J. Plus}\ }\textbf {\bibinfo {volume} {137}},\ \bibinfo {pages} {783} (\bibinfo {year} {2022})},\ \Eprint {http://arxiv.org/abs/2203.14464} {arXiv:2203.14464 [gr-qc]} \BibitemShut {NoStop}%
\bibitem [{\citenamefont {Bhardwaj}\ \emph {et~al.}(2022)\citenamefont {Bhardwaj}, \citenamefont {Garg}, \citenamefont {Pradhan},\ and\ \citenamefont {Krishnannair}}]{Bhardwaj:2022uhf}%
  \BibitemOpen
  \bibfield  {author} {\bibinfo {author} {\bibfnamefont {V.~K.}\ \bibnamefont {Bhardwaj}}, \bibinfo {author} {\bibfnamefont {P.}~\bibnamefont {Garg}}, \bibinfo {author} {\bibfnamefont {A.}~\bibnamefont {Pradhan}}, \ and\ \bibinfo {author} {\bibfnamefont {S.}~\bibnamefont {Krishnannair}},\ }\href {\doibase 10.1016/j.cjph.2022.06.028} {\bibfield  {journal} {\bibinfo  {journal} {Chin. J. Phys.}\ }\textbf {\bibinfo {volume} {79}},\ \bibinfo {pages} {471} (\bibinfo {year} {2022})},\ \Eprint {http://arxiv.org/abs/2207.00968} {arXiv:2207.00968 [gr-qc]} \BibitemShut {NoStop}%
\bibitem [{\citenamefont {Astashenok}\ and\ \citenamefont {Tepliakov}(2022)}]{Astashenok:2022pni}%
  \BibitemOpen
  \bibfield  {author} {\bibinfo {author} {\bibfnamefont {A.~V.}\ \bibnamefont {Astashenok}}\ and\ \bibinfo {author} {\bibfnamefont {A.~S.}\ \bibnamefont {Tepliakov}},\ }\href {\doibase 10.1142/S0219887822502346} {\bibfield  {journal} {\bibinfo  {journal} {Int. J. Geom. Meth. Mod. Phys.}\ }\textbf {\bibinfo {volume} {19}},\ \bibinfo {pages} {2250234} (\bibinfo {year} {2022})},\ \Eprint {http://arxiv.org/abs/2208.13320} {arXiv:2208.13320 [gr-qc]} \BibitemShut {NoStop}%
\bibitem [{\citenamefont {Astashenok}\ and\ \citenamefont {Tepliakov}(2023)}]{Astashenok:2023jfp}%
  \BibitemOpen
  \bibfield  {author} {\bibinfo {author} {\bibfnamefont {A.~V.}\ \bibnamefont {Astashenok}}\ and\ \bibinfo {author} {\bibfnamefont {A.~S.}\ \bibnamefont {Tepliakov}},\ }\href {\doibase 10.1142/S021827182350058X} {\bibfield  {journal} {\bibinfo  {journal} {Int. J. Mod. Phys. D}\ }\textbf {\bibinfo {volume} {32}},\ \bibinfo {pages} {2350058} (\bibinfo {year} {2023})},\ \Eprint {http://arxiv.org/abs/2305.10573} {arXiv:2305.10573 [gr-qc]} \BibitemShut {NoStop}%
\bibitem [{\citenamefont {Tita}\ \emph {et~al.}(2024)\citenamefont {Tita}, \citenamefont {Gumjudpai},\ and\ \citenamefont {Srisawad}}]{Tita:2024jzw}%
  \BibitemOpen
  \bibfield  {author} {\bibinfo {author} {\bibfnamefont {A.}~\bibnamefont {Tita}}, \bibinfo {author} {\bibfnamefont {B.}~\bibnamefont {Gumjudpai}}, \ and\ \bibinfo {author} {\bibfnamefont {P.}~\bibnamefont {Srisawad}},\ }\href@noop {} {\  (\bibinfo {year} {2024})},\ \Eprint {http://arxiv.org/abs/2402.18604} {arXiv:2402.18604 [gr-qc]} \BibitemShut {NoStop}%
\bibitem [{\citenamefont {Salzano}\ \emph {et~al.}(2017)\citenamefont {Salzano}, \citenamefont {Mota}, \citenamefont {Capozziello},\ and\ \citenamefont {Donahue}}]{Salzano:2017qac}%
  \BibitemOpen
  \bibfield  {author} {\bibinfo {author} {\bibfnamefont {V.}~\bibnamefont {Salzano}}, \bibinfo {author} {\bibfnamefont {D.~F.}\ \bibnamefont {Mota}}, \bibinfo {author} {\bibfnamefont {S.}~\bibnamefont {Capozziello}}, \ and\ \bibinfo {author} {\bibfnamefont {M.}~\bibnamefont {Donahue}},\ }\href {\doibase 10.1103/PhysRevD.95.044038} {\bibfield  {journal} {\bibinfo  {journal} {Phys. Rev. D}\ }\textbf {\bibinfo {volume} {95}},\ \bibinfo {pages} {044038} (\bibinfo {year} {2017})},\ \Eprint {http://arxiv.org/abs/1701.03517} {arXiv:1701.03517 [astro-ph.CO]} \BibitemShut {NoStop}%
\bibitem [{\citenamefont {Saridakis}(2020)}]{Saridakis_2020}%
  \BibitemOpen
  \bibfield  {author} {\bibinfo {author} {\bibfnamefont {E.~N.}\ \bibnamefont {Saridakis}},\ }\href {\doibase 10.1088/1475-7516/2020/07/031} {\bibfield  {journal} {\bibinfo  {journal} {Journal of Cosmology and Astroparticle Physics}\ }\textbf {\bibinfo {volume} {2020}},\ \bibinfo {pages} {031–031} (\bibinfo {year} {2020})}\BibitemShut {NoStop}%
\bibitem [{\citenamefont {Leon}\ \emph {et~al.}(2021)\citenamefont {Leon}, \citenamefont {Magaña}, \citenamefont {Hernández-Almada}, \citenamefont {García-Aspeitia}, \citenamefont {Verdugo},\ and\ \citenamefont {Motta}}]{Leon_2021}%
  \BibitemOpen
  \bibfield  {author} {\bibinfo {author} {\bibfnamefont {G.}~\bibnamefont {Leon}}, \bibinfo {author} {\bibfnamefont {J.}~\bibnamefont {Magaña}}, \bibinfo {author} {\bibfnamefont {A.}~\bibnamefont {Hernández-Almada}}, \bibinfo {author} {\bibfnamefont {M.~A.}\ \bibnamefont {García-Aspeitia}}, \bibinfo {author} {\bibfnamefont {T.}~\bibnamefont {Verdugo}}, \ and\ \bibinfo {author} {\bibfnamefont {V.}~\bibnamefont {Motta}},\ }\href {\doibase 10.1088/1475-7516/2021/12/032} {\bibfield  {journal} {\bibinfo  {journal} {Journal of Cosmology and Astroparticle Physics}\ }\textbf {\bibinfo {volume} {2021}},\ \bibinfo {pages} {032} (\bibinfo {year} {2021})}\BibitemShut {NoStop}%
\bibitem [{\citenamefont {Cai}\ and\ \citenamefont {Kim}(2005)}]{Cai:2005ra}%
  \BibitemOpen
  \bibfield  {author} {\bibinfo {author} {\bibfnamefont {R.-G.}\ \bibnamefont {Cai}}\ and\ \bibinfo {author} {\bibfnamefont {S.~P.}\ \bibnamefont {Kim}},\ }\href {\doibase 10.1088/1126-6708/2005/02/050} {\bibfield  {journal} {\bibinfo  {journal} {JHEP}\ }\textbf {\bibinfo {volume} {02}},\ \bibinfo {pages} {050} (\bibinfo {year} {2005})},\ \Eprint {http://arxiv.org/abs/hep-th/0501055} {arXiv:hep-th/0501055} \BibitemShut {NoStop}%
\bibitem [{\citenamefont {Asghari}\ \emph {et~al.}(2024)\citenamefont {Asghari}, \citenamefont {Allahyari},\ and\ \citenamefont {Mota}}]{Asghari:2024sbu}%
  \BibitemOpen
  \bibfield  {author} {\bibinfo {author} {\bibfnamefont {M.}~\bibnamefont {Asghari}}, \bibinfo {author} {\bibfnamefont {A.}~\bibnamefont {Allahyari}}, \ and\ \bibinfo {author} {\bibfnamefont {D.~F.}\ \bibnamefont {Mota}},\ }\href@noop {} {\  (\bibinfo {year} {2024})},\ \Eprint {http://arxiv.org/abs/2404.13025} {arXiv:2404.13025 [gr-qc]} \BibitemShut {NoStop}%
\bibitem [{\citenamefont {Hawking}(1976)}]{PhysRevD.13.191}%
  \BibitemOpen
  \bibfield  {author} {\bibinfo {author} {\bibfnamefont {S.~W.}\ \bibnamefont {Hawking}},\ }\href {\doibase 10.1103/PhysRevD.13.191} {\bibfield  {journal} {\bibinfo  {journal} {Phys. Rev. D}\ }\textbf {\bibinfo {volume} {13}},\ \bibinfo {pages} {191} (\bibinfo {year} {1976})}\BibitemShut {NoStop}%
\bibitem [{\citenamefont {Tsallis}\ and\ \citenamefont {Cirto}(2013)}]{Tsallis_2013}%
  \BibitemOpen
  \bibfield  {author} {\bibinfo {author} {\bibfnamefont {C.}~\bibnamefont {Tsallis}}\ and\ \bibinfo {author} {\bibfnamefont {L.~J.~L.}\ \bibnamefont {Cirto}},\ }\href {\doibase 10.1140/epjc/s10052-013-2487-6} {\bibfield  {journal} {\bibinfo  {journal} {The European Physical Journal C}\ }\textbf {\bibinfo {volume} {73}} (\bibinfo {year} {2013}),\ 10.1140/epjc/s10052-013-2487-6}\BibitemShut {NoStop}%
\bibitem [{\citenamefont {Tsallis}(1988)}]{tsallis1988possible}%
  \BibitemOpen
  \bibfield  {author} {\bibinfo {author} {\bibfnamefont {C.}~\bibnamefont {Tsallis}},\ }\href@noop {} {\bibfield  {journal} {\bibinfo  {journal} {Journal of statistical physics}\ }\textbf {\bibinfo {volume} {52}},\ \bibinfo {pages} {479} (\bibinfo {year} {1988})}\BibitemShut {NoStop}%
\bibitem [{\citenamefont {Rényi}(1960)}]{Rényi}%
  \BibitemOpen
  \bibfield  {author} {\bibinfo {author} {\bibfnamefont {A.}~\bibnamefont {Rényi}},\ }\href@noop {} {\bibfield  {journal} {\bibinfo  {journal} {Fourth Berkeley symposium on mathematical statistics and probability}\ }\textbf {\bibinfo {volume} {1}},\ \bibinfo {pages} {547} (\bibinfo {year} {1960})}\BibitemShut {NoStop}%
\bibitem [{\citenamefont {Barrow}(2020{\natexlab{a}})}]{Barrow:2020tzx}%
  \BibitemOpen
  \bibfield  {author} {\bibinfo {author} {\bibfnamefont {J.~D.}\ \bibnamefont {Barrow}},\ }\href {\doibase 10.1016/j.physletb.2020.135643} {\bibfield  {journal} {\bibinfo  {journal} {Phys. Lett. B}\ }\textbf {\bibinfo {volume} {808}},\ \bibinfo {pages} {135643} (\bibinfo {year} {2020}{\natexlab{a}})},\ \Eprint {http://arxiv.org/abs/2004.09444} {arXiv:2004.09444 [gr-qc]} \BibitemShut {NoStop}%
\bibitem [{\citenamefont {Sharma}\ and\ \citenamefont {Mittal}(1975)}]{sharma1975new}%
  \BibitemOpen
  \bibfield  {author} {\bibinfo {author} {\bibfnamefont {B.}~\bibnamefont {Sharma}}\ and\ \bibinfo {author} {\bibfnamefont {D.}~\bibnamefont {Mittal}},\ }\href@noop {} {\bibfield  {journal} {\bibinfo  {journal} {J. Math. Sci}\ }\textbf {\bibinfo {volume} {10}},\ \bibinfo {pages} {28} (\bibinfo {year} {1975})}\BibitemShut {NoStop}%
\bibitem [{\citenamefont {Sharma}\ and\ \citenamefont {Mittal}(1977)}]{sharma1977new}%
  \BibitemOpen
  \bibfield  {author} {\bibinfo {author} {\bibfnamefont {B.}~\bibnamefont {Sharma}}\ and\ \bibinfo {author} {\bibfnamefont {D.}~\bibnamefont {Mittal}},\ }\href@noop {} {\bibfield  {journal} {\bibinfo  {journal} {Journal of Combinatorics Information \& System Sciences}\ }\textbf {\bibinfo {volume} {2}},\ \bibinfo {pages} {122} (\bibinfo {year} {1977})}\BibitemShut {NoStop}%
\bibitem [{\citenamefont {Kaniadakis}(2005)}]{Kaniadakis:2005zk}%
  \BibitemOpen
  \bibfield  {author} {\bibinfo {author} {\bibfnamefont {G.}~\bibnamefont {Kaniadakis}},\ }\href {\doibase 10.1103/PhysRevE.72.036108} {\bibfield  {journal} {\bibinfo  {journal} {Phys. Rev. E}\ }\textbf {\bibinfo {volume} {72}},\ \bibinfo {pages} {036108} (\bibinfo {year} {2005})},\ \Eprint {http://arxiv.org/abs/cond-mat/0507311} {arXiv:cond-mat/0507311} \BibitemShut {NoStop}%
\bibitem [{\citenamefont {Nojiri}\ \emph {et~al.}(2022{\natexlab{a}})\citenamefont {Nojiri}, \citenamefont {Odintsov},\ and\ \citenamefont {Faraoni}}]{Nojiri:2022aof}%
  \BibitemOpen
  \bibfield  {author} {\bibinfo {author} {\bibfnamefont {S.}~\bibnamefont {Nojiri}}, \bibinfo {author} {\bibfnamefont {S.~D.}\ \bibnamefont {Odintsov}}, \ and\ \bibinfo {author} {\bibfnamefont {V.}~\bibnamefont {Faraoni}},\ }\href {\doibase 10.1103/PhysRevD.105.044042} {\bibfield  {journal} {\bibinfo  {journal} {Phys. Rev. D}\ }\textbf {\bibinfo {volume} {105}},\ \bibinfo {pages} {044042} (\bibinfo {year} {2022}{\natexlab{a}})},\ \Eprint {http://arxiv.org/abs/2201.02424} {arXiv:2201.02424 [gr-qc]} \BibitemShut {NoStop}%
\bibitem [{\citenamefont {Nojiri}\ \emph {et~al.}(2023{\natexlab{a}})\citenamefont {Nojiri}, \citenamefont {Odintsov},\ and\ \citenamefont {Paul}}]{Nojiri:2023bom}%
  \BibitemOpen
  \bibfield  {author} {\bibinfo {author} {\bibfnamefont {S.}~\bibnamefont {Nojiri}}, \bibinfo {author} {\bibfnamefont {S.~D.}\ \bibnamefont {Odintsov}}, \ and\ \bibinfo {author} {\bibfnamefont {T.}~\bibnamefont {Paul}},\ }\href {\doibase 10.1016/j.physletb.2023.138321} {\bibfield  {journal} {\bibinfo  {journal} {Phys. Lett. B}\ }\textbf {\bibinfo {volume} {847}},\ \bibinfo {pages} {138321} (\bibinfo {year} {2023}{\natexlab{a}})},\ \Eprint {http://arxiv.org/abs/2311.03848} {arXiv:2311.03848 [gr-qc]} \BibitemShut {NoStop}%
\bibitem [{\citenamefont {Ourabah}(2024)}]{Ourabah:2023rsm}%
  \BibitemOpen
  \bibfield  {author} {\bibinfo {author} {\bibfnamefont {K.}~\bibnamefont {Ourabah}},\ }\href {\doibase 10.1088/1361-6382/ad0eeb} {\bibfield  {journal} {\bibinfo  {journal} {Class. Quant. Grav.}\ }\textbf {\bibinfo {volume} {41}},\ \bibinfo {pages} {015010} (\bibinfo {year} {2024})},\ \Eprint {http://arxiv.org/abs/2306.09540} {arXiv:2306.09540 [gr-qc]} \BibitemShut {NoStop}%
\bibitem [{\citenamefont {Odintsov}\ \emph {et~al.}(2023)\citenamefont {Odintsov}, \citenamefont {D'Onofrio},\ and\ \citenamefont {Paul}}]{Odintsov:2023vpj}%
  \BibitemOpen
  \bibfield  {author} {\bibinfo {author} {\bibfnamefont {S.~D.}\ \bibnamefont {Odintsov}}, \bibinfo {author} {\bibfnamefont {S.}~\bibnamefont {D'Onofrio}}, \ and\ \bibinfo {author} {\bibfnamefont {T.}~\bibnamefont {Paul}},\ }\href {\doibase 10.1016/j.dark.2023.101277} {\bibfield  {journal} {\bibinfo  {journal} {Phys. Dark Univ.}\ }\textbf {\bibinfo {volume} {42}},\ \bibinfo {pages} {101277} (\bibinfo {year} {2023})},\ \Eprint {http://arxiv.org/abs/2306.15225} {arXiv:2306.15225 [gr-qc]} \BibitemShut {NoStop}%
\bibitem [{\citenamefont {Naeem}\ and\ \citenamefont {Bibi}(2024)}]{Naeem:2023tcu}%
  \BibitemOpen
  \bibfield  {author} {\bibinfo {author} {\bibfnamefont {M.}~\bibnamefont {Naeem}}\ and\ \bibinfo {author} {\bibfnamefont {A.}~\bibnamefont {Bibi}},\ }\href {\doibase 10.1016/j.aop.2024.169618} {\bibfield  {journal} {\bibinfo  {journal} {Annals Phys.}\ }\textbf {\bibinfo {volume} {462}},\ \bibinfo {pages} {169618} (\bibinfo {year} {2024})},\ \Eprint {http://arxiv.org/abs/2308.02936} {arXiv:2308.02936 [gr-qc]} \BibitemShut {NoStop}%
\bibitem [{\citenamefont {Nakarachinda}\ \emph {et~al.}(2023)\citenamefont {Nakarachinda}, \citenamefont {Pongkitivanichkul}, \citenamefont {Samart}, \citenamefont {Tannukij},\ and\ \citenamefont {Wongjun}}]{Nakarachinda:2023jko}%
  \BibitemOpen
  \bibfield  {author} {\bibinfo {author} {\bibfnamefont {R.}~\bibnamefont {Nakarachinda}}, \bibinfo {author} {\bibfnamefont {C.}~\bibnamefont {Pongkitivanichkul}}, \bibinfo {author} {\bibfnamefont {D.}~\bibnamefont {Samart}}, \bibinfo {author} {\bibfnamefont {L.}~\bibnamefont {Tannukij}}, \ and\ \bibinfo {author} {\bibfnamefont {P.}~\bibnamefont {Wongjun}},\ }\href@noop {} {\  (\bibinfo {year} {2023})},\ \Eprint {http://arxiv.org/abs/2312.16901} {arXiv:2312.16901 [gr-qc]} \BibitemShut {NoStop}%
\bibitem [{\citenamefont {Fazlollahi}(2023)}]{Fazlollahi:2022bgf}%
  \BibitemOpen
  \bibfield  {author} {\bibinfo {author} {\bibfnamefont {H.~R.}\ \bibnamefont {Fazlollahi}},\ }\href {\doibase 10.1140/epjc/s10052-023-11183-w} {\bibfield  {journal} {\bibinfo  {journal} {Eur. Phys. J. C}\ }\textbf {\bibinfo {volume} {83}},\ \bibinfo {pages} {29} (\bibinfo {year} {2023})},\ \Eprint {http://arxiv.org/abs/2208.12048} {arXiv:2208.12048 [gr-qc]} \BibitemShut {NoStop}%
\bibitem [{\citenamefont {Kord~Zangeneh}\ and\ \citenamefont {Sheykhi}(2023)}]{KordZangeneh:2023syq}%
  \BibitemOpen
  \bibfield  {author} {\bibinfo {author} {\bibfnamefont {M.}~\bibnamefont {Kord~Zangeneh}}\ and\ \bibinfo {author} {\bibfnamefont {A.}~\bibnamefont {Sheykhi}},\ }\href@noop {} {\  (\bibinfo {year} {2023})},\ \Eprint {http://arxiv.org/abs/2311.01969} {arXiv:2311.01969 [gr-qc]} \BibitemShut {NoStop}%
\bibitem [{\citenamefont {Salehi}(2023)}]{Salehi:2023zqg}%
  \BibitemOpen
  \bibfield  {author} {\bibinfo {author} {\bibfnamefont {A.}~\bibnamefont {Salehi}},\ }\href@noop {} {\  (\bibinfo {year} {2023})},\ \Eprint {http://arxiv.org/abs/2309.15956} {arXiv:2309.15956 [gr-qc]} \BibitemShut {NoStop}%
\bibitem [{\citenamefont {Nojiri}\ \emph {et~al.}(2022{\natexlab{b}})\citenamefont {Nojiri}, \citenamefont {Odintsov},\ and\ \citenamefont {Paul}}]{Nojiri:2021jxf}%
  \BibitemOpen
  \bibfield  {author} {\bibinfo {author} {\bibfnamefont {S.}~\bibnamefont {Nojiri}}, \bibinfo {author} {\bibfnamefont {S.~D.}\ \bibnamefont {Odintsov}}, \ and\ \bibinfo {author} {\bibfnamefont {T.}~\bibnamefont {Paul}},\ }\href {\doibase 10.1016/j.physletb.2021.136844} {\bibfield  {journal} {\bibinfo  {journal} {Phys. Lett. B}\ }\textbf {\bibinfo {volume} {825}},\ \bibinfo {pages} {136844} (\bibinfo {year} {2022}{\natexlab{b}})},\ \Eprint {http://arxiv.org/abs/2112.10159} {arXiv:2112.10159 [gr-qc]} \BibitemShut {NoStop}%
\bibitem [{\citenamefont {Nojiri}\ \emph {et~al.}(2019)\citenamefont {Nojiri}, \citenamefont {Odintsov},\ and\ \citenamefont {Saridakis}}]{Nojiri:2019skr}%
  \BibitemOpen
  \bibfield  {author} {\bibinfo {author} {\bibfnamefont {S.}~\bibnamefont {Nojiri}}, \bibinfo {author} {\bibfnamefont {S.~D.}\ \bibnamefont {Odintsov}}, \ and\ \bibinfo {author} {\bibfnamefont {E.~N.}\ \bibnamefont {Saridakis}},\ }\href {\doibase 10.1140/epjc/s10052-019-6740-5} {\bibfield  {journal} {\bibinfo  {journal} {Eur. Phys. J. C}\ }\textbf {\bibinfo {volume} {79}},\ \bibinfo {pages} {242} (\bibinfo {year} {2019})},\ \Eprint {http://arxiv.org/abs/1903.03098} {arXiv:1903.03098 [gr-qc]} \BibitemShut {NoStop}%
\bibitem [{\citenamefont {Nojiri}\ \emph {et~al.}(2022{\natexlab{c}})\citenamefont {Nojiri}, \citenamefont {Odintsov},\ and\ \citenamefont {Paul}}]{Nojiri:2022dkr}%
  \BibitemOpen
  \bibfield  {author} {\bibinfo {author} {\bibfnamefont {S.}~\bibnamefont {Nojiri}}, \bibinfo {author} {\bibfnamefont {S.~D.}\ \bibnamefont {Odintsov}}, \ and\ \bibinfo {author} {\bibfnamefont {T.}~\bibnamefont {Paul}},\ }\href {\doibase 10.1016/j.physletb.2022.137189} {\bibfield  {journal} {\bibinfo  {journal} {Phys. Lett. B}\ }\textbf {\bibinfo {volume} {831}},\ \bibinfo {pages} {137189} (\bibinfo {year} {2022}{\natexlab{c}})},\ \Eprint {http://arxiv.org/abs/2205.08876} {arXiv:2205.08876 [gr-qc]} \BibitemShut {NoStop}%
\bibitem [{\citenamefont {Odintsov}\ and\ \citenamefont {Paul}(2023)}]{Odintsov:2022qnn}%
  \BibitemOpen
  \bibfield  {author} {\bibinfo {author} {\bibfnamefont {S.~D.}\ \bibnamefont {Odintsov}}\ and\ \bibinfo {author} {\bibfnamefont {T.}~\bibnamefont {Paul}},\ }\href {\doibase 10.1016/j.dark.2022.101159} {\bibfield  {journal} {\bibinfo  {journal} {Phys. Dark Univ.}\ }\textbf {\bibinfo {volume} {39}},\ \bibinfo {pages} {101159} (\bibinfo {year} {2023})},\ \Eprint {http://arxiv.org/abs/2212.05531} {arXiv:2212.05531 [gr-qc]} \BibitemShut {NoStop}%
\bibitem [{\citenamefont {Nojiri}\ \emph {et~al.}(2022{\natexlab{d}})\citenamefont {Nojiri}, \citenamefont {Odintsov},\ and\ \citenamefont {Paul}}]{Nojiri:2022nmu}%
  \BibitemOpen
  \bibfield  {author} {\bibinfo {author} {\bibfnamefont {S.}~\bibnamefont {Nojiri}}, \bibinfo {author} {\bibfnamefont {S.~D.}\ \bibnamefont {Odintsov}}, \ and\ \bibinfo {author} {\bibfnamefont {T.}~\bibnamefont {Paul}},\ }\href {\doibase 10.1016/j.physletb.2022.137553} {\bibfield  {journal} {\bibinfo  {journal} {Phys. Lett. B}\ }\textbf {\bibinfo {volume} {835}},\ \bibinfo {pages} {137553} (\bibinfo {year} {2022}{\natexlab{d}})},\ \Eprint {http://arxiv.org/abs/2211.02822} {arXiv:2211.02822 [gr-qc]} \BibitemShut {NoStop}%
\bibitem [{\citenamefont {Nojiri}\ \emph {et~al.}(2023{\natexlab{b}})\citenamefont {Nojiri}, \citenamefont {Odintsov},\ and\ \citenamefont {Paul}}]{Nojiri:2023nop}%
  \BibitemOpen
  \bibfield  {author} {\bibinfo {author} {\bibfnamefont {S.}~\bibnamefont {Nojiri}}, \bibinfo {author} {\bibfnamefont {S.~D.}\ \bibnamefont {Odintsov}}, \ and\ \bibinfo {author} {\bibfnamefont {T.}~\bibnamefont {Paul}},\ }\href {\doibase 10.1016/j.physletb.2023.137926} {\bibfield  {journal} {\bibinfo  {journal} {Phys. Lett. B}\ }\textbf {\bibinfo {volume} {841}},\ \bibinfo {pages} {137926} (\bibinfo {year} {2023}{\natexlab{b}})},\ \Eprint {http://arxiv.org/abs/2304.09436} {arXiv:2304.09436 [gr-qc]} \BibitemShut {NoStop}%
\bibitem [{\citenamefont {Nojiri}\ \emph {et~al.}(2024)\citenamefont {Nojiri}, \citenamefont {Odintsov}, \citenamefont {Paul},\ and\ \citenamefont {SenGupta}}]{Nojiri:2023wzz}%
  \BibitemOpen
  \bibfield  {author} {\bibinfo {author} {\bibfnamefont {S.}~\bibnamefont {Nojiri}}, \bibinfo {author} {\bibfnamefont {S.~D.}\ \bibnamefont {Odintsov}}, \bibinfo {author} {\bibfnamefont {T.}~\bibnamefont {Paul}}, \ and\ \bibinfo {author} {\bibfnamefont {S.}~\bibnamefont {SenGupta}},\ }\href {\doibase 10.1103/PhysRevD.109.043532} {\bibfield  {journal} {\bibinfo  {journal} {Phys. Rev. D}\ }\textbf {\bibinfo {volume} {109}},\ \bibinfo {pages} {043532} (\bibinfo {year} {2024})},\ \Eprint {http://arxiv.org/abs/2307.05011} {arXiv:2307.05011 [gr-qc]} \BibitemShut {NoStop}%
\bibitem [{\citenamefont {Manoharan}(2024)}]{Manoharan:2024thb}%
  \BibitemOpen
  \bibfield  {author} {\bibinfo {author} {\bibfnamefont {M.~T.}\ \bibnamefont {Manoharan}},\ }\href {\doibase 10.1140/epjc/s10052-024-12926-z} {\bibfield  {journal} {\bibinfo  {journal} {Eur. Phys. J. C}\ }\textbf {\bibinfo {volume} {84}},\ \bibinfo {pages} {552} (\bibinfo {year} {2024})}\BibitemShut {NoStop}%
\bibitem [{\citenamefont {Manoharan}\ \emph {et~al.}(2023)\citenamefont {Manoharan}, \citenamefont {Shaji},\ and\ \citenamefont {Mathew}}]{Manoharan:2022qll}%
  \BibitemOpen
  \bibfield  {author} {\bibinfo {author} {\bibfnamefont {M.~T.}\ \bibnamefont {Manoharan}}, \bibinfo {author} {\bibfnamefont {N.}~\bibnamefont {Shaji}}, \ and\ \bibinfo {author} {\bibfnamefont {T.~K.}\ \bibnamefont {Mathew}},\ }\href {\doibase 10.1140/epjc/s10052-023-11202-w} {\bibfield  {journal} {\bibinfo  {journal} {Eur. Phys. J. C}\ }\textbf {\bibinfo {volume} {83}},\ \bibinfo {pages} {19} (\bibinfo {year} {2023})},\ \Eprint {http://arxiv.org/abs/2208.08736} {arXiv:2208.08736 [gr-qc]} \BibitemShut {NoStop}%
\bibitem [{\citenamefont {Barrow}(2020{\natexlab{b}})}]{Barrow_2020}%
  \BibitemOpen
  \bibfield  {author} {\bibinfo {author} {\bibfnamefont {J.~D.}\ \bibnamefont {Barrow}},\ }\href {\doibase 10.1016/j.physletb.2020.135643} {\bibfield  {journal} {\bibinfo  {journal} {Physics Letters B}\ }\textbf {\bibinfo {volume} {808}},\ \bibinfo {pages} {135643} (\bibinfo {year} {2020}{\natexlab{b}})}\BibitemShut {NoStop}%
\bibitem [{\citenamefont {Srednicki}(1993)}]{Srednicki:1993im}%
  \BibitemOpen
  \bibfield  {author} {\bibinfo {author} {\bibfnamefont {M.}~\bibnamefont {Srednicki}},\ }\href {\doibase 10.1103/PhysRevLett.71.666} {\bibfield  {journal} {\bibinfo  {journal} {Phys. Rev. Lett.}\ }\textbf {\bibinfo {volume} {71}},\ \bibinfo {pages} {666} (\bibinfo {year} {1993})},\ \Eprint {http://arxiv.org/abs/hep-th/9303048} {arXiv:hep-th/9303048} \BibitemShut {NoStop}%
\bibitem [{\citenamefont {Jacobson}(2016)}]{Jacobson:2015hqa}%
  \BibitemOpen
  \bibfield  {author} {\bibinfo {author} {\bibfnamefont {T.}~\bibnamefont {Jacobson}},\ }\href {\doibase 10.1103/PhysRevLett.116.201101} {\bibfield  {journal} {\bibinfo  {journal} {Phys. Rev. Lett.}\ }\textbf {\bibinfo {volume} {116}},\ \bibinfo {pages} {201101} (\bibinfo {year} {2016})},\ \Eprint {http://arxiv.org/abs/1505.04753} {arXiv:1505.04753 [gr-qc]} \BibitemShut {NoStop}%
\bibitem [{\citenamefont {Jacobson}(1995)}]{Jacobson:1995ab}%
  \BibitemOpen
  \bibfield  {author} {\bibinfo {author} {\bibfnamefont {T.}~\bibnamefont {Jacobson}},\ }\href {\doibase 10.1103/PhysRevLett.75.1260} {\bibfield  {journal} {\bibinfo  {journal} {Phys. Rev. Lett.}\ }\textbf {\bibinfo {volume} {75}},\ \bibinfo {pages} {1260} (\bibinfo {year} {1995})},\ \Eprint {http://arxiv.org/abs/gr-qc/9504004} {arXiv:gr-qc/9504004} \BibitemShut {NoStop}%
\bibitem [{\citenamefont {Padmanabhan}(2005)}]{Padmanabhan:2003gd}%
  \BibitemOpen
  \bibfield  {author} {\bibinfo {author} {\bibfnamefont {T.}~\bibnamefont {Padmanabhan}},\ }\href {\doibase 10.1016/j.physrep.2004.10.003} {\bibfield  {journal} {\bibinfo  {journal} {Phys. Rept.}\ }\textbf {\bibinfo {volume} {406}},\ \bibinfo {pages} {49} (\bibinfo {year} {2005})},\ \Eprint {http://arxiv.org/abs/gr-qc/0311036} {arXiv:gr-qc/0311036} \BibitemShut {NoStop}%
\bibitem [{\citenamefont {Carroll}\ and\ \citenamefont {Remmen}(2016)}]{Carroll:2016lku}%
  \BibitemOpen
  \bibfield  {author} {\bibinfo {author} {\bibfnamefont {S.~M.}\ \bibnamefont {Carroll}}\ and\ \bibinfo {author} {\bibfnamefont {G.~N.}\ \bibnamefont {Remmen}},\ }\href {\doibase 10.1103/PhysRevD.93.124052} {\bibfield  {journal} {\bibinfo  {journal} {Phys. Rev. D}\ }\textbf {\bibinfo {volume} {93}},\ \bibinfo {pages} {124052} (\bibinfo {year} {2016})},\ \Eprint {http://arxiv.org/abs/1601.07558} {arXiv:1601.07558 [hep-th]} \BibitemShut {NoStop}%
\bibitem [{\citenamefont {Li}(2004)}]{Li_2004}%
  \BibitemOpen
  \bibfield  {author} {\bibinfo {author} {\bibfnamefont {M.}~\bibnamefont {Li}},\ }\href {\doibase 10.1016/j.physletb.2004.10.014} {\bibfield  {journal} {\bibinfo  {journal} {Physics Letters B}\ }\textbf {\bibinfo {volume} {603}},\ \bibinfo {pages} {1–5} (\bibinfo {year} {2004})}\BibitemShut {NoStop}%
\bibitem [{\citenamefont {Wang}\ \emph {et~al.}(2017{\natexlab{b}})\citenamefont {Wang}, \citenamefont {Wang},\ and\ \citenamefont {Li}}]{WANG20171}%
  \BibitemOpen
  \bibfield  {author} {\bibinfo {author} {\bibfnamefont {S.}~\bibnamefont {Wang}}, \bibinfo {author} {\bibfnamefont {Y.}~\bibnamefont {Wang}}, \ and\ \bibinfo {author} {\bibfnamefont {M.}~\bibnamefont {Li}},\ }\href {\doibase https://doi.org/10.1016/j.physrep.2017.06.003} {\bibfield  {journal} {\bibinfo  {journal} {Physics Reports}\ }\textbf {\bibinfo {volume} {696}},\ \bibinfo {pages} {1} (\bibinfo {year} {2017}{\natexlab{b}})},\ \bibinfo {note} {holographic Dark Energy}\BibitemShut {NoStop}%
\bibitem [{\citenamefont {Padmanabhan}(2010)}]{Padmanabhan:2009vy}%
  \BibitemOpen
  \bibfield  {author} {\bibinfo {author} {\bibfnamefont {T.}~\bibnamefont {Padmanabhan}},\ }\href {\doibase 10.1088/0034-4885/73/4/046901} {\bibfield  {journal} {\bibinfo  {journal} {Rept. Prog. Phys.}\ }\textbf {\bibinfo {volume} {73}},\ \bibinfo {pages} {046901} (\bibinfo {year} {2010})},\ \Eprint {http://arxiv.org/abs/0911.5004} {arXiv:0911.5004 [gr-qc]} \BibitemShut {NoStop}%
\bibitem [{\citenamefont {Adame}\ \emph {et~al.}(2024)\citenamefont {Adame} \emph {et~al.}}]{DESI:2024mwx}%
  \BibitemOpen
  \bibfield  {author} {\bibinfo {author} {\bibfnamefont {A.~G.}\ \bibnamefont {Adame}} \emph {et~al.} (\bibinfo {collaboration} {DESI}),\ }\href@noop {} {\  (\bibinfo {year} {2024})},\ \Eprint {http://arxiv.org/abs/2404.03002} {arXiv:2404.03002 [astro-ph.CO]} \BibitemShut {NoStop}%
\bibitem [{\citenamefont {Barrow}(2020{\natexlab{c}})}]{BARROW2020135643}%
  \BibitemOpen
  \bibfield  {author} {\bibinfo {author} {\bibfnamefont {J.~D.}\ \bibnamefont {Barrow}},\ }\href {\doibase https://doi.org/10.1016/j.physletb.2020.135643} {\bibfield  {journal} {\bibinfo  {journal} {Physics Letters B}\ }\textbf {\bibinfo {volume} {808}},\ \bibinfo {pages} {135643} (\bibinfo {year} {2020}{\natexlab{c}})}\BibitemShut {NoStop}%
\bibitem [{\citenamefont {Denkiewicz}\ \emph {et~al.}(2023)\citenamefont {Denkiewicz}, \citenamefont {Salzano},\ and\ \citenamefont {Dabrowski}}]{Denkiewicz:2023hyj}%
  \BibitemOpen
  \bibfield  {author} {\bibinfo {author} {\bibfnamefont {T.}~\bibnamefont {Denkiewicz}}, \bibinfo {author} {\bibfnamefont {V.}~\bibnamefont {Salzano}}, \ and\ \bibinfo {author} {\bibfnamefont {M.~P.}\ \bibnamefont {Dabrowski}},\ }\href {\doibase 10.1103/PhysRevD.108.103533} {\bibfield  {journal} {\bibinfo  {journal} {Phys. Rev. D}\ }\textbf {\bibinfo {volume} {108}},\ \bibinfo {pages} {103533} (\bibinfo {year} {2023})},\ \Eprint {http://arxiv.org/abs/2303.11680} {arXiv:2303.11680 [astro-ph.CO]} \BibitemShut {NoStop}%
\bibitem [{\citenamefont {Hsu}(2004{\natexlab{b}})}]{Hsu_2004}%
  \BibitemOpen
  \bibfield  {author} {\bibinfo {author} {\bibfnamefont {S.~D.}\ \bibnamefont {Hsu}},\ }\href {\doibase 10.1016/j.physletb.2004.05.020} {\bibfield  {journal} {\bibinfo  {journal} {Physics Letters B}\ }\textbf {\bibinfo {volume} {594}},\ \bibinfo {pages} {13–16} (\bibinfo {year} {2004}{\natexlab{b}})}\BibitemShut {NoStop}%
\bibitem [{\citenamefont {Davis}\ and\ \citenamefont {Lineweaver}(2004)}]{Davis:2003ad}%
  \BibitemOpen
  \bibfield  {author} {\bibinfo {author} {\bibfnamefont {T.~M.}\ \bibnamefont {Davis}}\ and\ \bibinfo {author} {\bibfnamefont {C.~H.}\ \bibnamefont {Lineweaver}},\ }\href {\doibase 10.1071/AS03040} {\bibfield  {journal} {\bibinfo  {journal} {Publ. Astron. Soc. Austral.}\ }\textbf {\bibinfo {volume} {21}},\ \bibinfo {pages} {97} (\bibinfo {year} {2004})},\ \Eprint {http://arxiv.org/abs/astro-ph/0310808} {arXiv:astro-ph/0310808} \BibitemShut {NoStop}%
\bibitem [{\citenamefont {Dabrowski}\ and\ \citenamefont {Salzano}(2020)}]{Dabrowski:2020atl}%
  \BibitemOpen
  \bibfield  {author} {\bibinfo {author} {\bibfnamefont {M.~P.}\ \bibnamefont {Dabrowski}}\ and\ \bibinfo {author} {\bibfnamefont {V.}~\bibnamefont {Salzano}},\ }\href {\doibase 10.1103/PhysRevD.102.064047} {\bibfield  {journal} {\bibinfo  {journal} {Phys. Rev. D}\ }\textbf {\bibinfo {volume} {102}},\ \bibinfo {pages} {064047} (\bibinfo {year} {2020})},\ \Eprint {http://arxiv.org/abs/2009.08306} {arXiv:2009.08306 [astro-ph.CO]} \BibitemShut {NoStop}%
\bibitem [{\citenamefont {Pavon}\ and\ \citenamefont {Zimdahl}(2005)}]{Pavon:2005yx}%
  \BibitemOpen
  \bibfield  {author} {\bibinfo {author} {\bibfnamefont {D.}~\bibnamefont {Pavon}}\ and\ \bibinfo {author} {\bibfnamefont {W.}~\bibnamefont {Zimdahl}},\ }\href {\doibase 10.1016/j.physletb.2005.08.134} {\bibfield  {journal} {\bibinfo  {journal} {Phys. Lett. B}\ }\textbf {\bibinfo {volume} {628}},\ \bibinfo {pages} {206} (\bibinfo {year} {2005})},\ \Eprint {http://arxiv.org/abs/gr-qc/0505020} {arXiv:gr-qc/0505020} \BibitemShut {NoStop}%
\bibitem [{\citenamefont {Haridasu}\ \emph {et~al.}(2018{\natexlab{b}})\citenamefont {Haridasu}, \citenamefont {Lukovi{\'c}}, \citenamefont {Moresco},\ and\ \citenamefont {Vittorio}}]{Haridasu18}%
  \BibitemOpen
  \bibfield  {author} {\bibinfo {author} {\bibfnamefont {B.~S.}\ \bibnamefont {Haridasu}}, \bibinfo {author} {\bibfnamefont {V.~V.}\ \bibnamefont {Lukovi{\'c}}}, \bibinfo {author} {\bibfnamefont {M.}~\bibnamefont {Moresco}}, \ and\ \bibinfo {author} {\bibfnamefont {N.}~\bibnamefont {Vittorio}},\ }\href {\doibase 10.1088/1475-7516/2018/10/015} {\bibfield  {journal} {\bibinfo  {journal} {JCAP}\ }\textbf {\bibinfo {volume} {1810}},\ \bibinfo {pages} {015} (\bibinfo {year} {2018}{\natexlab{b}})},\ \Eprint {http://arxiv.org/abs/1805.03595} {arXiv:1805.03595 [astro-ph.CO]} \BibitemShut {NoStop}%
\bibitem [{\citenamefont {{G{\'o}mez-Valent}}\ and\ \citenamefont {{Amendola}}(2018)}]{Gomez-Valent18}%
  \BibitemOpen
  \bibfield  {author} {\bibinfo {author} {\bibfnamefont {A.}~\bibnamefont {{G{\'o}mez-Valent}}}\ and\ \bibinfo {author} {\bibfnamefont {L.}~\bibnamefont {{Amendola}}},\ }\href {\doibase 10.1088/1475-7516/2018/04/051} {\bibfield  {journal} {\bibinfo  {journal} {\jcap}\ }\textbf {\bibinfo {volume} {2018}},\ \bibinfo {eid} {051} (\bibinfo {year} {2018})},\ \Eprint {http://arxiv.org/abs/1802.01505} {arXiv:1802.01505 [astro-ph.CO]} \BibitemShut {NoStop}%
\bibitem [{\citenamefont {Farooq}\ and\ \citenamefont {Ratra}(2013)}]{Farooq:2013hq}%
  \BibitemOpen
  \bibfield  {author} {\bibinfo {author} {\bibfnamefont {O.}~\bibnamefont {Farooq}}\ and\ \bibinfo {author} {\bibfnamefont {B.}~\bibnamefont {Ratra}},\ }\href {\doibase 10.1088/2041-8205/766/1/L7} {\bibfield  {journal} {\bibinfo  {journal} {Astrophys. J. Lett.}\ }\textbf {\bibinfo {volume} {766}},\ \bibinfo {pages} {L7} (\bibinfo {year} {2013})},\ \Eprint {http://arxiv.org/abs/1301.5243} {arXiv:1301.5243 [astro-ph.CO]} \BibitemShut {NoStop}%
\bibitem [{\citenamefont {Farooq}\ \emph {et~al.}(2017)\citenamefont {Farooq}, \citenamefont {Madiyar}, \citenamefont {Crandall},\ and\ \citenamefont {Ratra}}]{Farooq:2016zwm}%
  \BibitemOpen
  \bibfield  {author} {\bibinfo {author} {\bibfnamefont {O.}~\bibnamefont {Farooq}}, \bibinfo {author} {\bibfnamefont {F.~R.}\ \bibnamefont {Madiyar}}, \bibinfo {author} {\bibfnamefont {S.}~\bibnamefont {Crandall}}, \ and\ \bibinfo {author} {\bibfnamefont {B.}~\bibnamefont {Ratra}},\ }\href {\doibase 10.3847/1538-4357/835/1/26} {\bibfield  {journal} {\bibinfo  {journal} {Astrophys. J.}\ }\textbf {\bibinfo {volume} {835}},\ \bibinfo {pages} {26} (\bibinfo {year} {2017})},\ \Eprint {http://arxiv.org/abs/1607.03537} {arXiv:1607.03537 [astro-ph.CO]} \BibitemShut {NoStop}%
\bibitem [{\citenamefont {Yu}\ \emph {et~al.}(2018)\citenamefont {Yu}, \citenamefont {Ratra},\ and\ \citenamefont {Wang}}]{Yu:2017iju}%
  \BibitemOpen
  \bibfield  {author} {\bibinfo {author} {\bibfnamefont {H.}~\bibnamefont {Yu}}, \bibinfo {author} {\bibfnamefont {B.}~\bibnamefont {Ratra}}, \ and\ \bibinfo {author} {\bibfnamefont {F.-Y.}\ \bibnamefont {Wang}},\ }\href {\doibase 10.3847/1538-4357/aab0a2} {\bibfield  {journal} {\bibinfo  {journal} {Astrophys. J.}\ }\textbf {\bibinfo {volume} {856}},\ \bibinfo {pages} {3} (\bibinfo {year} {2018})},\ \Eprint {http://arxiv.org/abs/1711.03437} {arXiv:1711.03437 [astro-ph.CO]} \BibitemShut {NoStop}%
\bibitem [{\citenamefont {Brout}\ \emph {et~al.}(2022)\citenamefont {Brout} \emph {et~al.}}]{Brout:2022vxf}%
  \BibitemOpen
  \bibfield  {author} {\bibinfo {author} {\bibfnamefont {D.}~\bibnamefont {Brout}} \emph {et~al.},\ }\href {\doibase 10.3847/1538-4357/ac8e04} {\bibfield  {journal} {\bibinfo  {journal} {Astrophys. J.}\ }\textbf {\bibinfo {volume} {938}},\ \bibinfo {pages} {110} (\bibinfo {year} {2022})},\ \Eprint {http://arxiv.org/abs/2202.04077} {arXiv:2202.04077 [astro-ph.CO]} \BibitemShut {NoStop}%
\bibitem [{\citenamefont {Peterson}\ \emph {et~al.}(2022)\citenamefont {Peterson} \emph {et~al.}}]{Peterson:2021hel}%
  \BibitemOpen
  \bibfield  {author} {\bibinfo {author} {\bibfnamefont {E.~R.}\ \bibnamefont {Peterson}} \emph {et~al.},\ }\href {\doibase 10.3847/1538-4357/ac4698} {\bibfield  {journal} {\bibinfo  {journal} {Astrophys. J.}\ }\textbf {\bibinfo {volume} {938}},\ \bibinfo {pages} {112} (\bibinfo {year} {2022})},\ \Eprint {http://arxiv.org/abs/2110.03487} {arXiv:2110.03487 [astro-ph.CO]} \BibitemShut {NoStop}%
\bibitem [{\citenamefont {M\"oller}\ \emph {et~al.}(2024)\citenamefont {M\"oller} \emph {et~al.}}]{DES:2024haq}%
  \BibitemOpen
  \bibfield  {author} {\bibinfo {author} {\bibfnamefont {A.}~\bibnamefont {M\"oller}} \emph {et~al.} (\bibinfo {collaboration} {DES}),\ }\href@noop {} {\  (\bibinfo {year} {2024})},\ \Eprint {http://arxiv.org/abs/2402.18690} {arXiv:2402.18690 [astro-ph.CO]} \BibitemShut {NoStop}%
\bibitem [{\citenamefont {{Wang}}\ and\ \citenamefont {{Mukherjee}}(2006)}]{Wang06}%
  \BibitemOpen
  \bibfield  {author} {\bibinfo {author} {\bibfnamefont {Y.}~\bibnamefont {{Wang}}}\ and\ \bibinfo {author} {\bibfnamefont {P.}~\bibnamefont {{Mukherjee}}},\ }\href {\doibase 10.1086/507091} {\bibfield  {journal} {\bibinfo  {journal} {\apj}\ }\textbf {\bibinfo {volume} {650}},\ \bibinfo {pages} {1} (\bibinfo {year} {2006})},\ \Eprint {http://arxiv.org/abs/astro-ph/0604051} {arXiv:astro-ph/0604051 [astro-ph]} \BibitemShut {NoStop}%
\bibitem [{\citenamefont {{Wang}}\ and\ \citenamefont {{Mukherjee}}(2007)}]{Wang07}%
  \BibitemOpen
  \bibfield  {author} {\bibinfo {author} {\bibfnamefont {Y.}~\bibnamefont {{Wang}}}\ and\ \bibinfo {author} {\bibfnamefont {P.}~\bibnamefont {{Mukherjee}}},\ }\href {\doibase 10.1103/PhysRevD.76.103533} {\bibfield  {journal} {\bibinfo  {journal} {\prd}\ }\textbf {\bibinfo {volume} {76}},\ \bibinfo {eid} {103533} (\bibinfo {year} {2007})},\ \Eprint {http://arxiv.org/abs/astro-ph/0703780} {arXiv:astro-ph/0703780 [astro-ph]} \BibitemShut {NoStop}%
\bibitem [{\citenamefont {{Verde}}\ \emph {et~al.}(2017)\citenamefont {{Verde}}, \citenamefont {{Bellini}}, \citenamefont {{Pigozzo}}, \citenamefont {{Heavens}},\ and\ \citenamefont {{Jimenez}}}]{Verde17}%
  \BibitemOpen
  \bibfield  {author} {\bibinfo {author} {\bibfnamefont {L.}~\bibnamefont {{Verde}}}, \bibinfo {author} {\bibfnamefont {E.}~\bibnamefont {{Bellini}}}, \bibinfo {author} {\bibfnamefont {C.}~\bibnamefont {{Pigozzo}}}, \bibinfo {author} {\bibfnamefont {A.~F.}\ \bibnamefont {{Heavens}}}, \ and\ \bibinfo {author} {\bibfnamefont {R.}~\bibnamefont {{Jimenez}}},\ }\href {\doibase 10.1088/1475-7516/2017/04/023} {\bibfield  {journal} {\bibinfo  {journal} {Journal of Cosmology and Astro-Particle Physics}\ }\textbf {\bibinfo {volume} {2017}},\ \bibinfo {eid} {023} (\bibinfo {year} {2017})},\ \Eprint {http://arxiv.org/abs/1611.00376} {arXiv:1611.00376 [astro-ph.CO]} \BibitemShut {NoStop}%
\bibitem [{\citenamefont {Haridasu}\ \emph {et~al.}(2021)\citenamefont {Haridasu}, \citenamefont {Viel},\ and\ \citenamefont {Vittorio}}]{Haridasu:2020pms}%
  \BibitemOpen
  \bibfield  {author} {\bibinfo {author} {\bibfnamefont {B.~S.}\ \bibnamefont {Haridasu}}, \bibinfo {author} {\bibfnamefont {M.}~\bibnamefont {Viel}}, \ and\ \bibinfo {author} {\bibfnamefont {N.}~\bibnamefont {Vittorio}},\ }\href {\doibase 10.1103/PhysRevD.103.063539} {\bibfield  {journal} {\bibinfo  {journal} {Phys. Rev. D}\ }\textbf {\bibinfo {volume} {103}},\ \bibinfo {pages} {063539} (\bibinfo {year} {2021})},\ \Eprint {http://arxiv.org/abs/2012.10324} {arXiv:2012.10324 [astro-ph.CO]} \BibitemShut {NoStop}%
\bibitem [{\citenamefont {{Fixsen}}(2009)}]{Fixsen09}%
  \BibitemOpen
  \bibfield  {author} {\bibinfo {author} {\bibfnamefont {D.~J.}\ \bibnamefont {{Fixsen}}},\ }\href {\doibase 10.1088/0004-637X/707/2/916} {\bibfield  {journal} {\bibinfo  {journal} {\apj}\ }\textbf {\bibinfo {volume} {707}},\ \bibinfo {pages} {916} (\bibinfo {year} {2009})},\ \Eprint {http://arxiv.org/abs/0911.1955} {arXiv:0911.1955 [astro-ph.CO]} \BibitemShut {NoStop}%
\bibitem [{\citenamefont {{Foreman-Mackey}}\ \emph {et~al.}(2013)\citenamefont {{Foreman-Mackey}}, \citenamefont {{Hogg}}, \citenamefont {{Lang}},\ and\ \citenamefont {{Goodman}}}]{Foreman-Mackey13}%
  \BibitemOpen
  \bibfield  {author} {\bibinfo {author} {\bibfnamefont {D.}~\bibnamefont {{Foreman-Mackey}}}, \bibinfo {author} {\bibfnamefont {D.~W.}\ \bibnamefont {{Hogg}}}, \bibinfo {author} {\bibfnamefont {D.}~\bibnamefont {{Lang}}}, \ and\ \bibinfo {author} {\bibfnamefont {J.}~\bibnamefont {{Goodman}}},\ }\href {\doibase 10.1086/670067} {\bibfield  {journal} {\bibinfo  {journal} {\pasp}\ }\textbf {\bibinfo {volume} {125}},\ \bibinfo {pages} {306} (\bibinfo {year} {2013})},\ \Eprint {http://arxiv.org/abs/1202.3665} {arXiv:1202.3665 [astro-ph.IM]} \BibitemShut {NoStop}%
\bibitem [{\citenamefont {Lewis}(2019)}]{Lewis:2019xzd}%
  \BibitemOpen
  \bibfield  {author} {\bibinfo {author} {\bibfnamefont {A.}~\bibnamefont {Lewis}},\ }\href@noop {} {\  (\bibinfo {year} {2019})},\ \Eprint {http://arxiv.org/abs/1910.13970} {arXiv:1910.13970 [astro-ph.IM]} \BibitemShut {NoStop}%
\bibitem [{\citenamefont {Trotta}(2017)}]{Trotta:2017wnx}%
  \BibitemOpen
  \bibfield  {author} {\bibinfo {author} {\bibfnamefont {R.}~\bibnamefont {Trotta}}\ }(\bibinfo {year} {2017})\ \Eprint {http://arxiv.org/abs/1701.01467} {arXiv:1701.01467 [astro-ph.CO]} \BibitemShut {NoStop}%
\bibitem [{\citenamefont {Trotta}(2008)}]{Trotta:2008qt}%
  \BibitemOpen
  \bibfield  {author} {\bibinfo {author} {\bibfnamefont {R.}~\bibnamefont {Trotta}},\ }\href {\doibase 10.1080/00107510802066753} {\bibfield  {journal} {\bibinfo  {journal} {Contemp. Phys.}\ }\textbf {\bibinfo {volume} {49}},\ \bibinfo {pages} {71} (\bibinfo {year} {2008})},\ \Eprint {http://arxiv.org/abs/0803.4089} {arXiv:0803.4089 [astro-ph]} \BibitemShut {NoStop}%
\bibitem [{\citenamefont {Haridasu}\ \emph {et~al.}(2018{\natexlab{c}})\citenamefont {Haridasu}, \citenamefont {Lukovi{\'{c}}},\ and\ \citenamefont {Vittorio}}]{Haridasu17_bao}%
  \BibitemOpen
  \bibfield  {author} {\bibinfo {author} {\bibfnamefont {B.~S.}\ \bibnamefont {Haridasu}}, \bibinfo {author} {\bibfnamefont {V.~V.}\ \bibnamefont {Lukovi{\'{c}}}}, \ and\ \bibinfo {author} {\bibfnamefont {N.}~\bibnamefont {Vittorio}},\ }\href {\doibase 10.1088/1475-7516/2018/05/033} {\bibfield  {journal} {\bibinfo  {journal} {JCAP}\ }\textbf {\bibinfo {volume} {1805}},\ \bibinfo {pages} {033} (\bibinfo {year} {2018}{\natexlab{c}})},\ \Eprint {http://arxiv.org/abs/1711.03929} {arXiv:1711.03929 [astro-ph.CO]} \BibitemShut {NoStop}%
\bibitem [{\citenamefont {Heavens}\ \emph {et~al.}(2017)\citenamefont {Heavens}, \citenamefont {Fantaye}, \citenamefont {Mootoovaloo}, \citenamefont {Eggers}, \citenamefont {Hosenie}, \citenamefont {Kroon},\ and\ \citenamefont {Sellentin}}]{Heavens:2017afc}%
  \BibitemOpen
  \bibfield  {author} {\bibinfo {author} {\bibfnamefont {A.}~\bibnamefont {Heavens}}, \bibinfo {author} {\bibfnamefont {Y.}~\bibnamefont {Fantaye}}, \bibinfo {author} {\bibfnamefont {A.}~\bibnamefont {Mootoovaloo}}, \bibinfo {author} {\bibfnamefont {H.}~\bibnamefont {Eggers}}, \bibinfo {author} {\bibfnamefont {Z.}~\bibnamefont {Hosenie}}, \bibinfo {author} {\bibfnamefont {S.}~\bibnamefont {Kroon}}, \ and\ \bibinfo {author} {\bibfnamefont {E.}~\bibnamefont {Sellentin}},\ }\href@noop {} {\  (\bibinfo {year} {2017})},\ \Eprint {http://arxiv.org/abs/1704.03472} {arXiv:1704.03472 [stat.CO]} \BibitemShut {NoStop}%
\bibitem [{\citenamefont {Jeffreys}(1998)}]{jeffreys1998theory}%
  \BibitemOpen
  \bibfield  {author} {\bibinfo {author} {\bibfnamefont {H.}~\bibnamefont {Jeffreys}},\ }\href {https://books.google.it/books?id=vh9Act9rtzQC} {\emph {\bibinfo {title} {The Theory of Probability}}},\ Oxford Classic Texts in the Physical Sciences\ (\bibinfo  {publisher} {OUP Oxford},\ \bibinfo {year} {1998})\BibitemShut {NoStop}%
\bibitem [{\citenamefont {Kass}\ and\ \citenamefont {Raftery}(1995)}]{Kass95}%
  \BibitemOpen
  \bibfield  {author} {\bibinfo {author} {\bibfnamefont {R.~E.}\ \bibnamefont {Kass}}\ and\ \bibinfo {author} {\bibfnamefont {A.~E.}\ \bibnamefont {Raftery}},\ }\href {\doibase 10.1080/01621459.1995.10476572} {\bibfield  {journal} {\bibinfo  {journal} {Journal of the American Statistical Association}\ }\textbf {\bibinfo {volume} {90}},\ \bibinfo {pages} {773} (\bibinfo {year} {1995})}\BibitemShut {NoStop}%
\bibitem [{\citenamefont {Nesseris}\ and\ \citenamefont {Garcia-Bellido}(2013)}]{Nesseris:2012cq}%
  \BibitemOpen
  \bibfield  {author} {\bibinfo {author} {\bibfnamefont {S.}~\bibnamefont {Nesseris}}\ and\ \bibinfo {author} {\bibfnamefont {J.}~\bibnamefont {Garcia-Bellido}},\ }\href {\doibase 10.1088/1475-7516/2013/08/036} {\bibfield  {journal} {\bibinfo  {journal} {JCAP}\ }\textbf {\bibinfo {volume} {08}},\ \bibinfo {pages} {036} (\bibinfo {year} {2013})},\ \Eprint {http://arxiv.org/abs/1210.7652} {arXiv:1210.7652 [astro-ph.CO]} \BibitemShut {NoStop}%
\bibitem [{\citenamefont {Mendoza-Mart\'\i{}nez}\ \emph {et~al.}(2024)\citenamefont {Mendoza-Mart\'\i{}nez}, \citenamefont {Cervantes-Contreras}, \citenamefont {Trejo-Alonso},\ and\ \citenamefont {Hernandez-Almada}}]{Mendoza-Martinez:2024gbx}%
  \BibitemOpen
  \bibfield  {author} {\bibinfo {author} {\bibfnamefont {M.~L.}\ \bibnamefont {Mendoza-Mart\'\i{}nez}}, \bibinfo {author} {\bibfnamefont {A.}~\bibnamefont {Cervantes-Contreras}}, \bibinfo {author} {\bibfnamefont {J.~J.}\ \bibnamefont {Trejo-Alonso}}, \ and\ \bibinfo {author} {\bibfnamefont {A.}~\bibnamefont {Hernandez-Almada}},\ }\href@noop {} {\  (\bibinfo {year} {2024})},\ \Eprint {http://arxiv.org/abs/2404.18346} {arXiv:2404.18346 [astro-ph.CO]} \BibitemShut {NoStop}%
\bibitem [{\citenamefont {D'Agostino}(2019)}]{DAgostino:2019wko}%
  \BibitemOpen
  \bibfield  {author} {\bibinfo {author} {\bibfnamefont {R.}~\bibnamefont {D'Agostino}},\ }\href {\doibase 10.1103/PhysRevD.99.103524} {\bibfield  {journal} {\bibinfo  {journal} {Phys. Rev. D}\ }\textbf {\bibinfo {volume} {99}},\ \bibinfo {pages} {103524} (\bibinfo {year} {2019})},\ \Eprint {http://arxiv.org/abs/1903.03836} {arXiv:1903.03836 [gr-qc]} \BibitemShut {NoStop}%
\bibitem [{\citenamefont {Moresco}\ \emph {et~al.}(2016{\natexlab{a}})\citenamefont {Moresco}, \citenamefont {Jimenez}, \citenamefont {Verde}, \citenamefont {Cimatti}, \citenamefont {Pozzetti}, \citenamefont {Maraston},\ and\ \citenamefont {Thomas}}]{Moresco16a}%
  \BibitemOpen
  \bibfield  {author} {\bibinfo {author} {\bibfnamefont {M.}~\bibnamefont {Moresco}}, \bibinfo {author} {\bibfnamefont {R.}~\bibnamefont {Jimenez}}, \bibinfo {author} {\bibfnamefont {L.}~\bibnamefont {Verde}}, \bibinfo {author} {\bibfnamefont {A.}~\bibnamefont {Cimatti}}, \bibinfo {author} {\bibfnamefont {L.}~\bibnamefont {Pozzetti}}, \bibinfo {author} {\bibfnamefont {C.}~\bibnamefont {Maraston}}, \ and\ \bibinfo {author} {\bibfnamefont {D.}~\bibnamefont {Thomas}},\ }\href {\doibase 10.1088/1475-7516/2016/12/039} {\bibfield  {journal} {\bibinfo  {journal} {\jcap}\ }\textbf {\bibinfo {volume} {12}},\ \bibinfo {eid} {039} (\bibinfo {year} {2016}{\natexlab{a}})},\ \Eprint {http://arxiv.org/abs/1604.00183} {arXiv:1604.00183} \BibitemShut {NoStop}%
\bibitem [{\citenamefont {Moresco}\ \emph {et~al.}(2016{\natexlab{b}})\citenamefont {Moresco}, \citenamefont {Pozzetti}, \citenamefont {Cimatti}, \citenamefont {Jimenez}, \citenamefont {Maraston}, \citenamefont {Verde}, \citenamefont {Thomas}, \citenamefont {Citro}, \citenamefont {Tojeiro},\ and\ \citenamefont {Wilkinson}}]{Moresco16}%
  \BibitemOpen
  \bibfield  {author} {\bibinfo {author} {\bibfnamefont {M.}~\bibnamefont {Moresco}}, \bibinfo {author} {\bibfnamefont {L.}~\bibnamefont {Pozzetti}}, \bibinfo {author} {\bibfnamefont {A.}~\bibnamefont {Cimatti}}, \bibinfo {author} {\bibfnamefont {R.}~\bibnamefont {Jimenez}}, \bibinfo {author} {\bibfnamefont {C.}~\bibnamefont {Maraston}}, \bibinfo {author} {\bibfnamefont {L.}~\bibnamefont {Verde}}, \bibinfo {author} {\bibfnamefont {D.}~\bibnamefont {Thomas}}, \bibinfo {author} {\bibfnamefont {A.}~\bibnamefont {Citro}}, \bibinfo {author} {\bibfnamefont {R.}~\bibnamefont {Tojeiro}}, \ and\ \bibinfo {author} {\bibfnamefont {D.}~\bibnamefont {Wilkinson}},\ }\href {\doibase 10.1088/1475-7516/2016/05/014} {\bibfield  {journal} {\bibinfo  {journal} {JCAP}\ }\textbf {\bibinfo {volume} {1605}},\ \bibinfo {pages} {014} (\bibinfo {year} {2016}{\natexlab{b}})},\ \Eprint {http://arxiv.org/abs/1601.01701} {arXiv:1601.01701 [astro-ph.CO]} \BibitemShut {NoStop}%
\bibitem [{\citenamefont {Birrer}\ \emph {et~al.}(2019)\citenamefont {Birrer} \emph {et~al.}}]{Birrer:2018vtm}%
  \BibitemOpen
  \bibfield  {author} {\bibinfo {author} {\bibfnamefont {S.}~\bibnamefont {Birrer}} \emph {et~al.},\ }\href {\doibase 10.1093/mnras/stz200} {\bibfield  {journal} {\bibinfo  {journal} {Mon. Not. Roy. Astron. Soc.}\ }\textbf {\bibinfo {volume} {484}},\ \bibinfo {pages} {4726} (\bibinfo {year} {2019})},\ \Eprint {http://arxiv.org/abs/1809.01274} {arXiv:1809.01274 [astro-ph.CO]} \BibitemShut {NoStop}%
\bibitem [{\citenamefont {Colg\'ain}\ \emph {et~al.}(2022)\citenamefont {Colg\'ain}, \citenamefont {Sheikh-Jabbari}, \citenamefont {Solomon}, \citenamefont {Bargiacchi}, \citenamefont {Capozziello}, \citenamefont {Dainotti},\ and\ \citenamefont {Stojkovic}}]{Colgain:2022nlb}%
  \BibitemOpen
  \bibfield  {author} {\bibinfo {author} {\bibfnamefont {E.~O.}\ \bibnamefont {Colg\'ain}}, \bibinfo {author} {\bibfnamefont {M.~M.}\ \bibnamefont {Sheikh-Jabbari}}, \bibinfo {author} {\bibfnamefont {R.}~\bibnamefont {Solomon}}, \bibinfo {author} {\bibfnamefont {G.}~\bibnamefont {Bargiacchi}}, \bibinfo {author} {\bibfnamefont {S.}~\bibnamefont {Capozziello}}, \bibinfo {author} {\bibfnamefont {M.~G.}\ \bibnamefont {Dainotti}}, \ and\ \bibinfo {author} {\bibfnamefont {D.}~\bibnamefont {Stojkovic}},\ }\href {\doibase 10.1103/PhysRevD.106.L041301} {\bibfield  {journal} {\bibinfo  {journal} {Phys. Rev. D}\ }\textbf {\bibinfo {volume} {106}},\ \bibinfo {pages} {L041301} (\bibinfo {year} {2022})},\ \Eprint {http://arxiv.org/abs/2203.10558} {arXiv:2203.10558 [astro-ph.CO]} \BibitemShut {NoStop}%
\bibitem [{\citenamefont {Chevallier}\ and\ \citenamefont {Polarski}(2001)}]{Chevallier:2000qy}%
  \BibitemOpen
  \bibfield  {author} {\bibinfo {author} {\bibfnamefont {M.}~\bibnamefont {Chevallier}}\ and\ \bibinfo {author} {\bibfnamefont {D.}~\bibnamefont {Polarski}},\ }\href {\doibase 10.1142/S0218271801000822} {\bibfield  {journal} {\bibinfo  {journal} {Int. J. Mod. Phys. D}\ }\textbf {\bibinfo {volume} {10}},\ \bibinfo {pages} {213} (\bibinfo {year} {2001})},\ \Eprint {http://arxiv.org/abs/gr-qc/0009008} {arXiv:gr-qc/0009008} \BibitemShut {NoStop}%
\bibitem [{\citenamefont {Linder}(2003)}]{Linder:2002et}%
  \BibitemOpen
  \bibfield  {author} {\bibinfo {author} {\bibfnamefont {E.~V.}\ \bibnamefont {Linder}},\ }\href {\doibase 10.1103/PhysRevLett.90.091301} {\bibfield  {journal} {\bibinfo  {journal} {Phys. Rev. Lett.}\ }\textbf {\bibinfo {volume} {90}},\ \bibinfo {pages} {091301} (\bibinfo {year} {2003})},\ \Eprint {http://arxiv.org/abs/astro-ph/0208512} {arXiv:astro-ph/0208512} \BibitemShut {NoStop}%
\bibitem [{\citenamefont {Barrow}\ \emph {et~al.}(2021)\citenamefont {Barrow}, \citenamefont {Basilakos},\ and\ \citenamefont {Saridakis}}]{Barrow:2020kug}%
  \BibitemOpen
  \bibfield  {author} {\bibinfo {author} {\bibfnamefont {J.~D.}\ \bibnamefont {Barrow}}, \bibinfo {author} {\bibfnamefont {S.}~\bibnamefont {Basilakos}}, \ and\ \bibinfo {author} {\bibfnamefont {E.~N.}\ \bibnamefont {Saridakis}},\ }\href {\doibase 10.1016/j.physletb.2021.136134} {\bibfield  {journal} {\bibinfo  {journal} {Phys. Lett. B}\ }\textbf {\bibinfo {volume} {815}},\ \bibinfo {pages} {136134} (\bibinfo {year} {2021})},\ \Eprint {http://arxiv.org/abs/2010.00986} {arXiv:2010.00986 [gr-qc]} \BibitemShut {NoStop}%
\bibitem [{\citenamefont {Ghoshal}\ and\ \citenamefont {Lambiase}(2021)}]{Ghoshal:2021ief}%
  \BibitemOpen
  \bibfield  {author} {\bibinfo {author} {\bibfnamefont {A.}~\bibnamefont {Ghoshal}}\ and\ \bibinfo {author} {\bibfnamefont {G.}~\bibnamefont {Lambiase}},\ }\href@noop {} {\  (\bibinfo {year} {2021})},\ \Eprint {http://arxiv.org/abs/2104.11296} {arXiv:2104.11296 [astro-ph.CO]} \BibitemShut {NoStop}%
\bibitem [{\citenamefont {Jizba}\ and\ \citenamefont {Lambiase}(2023)}]{Jizba:2023fkp}%
  \BibitemOpen
  \bibfield  {author} {\bibinfo {author} {\bibfnamefont {P.}~\bibnamefont {Jizba}}\ and\ \bibinfo {author} {\bibfnamefont {G.}~\bibnamefont {Lambiase}},\ }\href {\doibase 10.3390/e25111495} {\bibfield  {journal} {\bibinfo  {journal} {Entropy}\ }\textbf {\bibinfo {volume} {25}},\ \bibinfo {pages} {1495} (\bibinfo {year} {2023})},\ \Eprint {http://arxiv.org/abs/2310.19045} {arXiv:2310.19045 [gr-qc]} \BibitemShut {NoStop}%
\bibitem [{\citenamefont {Luciano}(2023)}]{Luciano:2023roh}%
  \BibitemOpen
  \bibfield  {author} {\bibinfo {author} {\bibfnamefont {G.~G.}\ \bibnamefont {Luciano}},\ }\href {\doibase 10.1140/epjc/s10052-023-11499-7} {\bibfield  {journal} {\bibinfo  {journal} {Eur. Phys. J. C}\ }\textbf {\bibinfo {volume} {83}},\ \bibinfo {pages} {329} (\bibinfo {year} {2023})},\ \Eprint {http://arxiv.org/abs/2301.12509} {arXiv:2301.12509 [gr-qc]} \BibitemShut {NoStop}%
\bibitem [{\citenamefont {Luciano}\ and\ \citenamefont {Gin\'e}(2023)}]{Luciano:2022hhy}%
  \BibitemOpen
  \bibfield  {author} {\bibinfo {author} {\bibfnamefont {G.~G.}\ \bibnamefont {Luciano}}\ and\ \bibinfo {author} {\bibfnamefont {J.}~\bibnamefont {Gin\'e}},\ }\href {\doibase 10.1016/j.dark.2023.101256} {\bibfield  {journal} {\bibinfo  {journal} {Phys. Dark Univ.}\ }\textbf {\bibinfo {volume} {41}},\ \bibinfo {pages} {101256} (\bibinfo {year} {2023})},\ \Eprint {http://arxiv.org/abs/2210.09755} {arXiv:2210.09755 [gr-qc]} \BibitemShut {NoStop}%
\bibitem [{\citenamefont {Sen}\ \emph {et~al.}(2022)\citenamefont {Sen}, \citenamefont {Adil},\ and\ \citenamefont {Sen}}]{Sen:2021wld}%
  \BibitemOpen
  \bibfield  {author} {\bibinfo {author} {\bibfnamefont {A.~A.}\ \bibnamefont {Sen}}, \bibinfo {author} {\bibfnamefont {S.~A.}\ \bibnamefont {Adil}}, \ and\ \bibinfo {author} {\bibfnamefont {S.}~\bibnamefont {Sen}},\ }\href {\doibase 10.1093/mnras/stac2796} {\bibfield  {journal} {\bibinfo  {journal} {Mon. Not. Roy. Astron. Soc.}\ }\textbf {\bibinfo {volume} {518}},\ \bibinfo {pages} {1098} (\bibinfo {year} {2022})},\ \Eprint {http://arxiv.org/abs/2112.10641} {arXiv:2112.10641 [astro-ph.CO]} \BibitemShut {NoStop}%
\bibitem [{\citenamefont {Akarsu}\ \emph {et~al.}(2023)\citenamefont {Akarsu}, \citenamefont {Di~Valentino}, \citenamefont {Kumar}, \citenamefont {Nunes}, \citenamefont {Vazquez},\ and\ \citenamefont {Yadav}}]{Akarsu:2023mfb}%
  \BibitemOpen
  \bibfield  {author} {\bibinfo {author} {\bibfnamefont {O.}~\bibnamefont {Akarsu}}, \bibinfo {author} {\bibfnamefont {E.}~\bibnamefont {Di~Valentino}}, \bibinfo {author} {\bibfnamefont {S.}~\bibnamefont {Kumar}}, \bibinfo {author} {\bibfnamefont {R.~C.}\ \bibnamefont {Nunes}}, \bibinfo {author} {\bibfnamefont {J.~A.}\ \bibnamefont {Vazquez}}, \ and\ \bibinfo {author} {\bibfnamefont {A.}~\bibnamefont {Yadav}},\ }\href@noop {} {\  (\bibinfo {year} {2023})},\ \Eprint {http://arxiv.org/abs/2307.10899} {arXiv:2307.10899 [astro-ph.CO]} \BibitemShut {NoStop}%
\bibitem [{\citenamefont {Linder}(2006)}]{Linder06}%
  \BibitemOpen
  \bibfield  {author} {\bibinfo {author} {\bibfnamefont {E.~V.}\ \bibnamefont {Linder}},\ }\href {\doibase 10.1103/PhysRevD.73.063010} {\bibfield  {journal} {\bibinfo  {journal} {Phys. Rev.}\ }\textbf {\bibinfo {volume} {D73}},\ \bibinfo {pages} {063010} (\bibinfo {year} {2006})},\ \Eprint {http://arxiv.org/abs/astro-ph/0601052} {arXiv:astro-ph/0601052 [astro-ph]} \BibitemShut {NoStop}%
\bibitem [{\citenamefont {Park}\ \emph {et~al.}(2024)\citenamefont {Park}, \citenamefont {de~Cruz~Perez},\ and\ \citenamefont {Ratra}}]{Park:2024jns}%
  \BibitemOpen
  \bibfield  {author} {\bibinfo {author} {\bibfnamefont {C.-G.}\ \bibnamefont {Park}}, \bibinfo {author} {\bibfnamefont {J.}~\bibnamefont {de~Cruz~Perez}}, \ and\ \bibinfo {author} {\bibfnamefont {B.}~\bibnamefont {Ratra}},\ }\href@noop {} {\  (\bibinfo {year} {2024})},\ \Eprint {http://arxiv.org/abs/2405.00502} {arXiv:2405.00502 [astro-ph.CO]} \BibitemShut {NoStop}%
\bibitem [{\citenamefont {Amendola}\ \emph {et~al.}(2018)\citenamefont {Amendola} \emph {et~al.}}]{Amendola:2016saw}%
  \BibitemOpen
  \bibfield  {author} {\bibinfo {author} {\bibfnamefont {L.}~\bibnamefont {Amendola}} \emph {et~al.},\ }\href {\doibase 10.1007/s41114-017-0010-3} {\bibfield  {journal} {\bibinfo  {journal} {Living Rev. Rel.}\ }\textbf {\bibinfo {volume} {21}},\ \bibinfo {pages} {2} (\bibinfo {year} {2018})},\ \Eprint {http://arxiv.org/abs/1606.00180} {arXiv:1606.00180 [astro-ph.CO]} \BibitemShut {NoStop}%
\bibitem [{\citenamefont {Mellier}\ \emph {et~al.}(2024)\citenamefont {Mellier} \emph {et~al.}}]{Euclid:2024yrr}%
  \BibitemOpen
  \bibfield  {author} {\bibinfo {author} {\bibfnamefont {Y.}~\bibnamefont {Mellier}} \emph {et~al.} (\bibinfo {collaboration} {Euclid, Isaac Newton Group of Telescopes, Apartado 321, 38700 Santa Cruz de La Palma, Spain, Astrophysics Group, Blackett Laboratory, Imperial College London, London SW7 2AZ, UK, ASDEX Upgrade Team, Max-Planck-Institut fur Plasmaphysik, Garching, Germany, MTA-CSFK Lend\"ulet Large-Scale Structure Research Group, Konkoly-Thege Mikl\'os \'ut 15-17, H-1121 Budapest, Hungary, NOVA optical infrared instrumentation group at ASTRON, Oude Hoogeveensedijk 4, 7991PD, Dwingeloo, The Netherlands}),\ }\href@noop {} {\  (\bibinfo {year} {2024})},\ \Eprint {http://arxiv.org/abs/2405.13491} {arXiv:2405.13491 [astro-ph.CO]} \BibitemShut {NoStop}%
\bibitem [{\citenamefont {Ivezic}\ \emph {et~al.}(2019)\citenamefont {Ivezic} \emph {et~al.}}]{Ivezic:2008fe}%
  \BibitemOpen
  \bibfield  {author} {\bibinfo {author} {\bibfnamefont {e.}~\bibnamefont {Ivezic}} \emph {et~al.} (\bibinfo {collaboration} {LSST}),\ }\href {\doibase 10.3847/1538-4357/ab042c} {\bibfield  {journal} {\bibinfo  {journal} {Astrophys. J.}\ }\textbf {\bibinfo {volume} {873}},\ \bibinfo {pages} {111} (\bibinfo {year} {2019})},\ \Eprint {http://arxiv.org/abs/0805.2366} {arXiv:0805.2366 [astro-ph]} \BibitemShut {NoStop}%
\end{thebibliography}%

\end{document}